\documentclass[useAMS,usenatbib,a4paper]{mn2e}

\usepackage{amssymb}
\usepackage{amsmath}
\usepackage{graphicx}
\usepackage{subfigure}
\usepackage{booktabs}
\usepackage{txfonts}
\usepackage{ifpdf}
\ifpdf
  \usepackage[pdftex]{hyperref}
\else
  \usepackage[ps2pdf,colorlinks=true]{hyperref}
\fi

\newif\ifbw
\bwfalse

\newcommand{\eqn}[1]{(#1)}

\newcommand{\tbl}[1]{Table~#1}

\newcommand{\fig}[1]{Fig.~#1}

\newcommand{\sectn}[1]{Sec.~#1}

\newcommand{\eg}{\mbox{\it e.g.}}
\newcommand{\ie}{\mbox{\it i.e.}}

\newcommand{\cmbtext}{{cosmic microwave background}}

\newcommand{\ska}{{SKA}}
\newcommand{\skatext}{Square Kilometre Array}

\newcommand{\healpix}{{\tt HEALPix}}

\newcommand{\stwo}{{\tt S2}}

\newcommand{\comb}{{\tt COMB}}

\newcommand{\lambdaarch}{{LAMBDA}}
\newcommand{\lambdaarchtext}{{Legacy Archive for Microwave Background Data Analysis}}

\newcommand{\fwhm}{{FWHM}}
\newcommand{\fwhmtext}{{full-width-half-maximum}}
\newcommand{\snr}{{\rm SNR}}
\newcommand{\snrp}{\ensuremath{\snr_{\pind}}}
\newcommand{\snrs}{\ensuremath{\snr_{\sind}}}

\newcommand{\fov}{FOV}
\newcommand{\wfov}{WFOV}
\newcommand{\wterm}{\mbox{$w$-term}}

\newcommand{\spcend}{\ensuremath{\:}}
\newcommand{\img}{\ensuremath{{\rm i}}}
 
\newcommand{\reals}{\ensuremath{\mathbb{R}}}

\newcommand{\complex}{\ensuremath{\mathbb{C}}}

\newcommand{\sphere}{\ensuremath{{\mathrm{S}^2}}}

\newcommand{\disk}{\ensuremath{{D^2}}}
\newcommand{\vect}[1]{\ensuremath{\mbox{\boldmath ${#1}$}}}

\newcommand{\dx}{\ensuremath{\mathrm{\,d}}}
\newcommand{\dmu}[1]{\ensuremath{\dx \Omega(#1)}}

\newcommand{\w}{\ensuremath{w}}

\newcommand{\sa}{\ensuremath{\vect{\hat{s}}}}
\newcommand{\saa}{\ensuremath{\theta}}
\newcommand{\sab}{\ensuremath{\varphi}}
\newcommand{\sas}{\ensuremath{\saa, \sab}}

\newcommand{\el}{\ensuremath{\ell}}

\newcommand{\elmax}{\ensuremath{{\el_{\rm max}}}}

\newcommand{\pind}{\ensuremath{{\rm p}}}
\newcommand{\sind}{\ensuremath{{\rm s}}}

\newcommand{\tfov}{\ensuremath{\saa_{\rm \fov}}}

\renewcommand{\exp}[1]{\ensuremath{{\rm e}^{#1}}}

\newcommand{\nside}{\ensuremath{{N_{\rm{side}}}}}
\newcommand{\npix}{\ensuremath{{N_{\rm{pix}}}}}
\newcommand{\nplane}{\ensuremath{{N_{\pind}}}}
\newcommand{\nsphere}{\ensuremath{{N_{\sind}}}}

\newcommand{\noise}{\ensuremath{n}}

\newcommand{\sao}{\ensuremath{\sa_0}}

\newcommand{\salcoordfree}{\ensuremath{\vect{\tau}}}
\newcommand{\vis}{\ensuremath{y}}
\newcommand{\visdis}{\ensuremath{\vect{y}}}
\newcommand{\beam}{\ensuremath{A}}

\newcommand{\beamdis}{\ensuremath{\vect{\beam}}}
\newcommand{\im}{\ensuremath{x}}
\newcommand{\imdis}{\ensuremath{\vect{\im}}}
\newcommand{\noisedis}{\ensuremath{\vect{\noise}}}
\newcommand{\imp}{\ensuremath{\im_{\pind}}}
\newcommand{\ims}{\ensuremath{\im_{\sind}}}

\newcommand{\nim}{\ensuremath{n}}
\newcommand{\visvect}{\ensuremath{\vect{\vis}}}

\newcommand{\impvect}{\ensuremath{\vect{\imp}}}
\newcommand{\imsvect}{\ensuremath{\vect{\ims}}}

\newcommand{\impvectrecon}{\ensuremath{\vect{\imp}^{\star}}}
\newcommand{\imsvectrecon}{\ensuremath{\vect{\ims}^{\star}}}
\newcommand{\nimvect}{\ensuremath{\vect{\nim}}}
\newcommand{\bu}{\ensuremath{u}}
\newcommand{\bv}{\ensuremath{v}}
\newcommand{\bw}{\ensuremath{w}}
\newcommand{\bwd}{\ensuremath{w_{\rm d}}}
\newcommand{\buvect}{\ensuremath{\vect{\bu}}}
\newcommand{\buvectfull}{\ensuremath{\vect{b}}}
\newcommand{\bumax}{\ensuremath{\bu_{\rm max}}}
\newcommand{\bumaxfov}{\ensuremath{\bumax}}
\newcommand{\elmaxfov}{\ensuremath{\elmax}}
\newcommand{\lx}{\ensuremath{l}}
\newcommand{\mx}{\ensuremath{m}}
\newcommand{\nx}{\ensuremath{n}}
\newcommand{\lxvect}{\ensuremath{\vect{\lx}}}
\newcommand{\lxvectdis}{\ensuremath{\vect{\lx}_{\ispar}}}

\newcommand{\chirp}{\ensuremath{C}}
\newcommand{\chirpfull}{\ensuremath{\chirp(\| \lxvect \|_2)}}

\newcommand{\chirpfullb}{\ensuremath{\chirp_{1}(\| \lxvect \|_2)}}
\newcommand{\chirpdis}{\ensuremath{\vect{\chirp}}}

\newcommand{\chirpfullbl}{\ensuremath{\bumax^{\chirp^{(\bw)}}}}
\newcommand{\chirpfullbbl}{\ensuremath{\bumax^{\chirp^{(\bw)}_1}}}

\newcommand{\objsize}{\ensuremath{\sigma_{\rm S}}}
\newcommand{\plevel}{\ensuremath{p}}
\newcommand{\pfov}{\ensuremath{L}}

\newcommand{\nmeas}{\ensuremath{M}}
\newcommand{\ndim}{\ensuremath{N}}
\newcommand{\sparsity}{\ensuremath{K}}
\newcommand{\ssmat}{\ensuremath{\Theta}}
\newcommand{\ispar}{\ensuremath{i}}
\newcommand{\isens}{\ensuremath{r}}
\newcommand{\sparatom}{\ensuremath{\vect{\psi}_\ispar}}
\newcommand{\sensatom}{\ensuremath{\vect{\phi}_\isens}}
\newcommand{\sparmat}{\ensuremath{\Psi}}
\newcommand{\sensmat}{\ensuremath{\Phi}}

\newcommand{\sensmatp}{\ensuremath{\Phi_{\pind}}}
\newcommand{\sensmats}{\ensuremath{\Phi_{\sind}}}
\newcommand{\coherence}{\ensuremath{\mu}}
\newcommand{\ccoherence}{\ensuremath{\nu}}

\newcommand{\ccoherences}{\ensuremath{\ccoherence _{\rm s}}}

\newcommand{\ccoherencep}{\ensuremath{\ccoherence _{\rm p}}}

\newcommand{\opmask}{\ensuremath{\mathbfss{M}}}
\newcommand{\opbeam}{\ensuremath{\mathbfss{A}}}
\newcommand{\opfourier}{\ensuremath{\mathbfss{F}}}
\newcommand{\opwhite}{\ensuremath{\mathbfss{W}}}
\newcommand{\opchirp}{\ensuremath{\mathbfss{C}}}

\newcommand{\opproj}{\ensuremath{\mathbfss{P}}}

\renewcommand{\vect}{\bmath}

\title[Compressed sensing for wide-field imaging]
   {Compressed sensing for wide-field radio interferometric imaging}

\author[J.~D.~McEwen \& Y.~Wiaux]
  {J.~D.~McEwen$^{1}$\thanks{E-mail: jason.mcewen@epfl.ch}
   and Y.~Wiaux$^{1,2,3}$\\ %\newauthor\\
  $^1$Institute of Electrical Engineering, Ecole Polytechnique F{\'e}d{\'e}rale de Lausanne (EPFL),
      CH-1015 Lausanne, Switzerland\\
  $^2$Institute of Bioengineering, Ecole Polytechnique F{\'e}d{\'e}rale de Lausanne (EPFL),
      CH-1015 Lausanne, Switzerland\\
  $^3$Department of Radiology and Medical Informatics, University of Geneva (UniGE), 
      CH-1211 Geneva, Switzerland\\
}

\date{Accepted 2010 December 16.  Received 2010 November 25; in original form 2010 October 18}
\pagerange{\pageref{sec:intro}--\pageref{lastpage}} 
\pubyear{2010}

\def\LaTeX{L\kern-.36em\raise.3ex\hbox{a}\kern-.15em
    T\kern-.1667em\lower.7ex\hbox{E}\kern-.125emX}

\begin{document}
\maketitle

%=============================================================================
\begin{abstract}
For the next generation of radio interferometric telescopes it is of paramount importance to incorporate wide field-of-view (\wfov) considerations in interferometric imaging, otherwise the fidelity of reconstructed images will suffer greatly.  We extend compressed sensing techniques for interferometric imaging to a \wfov\ and recover images in the spherical coordinate space in which they naturally live, eliminating any distorting projection.  The effectiveness of the spread spectrum phenomenon, highlighted recently by one of the authors, is enhanced when going to a \wfov, while sparsity is promoted by recovering images directly on the sphere.  Both of these properties act to improve the quality of reconstructed interferometric images.  We quantify the performance of compressed sensing reconstruction techniques through simulations, highlighting the superior reconstruction quality achieved by recovering interferometric images directly on the sphere rather than the plane.
\end{abstract}

%=============================================================================

%=============================================================================
\section{Introduction}
\label{sec:intro}
%=============================================================================

Incorporating wide field-of-view (\wfov) contributions in the images reconstructed from radio interferometric observations is becoming increasingly important.  Next-generation radio interferometers, such as the \skatext\footnote{\url{http://www.skatelescope.org/}} (\ska) \citep{carilli:2004}, will inherently observe very large fields of view about the pointing direction of the telescope.  Wide fields introduce two important distinctions to standard interferometric imaging: firstly, interferometric images are inherently spherical and planar projections necessarily introduce distortions; and, secondly, non-zero baseline components in the pointing direction of the telescope must be taken into account.  If these contributions are ignored, the fidelity of reconstructed images will suffer greatly.

\wfov\ contributions have been considered by \citet{mcewen:2008:fsi} in simulating the visibilities observed by an interferometer.  Full-sky interferometric formalisms were derived using a number of different signal representations, including representations in real, spherical harmonic and wavelet spaces.  Real and spherical harmonic space representations were shown to be numerically infeasible for simulating realistic observations, while a fast wavelet space method was developed, reducing the computational cost considerably and rendering realistic simulations feasible.  However, the forward problem only was considered in this work.  Very recently \citet{carozzi:2009} developed a generalised radio interferometric measurement equation that is valid for partially polarised sources over a \wfov, however this framework has not yet been applied in practice.  The inverse \wfov\ imaging problem has traditionally been tackled by faceting the sky into a number of regions which are sufficiently small that standard Fourier imaging is possible \citep{cornwell:1992}.  More recently, the $w$-projection algorithm has been developed by \citet{cornwell:2008:w} to incorporate  \wfov\ contributions by applying the modulating so-called \wterm\ that appears in the interferometric integral as a convolution in the Fourier plane.  This provides an order of magnitude speed enhancement over traditional facet-based approaches.  Nevertheless, both of these methods recover {\em planar} images in the space of directional cosines; this necessarily distorts the image.  None of the current methods recover images in the spherical coordinate space in which they live.  In this article we recover {\em spherical} interferometric images parameterised by colatitude and longitude on the celestial sphere.  By recovering images defined directly on the sphere, and thereby eliminating the distortion due to a projection to the plane, the performance of reconstruction is enhanced.  We develop reconstruction algorithms in the context of the theory of compressed sensing.  

Compressed sensing \citep{candes:2006a,candes:2006b,candes:2006c,donoho:2006,baraniuk:2007} is a recent development in the field of information theory, which goes beyond the usual Nyquist-Shannon sampling theorem.  It relies on the fact than many signals in Nature are {\em sparse} (or approximately so) and may be represented in a basis requiring many fewer non-zero coefficients than the dimensionality of the signal itself.  Compressed sensing theory shows that a sparse signal may be recovered from many fewer measurements than Nyquist-Shannon sampling would suggest and thus aims to merge data acquisition and compression.  These results also hold for signals that are approximately sparse only (\ie\ signals which contain many coefficients of small but non-zero value), so-called {\em compressible} signals.  

The first application of compressed sensing to radio interferometry was performed by \citet{wiaux:2009:cs}, where the problem of image reconstruction from incomplete visibility measurements was considered.  \citet{wiaux:2009:cs} demonstrated the versatility of the approach, through the ability to incorporate additional signal priors easily, and its superiority relative to standard interferometric imaging techniques, such as CLEAN \citep{hogbom:1974}.  The spread spectrum phenomenon, which arises by partially relaxing the small field-of-view (\fov) assumption and including a first order \wterm, was introduced by \citet{wiaux:2009:ss}, enhancing the performance of image reconstruction for sparsity bases that are not maximally incoherent with the measurement basis (see \sectn{\ref{sec:background:cs}} for a review of sparsity and measurement bases and their coherence).  Furthermore, a compressed sensing based approach was developed and evaluated by \citet{wiaux:2010:csstring} to recover the signal induced by cosmic strings in the \cmbtext, exploiting the sparse nature of line-like discontinuities due to the string signal.  All of these works consider uniformly random and discrete visibility coverage in order to remain as close to the theory of compressed sensing as possible.  First steps towards more realistic visibility coverages have been taken by \citet{suksmono:2009} and \citet{wenger:2010}, who consider coverages due to specific interferometer configurations but which remain discrete.  These preliminary works suggest that the performance of compressed sensing reconstructions is unlikely to deteriorate significantly for more realistic visibility coverages. 

In this article we generalise the compressed sensing imaging techniques developed by \citet{wiaux:2009:cs} and \citet{wiaux:2009:ss} to a \wfov.  To do so we recover interferometric images defined directly on the sphere, rather than a tangent plane.  Recovering images in the space in which the signal naturally lives eliminates any distortion due to projection.  Furthermore, projection effects also act to hamper the sparsity of the signal (\ie\ act to make it less sparse), impeding the performance of compressed sensing based reconstruction.  To remain close to the theoretical compressed sensing framework we follow \citet{wiaux:2009:cs} and \citet{wiaux:2009:ss} by considering uniformly random and discrete planar visibility coverage.  We also assume non-zero but constant baseline components in the pointing direction of the telescope (\ie\ non-zero but constant \bw, where \bw\ is defined explicitly in \sectn{\ref{sec:background:ri}}), in order to study the enhancement of reconstruction quality due to the spread spectrum phenomenon.  This assumption allows us to discard considerations related to specific interferometer configurations and to study the impact of the spread spectrum phenomenon at light computational load.  As we shall see, the effectiveness of the spread spectrum phenomenon is improved in the \wfov\ setting considered.  Due to all of these considerations, the quality of reconstruction on the sphere is enhanced considerably relative to planar reconstructions.

The remainder of this article is structured as follows.  In \sectn{\ref{sec:background}} we review radio interferometric imaging in the context of compressed sensing.  In \sectn{\ref{sec:wfov}} we generalise these techniques to a \wfov.  Simulations are presented in \sectn{\ref{sec:recon}} to evaluate the performance of the compressed sensing based reconstruction of spherical images compared to planar images.  Concluding remarks are made in \sectn{\ref{sec:conclusions}}.

%=============================================================================
\section{Planar interferometric imaging}
\label{sec:background}
%=============================================================================

In this section we review radio interferometric imaging in the context of compressed sensing.  Radio interferometry is reviewed in the \wfov\ setting, before small \fov\ assumptions are discussed explicitly.  A brief review of compressed sensing is given, highlighting various reconstruction techniques and the importance of sparsity and coherence on reconstruction performance.  Finally, these frameworks are merged and compressed sensing techniques for radio interferometric imaging are discussed.

%=============================================================================
\subsection{Radio interferometry}
\label{sec:background:ri}

In order to image a region on the sky, all radio telescopes of an interferometric array are oriented towards the same pointing direction $\sao$ on the unit celestial sphere $\sphere$.  The \fov\ observed by the interferometer is limited by the primary beam of the telescope $\beam(\salcoordfree)$, where the beam is defined relative to the pointing direction, \ie\ $\salcoordfree \equiv \sa - \sao$ for arbitrary directions \sa.  The sky intensity to be imaged $\im(\salcoordfree)$ is defined in the same coordinates.  The complex visibility measured by each telescope pair of the interferometer is given by \citep{thompson:2001}
\begin{equation}
\label{eqn:vis1}
\vis(\buvectfull) = \int_\sphere 
\beam(\salcoordfree) \: \im(\salcoordfree) \: 
\exp{-\img 2 \pi \buvectfull \cdot \salcoordfree } \:
\dmu{\salcoordfree}
\spcend ,
\end{equation}
where $\dmu{\sa}=\sin \saa \dx \saa \dx \sab$ is the usual rotation invariant measure on the sphere and $(\sas)$ denote the spherical coordinates of $\sa$, with colatitude $\saa \in [0,\pi]$ and longitude $\sab \in [0,2\pi)$.  The measured visibility depends only on the relative positions between telescope pairs, denoted in units of wavelengths by the baseline vector $\buvectfull=(\bu,\bv,\bw)$.  
%The baseline $\buvectfull$ is in a local coordinate system centred on the Earth and relative to $\sao$.  
%We define this local coordinate system by the orthogonal, right handed set of unit vectors $\{\eone,\etwo,\ethree\}$, where $\ethree$ is identified with $\sao$.  
As the Earth rotates relative to the celestial sphere the interferometer tracks the source position as it traverses the sky, with each position corresponding to a rotation of the baseline.  In this manner, visibility measurements are accumulated for various baselines, with each antenna pair generating an elliptical track of baseline values over the course of the observation.  Visibility coverage is therefore incomplete.

% \begin{figure}
% \centering
% \includegraphics[width=70mm]{figures/infer_obs_psfrag}
% \caption{Geometry of an observation of an extended source centred at
%   the interferometer pointing direction $\sao$, in the coordinate system 
%   of the celestial sky (defined by the orthogonal right handed set of unit vectors 
%   $\{\none,\ntwo,\nthree\}$).  The area element
%   $\dmun$ represents the contribution to the visibility integral from
%   point $\sa$.  The full visibility is obtained by summing all such
%   contributions over the sky.  Note that the sky has been mapped onto
%   the unit celestial sphere, hence $\sa$ and $\sao$ are unit vectors.}
% \label{fig:infer_obs}
% \end{figure}

The beam and sky intensity are typically expressed in the same Cartesian coordinate system as the baseline, which is centred on the Earth and aligned with \sao.  Consider the coordinates $(\lx,\mx,\nx)$ of \sa\ in this system, noting 
$\nx \equiv \nx(\lxvect) = ( 1-\| \lxvect \|_2 ^2)^{1/2}$ with $\lxvect=(\lx,\mx)$ and $\| \lxvect \|_2 ^2 = \lx^2 + \mx^2$.  The primary beam and sky intensity may then be seen as functions of $\lxvect$.  Following this change of coordinates the visibility integral of \eqn{\ref{eqn:vis1}} reads
\begin{equation}
\label{eqn:vis2}
\vis(\buvect, \bw) = \int_\disk 
\beam(\lxvect) \: \im(\lxvect) \: 
\exp{-\img 2 \pi [ \buvect \cdot \lxvect + \bw \, (\nx(\lxvect) - 1) ] } \:
\frac{\dx^2 \lxvect}{\nx (\lxvect)}
\spcend ,
\end{equation}
where $\buvect=(\bu,\bw)$ and we have noted that $\salcoordfree=(\lxvect,\nx(\lxvect)-1)$.  Through the change of coordinates the invariant measure on the sphere becomes $\dmu{\sa} = \nx^{-1}(\lxvect) \dx^2 \lxvect$, where $\dx^2 \lxvect = \dx\lx\dx\mx$ is the canonical invariant measure on the plane, and the integration is now performed over the unit disk $\disk$.  Note that \eqn{\ref{eqn:vis2}} remains general and does not rely on any small \fov\ assumption.  However, the directional cosines  $\lxvect=(\lx,\mx)$ express a projection onto the equatorial plane of the signal, which is inherently defined on the celestial sphere.  This projection is trivial in the continuous setting but will become important once we reach the discrete setting required for interferometric imaging.

Often small \fov\ assumptions are made, in which case it is convenient to represent the so-called \wterm\ 
\begin{equation}
\label{eqn:chirpfull}
\chirpfull \equiv \exp{
\img 2 \pi \bw \bigl ( 1 - \sqrt{1-\| \lxvect \|_2 ^2} \bigr )
}
\end{equation}
explicitly, from which it follows
\begin{equation}
\label{eqn:vis3}
\vis(\buvect, \bw) = \int_\disk 
\beam(\lxvect) \: \im(\lxvect) \: \chirpfull \:
\exp{-\img 2 \pi \buvect \cdot \lxvect } \:
\frac{\dx^2 \lxvect}{\nx (\lxvect)}
\spcend .
\end{equation}
Two types of approximation regarding small \fov s are used to relate \eqn{\ref{eqn:vis3}} to the Fourier transform of the beam-modulated intensity $\beam(\lxvect) \im(\lxvect)$.  Firstly, the assumption $\| \lxvect \|_2 ^2 \ll 1$ implies $\nx (\lxvect) \simeq 1$, so that the Jacobian term $\nx ^{-1}(\lxvect)$ is reduced to unity.  Alternatively, $\nx ^{-1}(\lxvect)$ may be absorbed into the beam.  Secondly, various assumptions are made to approximate the \wterm.  A zeroth order approximation of the \wterm\ is made by assuming $\| \lxvect \|_2 ^2 \: \bw \ll 1$, so that $\chirpfull \simeq  1$.  When incorporating both approximations, \eqn{\ref{eqn:vis3}} reduces to the Fourier transform and standard Fourier imaging may be used to recover images from measured visibilities.  Alternatively, a first order approximation $\| \lxvect \|_2 ^4 \: \bw \ll 1$ reduces the \wterm\ to $\chirpfull \simeq  \exp{ \img \pi \bw \| \lxvect \|_2 ^2 } \equiv \chirpfullb$.  This assumption gives rise to the linear chirp modulation responsible for the spread spectrum phenomenon \citep{wiaux:2009:ss}.  Although not considered here, direction dependent beam effects may also introduce a phase modulation, which would provide an alternative source of the spread spectrum phenomenon.  Since we intend to consider \wfov s, no small \fov\ assumptions will be made herein; we include the Jacobian term $\nx ^{-1}(\lxvect)$ and consider the full \wterm\ given by \eqn{\ref{eqn:chirpfull}}.

%=============================================================================
\subsection{Compressed sensing}
\label{sec:background:cs}

Compressed sensing \citep{candes:2006a,candes:2006b,candes:2006c,donoho:2006,baraniuk:2007} is concerned with the recovery of sparse or compressible signals from a small number of measurements.  The signal to be recovered is defined by its Nyquist-Shannon sampling, denoted by the vector $\vect{x}\in\complex^\ndim$, and is assumed to be sparse in some orthogonal basis $\{ \sparatom \}_{1\leq \ispar \leq \ndim}$, where $\sparatom \in \complex^\ndim$, $\forall \ispar$.  The signal may then be represented by its coefficients $\vect{\alpha}\in\complex^\ndim$ in this basis:
$\vect{x} = \sparmat \vect{\alpha}$, 
where $\sparmat \in \complex^{\ndim \times \ndim}$ is the $\ndim \times \ndim$ matrix with columns $\sparatom$.  Formally, $\vect{x}$ is said to be sparse or compressible if $\vect{\alpha}$ contains $\sparsity \ll \ndim$ non-zero or significant elements respectively.  Hereafter, we refer to enhancing (hampering) sparsity as synonymous with decreasing (increasing) the sparsity value $\sparsity$.
Measurement of $\vect{x}$ is assumed to be made by the projection onto measurement basis vectors $\{ \sensatom \}_{1\leq \isens \leq \nmeas}$, for $\nmeas$ measurements, belonging to an orthogonal sensing basis where $\sensatom \in \complex^\ndim$, $\forall \isens$.  This is a very flexible framework which allows a wide range of acquisition procedures to be modelled.  The vector of measurements $\vect{y} \in \complex^{\nmeas}$ may then be expressed by 
\begin{equation}
\label{eqn:measurement}
\vect{y} = \sensmat \vect{x} + \vect{n} = \sensmat \sparmat \vect{\alpha} + \vect{n} 
\spcend ,
\end{equation}
where $\sensmat \in \complex^{\nmeas \times \ndim}$ is the $\nmeas \times \ndim$ matrix withs rows $\sensatom$ and \mbox{$\vect{n} \in \complex^\nmeas$} represents noise.  Compressed sensing suggests that $\vect{x}$ can be recovered with a number of measurements \mbox{$\nmeas \sim \sparsity \ll \ndim$}.  However, recovering $\vect{x}$ in this setting involves solving the inverse problem \eqn{\ref{eqn:measurement}}, which becomes ill-posed for $\nmeas < \ndim$.

Compressed sensing techniques are generally based on global minimisation problems, which are solved by greedy algorithms or convex non-linear optimisation algorithms.  The ill-posed inverse problem described above can be defined by a constrained optimisation problem explicitly regularised by a sparsity or compressibility prior.  This results in the Basis Pursuit denoising (BP) problem\footnote{In the absence of noise the problem is simply called Basis Pursuit.}, which consists of minimising the $\el_1$-norm of the coefficients of the signal in the sparsity basis $\|\vect{\alpha} \|_1$ under a constraint on the $\el_2$-norm of the residual noise $\| \vect{y} - \sensmat \sparmat \vect{\alpha}\|_2$:
\begin{equation}
\label{eqn:min_bp}
\min_{\vect{\alpha}} \| \alpha \|_{1} \:\: \mbox{such that} \:\:
\| \vect{y} - \sensmat \sparmat \vect{\alpha}\|_2 \leq \epsilon
\spcend .
\end{equation}
Recall that the $\el_1$-norm is simply given by the sum of the absolute values of the elements of a vector $\|\vect{\alpha} \|_1=\sum_{i=1}^{\ndim} | \alpha_i |$, whereas the $\el_2$-norm is the standard norm defined previously.  The constraint $\epsilon$ may be related to a residual noise level estimator.  Assuming independent identically distributed Gaussian noise, a residual noise level estimator is given by twice the negative logarithm of the likelihood associated with the candidate reconstruction, which follows a $\chi^2$-distribution with $2M$ degrees of freedom.  The measurement constraint $\epsilon$ may then be chosen to correspond to some $(100\plevel)$th percentile of the $\chi^2$-distribution, for $0\leq \plevel \leq 1$ \citep{wiaux:2009:cs}.  The BP problem may be solved by the application of non-linear, iterative convex optimisation algorithms (\eg \citealt{combettes:2007}).  If the solution of the optimisation problem is denoted $\vect{\alpha}^\star$, then the signal is reconstructed through the synthesis $\vect{x}^{\star_{\rm BP}} = \sparmat \vect{\alpha}^\star$.

The BP denoising problem is appropriate for signals sparse or compressible in a given basis.  However, many signals in Nature are also sparse or compressible in the magnitude of their gradient, in which case the Total Variation (TV) minimisation problem applies \citep{candes:2006a}.  The TV problem involves replacing the \mbox{$\el_1$-norm} sparsity prior in the BP problem with a prior on the TV norm of the signal itself:
\begin{equation}
\label{eqn:min_tv}
\min_{\vect{x}} \| \vect{x} \|_{\rm TV} \:\: \mbox{such that} \:\:
\| \vect{y} - \sensmat \vect{x}\|_2 \leq \epsilon
\spcend ,
\end{equation}
where the TV norm is defined by the $\el_1$-norm of the gradient of the signal $\| \vect{x} \|_{\rm TV} = \| \nabla x \|_1$ \citep{rudin:1992,chambolle:2004}.  The TV problem may also be solved by the application of non-linear, iterative convex optimisation algorithms (\eg\ \citealt{chambolle:2004,durand:2010}).  The signal is directly recovered from the solution to the optimisation problem, denoted $\vect{x}^{\star_{\rm TV}}$.

Finally, we review the two fundamental criteria that drive the performance of compressed sensing reconstructions.  The first has already been central to our discussion of compressed sensing: sparsity.  The more sparse a signal the fewer measurements required to recover it, or similarly, the better the reconstruction quality for a given number of measurements.  In addition, the matrix $\ssmat = \sensmat \sparmat$ must satisfy the restricted isometry property (RIP) \citep{candes:2006a,candes:2006b,candes:2006c} to ensure accurate recovery.  It has been shown that this property can be satisfied by acquiring random measurements (\ie\ random visibility coverage in the terminology of interferometry), if the measurement and sparsity bases are {\em incoherent}.  Coherence is therefore the second criterion driving reconstruction performance; as the coherence between the two bases increases, the reconstruction performance degrades.  \mbox{Incoherence} ensures that the rows of $\sensmat$ cannot sparsely represent the columns of $\sparmat$, ensuring that signal content is sufficiently probed by random measurements.  Note that coherence is only defined strictly in the presence of a sparsity basis, hence it cannot be studied for the TV problem (nevertheless, approximate coherence arguments may still be made in this setting to gain intuition).  The {\em mutual coherence} between the measurement and sparsity bases is defined by the maximum absolute inner product of all combinations of their normalised basis vectors:
\begin{equation*}
\coherence(\sensmat, \sparmat) = \max_{\isens,\ispar} \: \bigl | \: \langle \sensatom \: | \: \sparatom \rangle \: \bigr |
\spcend ,
\end{equation*}
with the incoherence given by the reciprocal $\coherence ^{-1} (\sensmat, \sparmat)$.  Note that the Dirac and Fourier bases are maximally incoherent.  The number of measurements required to satisfy the RIP and ensure accurate recovery, satisfies the bound
\begin{equation*}
\nmeas \geq c \: \sparsity \: \coherence^2(\sensmat, \sparmat) \: \ndim \log^4 \ndim
\spcend ,
\end{equation*}
for some constant $c$ (note that the bound is not tight).  Although useful for theoretical and intuitive considerations, the mutual coherence is a blunt numerical instrument as it captures only the most extreme inner product between the measurement and sparsity bases \citep{tropp:2004}.  
An alternative measure of coherence is given by the {\em cumulative coherence} \citep{romberg:2009}
\begin{equation*}
\ccoherence(\sensmat, \sparmat, \Gamma) = \max_{\isens} \: \Biggl ( \sum_{\ispar\in\Gamma} \: \bigl | \: \langle \sensatom \: | \: \sparatom \rangle \: \bigr |^2 \Biggr )^{1/2}
\spcend ,
\end{equation*}
where $\Gamma$ represents the set of all sparsity basis vectors that contribute to the signal of interest.  The cumulative coherence is therefore signal dependent and incorporates sparsity information, which renders it more robust than mutual coherence to departures from the pure compressed sensing framework.  The number of measurements required for signal recovery may also be expressed in terms of cumulative coherence and can be shown to evolve as \citep{romberg:2009,rauhut:2009}
\begin{equation}
\label{eqn:ccoherence}
\nmeas \sim \ccoherence^2(\sensmat, \sparmat, \Gamma) \: \ndim
\spcend .
\end{equation}

%=============================================================================
\subsection{Interferometric imaging}
\label{sec:background:imaging}

We express the interferometric framework discussed in \sectn{\ref{sec:background:ri}} in a discrete setting which is amenable to the compressed sensing reconstruction techniques discussed in \sectn{\ref{sec:background:cs}}, following the approach taken by \citet{wiaux:2009:cs} and \citet{wiaux:2009:ss} previously.  Although we consider \wfov\ considerations we restrict ourselves to baselines with constant \bw, thus restricting visibilities to a Fourier plane.\footnote{The $\bw$-projection algorithm \citep{cornwell:2008:w} could be applied in future to remove this restriction, although the extension of the spread spectrum phenomenon to a varying $\bw$ would need to be studied extensively.}
In reality, \bw\ will vary and the impact of the spread spectrum phenomenon will lie somewhere between the extreme cases that we consider here.  Nevertheless, this assumption allows us to probe the expected impact of the spread spectrum phenomenon at light computational load and to discard considerations related to specific interferometer configurations.
Considering the signal projected onto the equatorial plane, we consider the $\lxvect=(\lx,\mx)$ coordinate system discretised on a uniform grid of $\ndim = \ndim^{1/2} \times \ndim^{1/2}$ points $\lxvectdis \in \reals^2$ in real space, with integer $1\leq\ispar\leq\ndim$, and the corresponding grid of discrete spatial frequencies $\buvect_\ispar$ defined by Nyquist-Shannon sampling.  The band-limited intensity and primary beam functions are defined on the spatial grid and denoted by $\imdis, \beamdis \in \complex^\ndim$, respectively.  The complex \wterm\ is also defined on this grid $\chirpdis\in\complex^\ndim$.  Since the \wterm\ is complex, the Fourier transform of $\chirpfull \beam(\lxvect) \im(\lxvect)$ does not bear any specific symmetry in the Fourier plane, even for a real beam and real source intensity.  
Following \citet{wiaux:2009:cs} and \citet{wiaux:2009:ss}, we assume that the spatial frequencies $\buvect$ probed by the interferometer fall on the discrete grid points $\buvect_\ispar$.  The Fourier coverage provided by the $\nmeas$ spatial frequencies probed $\buvect_\isens$, with integer $1\leq\isens\leq\nmeas$, can be identified by a binary mask in the Fourier plane equal to unity for each spatial frequency probed and zero otherwise.\footnote{Conventional gridding \citep{thompson:2001} of the visibilities measured at continuous spatial frequencies would result in a mask with non-binary weights, which could be incorporated easily in the framework described here.}  The visibilities measured are denoted by the vector $\visdis \in \complex^{\nmeas}$, which may be affected by complex noise $\noisedis \in \complex^{\nmeas}$.  
%The complex vectors $\visdis$ and $\noisedis$ may alternatively be represented by their real and imaginary parts, identified by vectors $\visdis,\noisedis \in \reals^{\nmeas}$ of $\nmeas$ real values.

The visibility integral \eqn{\ref{eqn:vis3}} may be represented in this discrete setting by the linear system 
\begin{equation}
\label{eqn:vis_linear_plane}
\visvect = \sensmatp \impvect + \nimvect
\spcend ,
\end{equation}
with
\begin{equation*}
\sensmatp  = \opwhite \, \opmask \, \opfourier \, \opchirp \, \opbeam
\spcend ,
\end{equation*}
where Gaussian noise and a whitening operator have also been included.
The measurement matrix $\sensmatp \in \complex ^{\nmeas \times \ndim}$ identifies the complete linear relation between the signal and the visibilities.  The matrix $\opbeam \in \complex^{\ndim \times \ndim}$ is the diagonal matrix implementing the primary beam, with the Jacobian term $\nx ^{-1}(\lxvect)$ due to the change to Cartesian coordinates absorbed.  The matrix $\opchirp \in \complex^{\ndim \times \ndim}$ is the diagonal matrix implementing the modulation by the \wterm .  The unitary matrix $\opfourier \in \complex^{\ndim \times \ndim}$ implements the discrete Fourier transform.  The matrix $\opmask \in \reals^{\nmeas \times \ndim}$ is the rectangular binary matrix implementing the mask that characterises visibility coverage, containing one non-zero value on each row only, at the index of the Fourier coefficient corresponding to the spatial frequency probed by each interferometric measurement.  We also augment the measurement matrix by a whitening operator $\opwhite \in \reals^{\nmeas \times \nmeas}$ \citep{wiaux:2009:cs}.  Whitening corresponds to dividing each measured visibility by the standard deviation of the corresponding noise component so that the final observed visibilities are then affected by independent identically distributed Gaussian noise with unit variance.  The subscript $\pind$ is used to denote the planar setting since we subsequently generalise this framework to functions defined on the sphere (throughout subscripts $\pind$ and $\sind$ are used to denote the plane and sphere respectively). 
For incomplete visibility coverage $\nmeas<\ndim$, \eqn{\ref{eqn:vis_linear_plane}} defines an ill-posed inverse problem, which we solve using the BP and TV reconstruction approaches described in \sectn{\ref{sec:background:cs}}.
Finally, let us note that the compressed sensing framework is defined strictly for orthogonal sparsity and sensing bases.  The application of the \wterm\ modulation necessitates an upsampling operation in practice to ensure that the modulated signal is Nyquist-Shannon sampled, which breaks the orthogonality of the sensing basis.  Compressed sensing techniques for interferometric imaging therefore depart from the theoretical compressed sensing framework but nevertheless have been demonstrated to perform very well.

We conclude this section with a review of the spread spectrum phenomenon.  As discussed in \sectn{\ref{sec:background:cs}}, the coherence between the measurement and sparsity bases is a key criterion effecting the performance of compressed sensing based reconstructions.  In the interferometric setting described above the measurement basis can essentially be identified with the Fourier basis, which will aid our intuitive understanding of the spread spectrum phenomenon.  In this case, the mutual coherence is given by the maximum modulus of the Fourier coefficient of the sparsity basis vectors.  Consequently, an operation that acts to reduce the maximum Fourier coefficient, reduces the coherence and thus improves the quality of compressed sensing reconstructions. 
Modulation by the \wterm\ corresponds to a norm-preserving convolution in the Fourier plane, spreading the spectrum of the sparsity basis vectors, and achieves exactly this.  Hence, the greater the frequency content of the \wterm, the more effective the spread spectrum phenomenon.

%=============================================================================
\section{Spherical interferometric imaging}
\label{sec:wfov}
%=============================================================================

In this section we generalise the compressed sensing imaging techniques developed by \citet{wiaux:2009:cs} and \citet{wiaux:2009:ss} to a \wfov.  We recover interferometric images defined directly on the sphere, rather than the equatorial plane, which enhances the effectiveness of compressed sensing reconstructions.

The measurement operator transforming the sky intensity defined on the sphere to visibilities, consists of augmenting the usual interferometric measurement operator with an initial projection from the sphere to the plane.  This initial projection corresponds to a change from spherical to Cartesian coordinates in much the same way \eqn{\ref{eqn:vis3}} is defined from \eqn{\ref{eqn:vis1}}.  Similarly, the framework remains general and does not rely on any smal \fov\ assumptions.  Practically, however, the projection which implements the change of variable is complicated by the discrete setting and the desire to recover a regular grid on the plane to allow the use of fast Fourier transforms (FFTs).\footnote{Fast algorithms have been developed to compute a discrete Fourier transform on non-equispaced spatial frequencies \citep{potts:2001}, which could in principle be used to avoid explicit gridding.}  In order to ensure information is not lost, we \mbox{define} the resolution of the planar grid so as to support the maximum projected frequency content of a band-limited signal on the sphere.  Although our \wfov\ framework still involves a projection to the plane, this is included in the measurement operator, and so will be regularised when solving the interferometric inverse problem.  Moreover, we recover the sky intensity directly on the sphere, where is it most sparse, and it is the signal space that is important for the impact of sparsity on the performance of compressed sensing reconstructions.

In the remainder of this section we first discuss considerations relating to band-limited signals on the sphere and the plane, operators used to project the sphere to the plane in a discrete setting and discrete gradient operators defined on the sphere (required for TV reconstructions on the sphere).  Finally, we define interferometric imaging in the \wfov\ setting, while commenting on the impact of sparsity, coherence and the spread spectrum phenomenon.

%=============================================================================
\subsection{Band-limited signals}
\label{sec:wfov:bandlimited}

We require a regular grid on the plane to enable the application of FFTs to reduce considerably the computational load of reconstructing images from visibility measurements.  To ensure information is not lost when projecting a signal from the sphere to the plane in this discrete setting, we define a planar grid that supports a band-limit corresponding to the maximum projected frequency content of a band-limited signal defined on the sphere.  The maximum projected frequency content arises at the extents of the \fov, where signal content defined on the surface of the sphere is projected onto much higher frequency content in the $\lxvect=(\lx,\mx)$ plane.  In the typical small \fov\ setting the relationship between the band-limit of a signal on the sphere $\elmax$ and its tangent plane $\bumax$ is given by $\elmax = 2 \pi \: \bumax$.  In the \wfov\ setting simple geometric considerations at the extent of the \fov\ lead to the relationship 
\begin{equation}
\label{eqn:bandlimit}
\elmaxfov = 2 \pi \: \cos(\tfov/2) \: \bumaxfov
\spcend ,
\end{equation}
where $\tfov$ denotes the angular opening of the \fov, corresponding to a planar \fov\ of $\pfov=2 \sin (\tfov / 2)$.  The band-limit relation is now dependent on the size of the \fov.  Note that for a \wfov\ a higher band-limit on the plane is required to support a given band-limit on the sphere, as expected intuitively, and as $\tfov \rightarrow 0$ the usual relationship for a small \fov\ is recovered.

\begin{figure}
\centering
\includegraphics[clip=,viewport=260 0 1100 520,width=80mm]{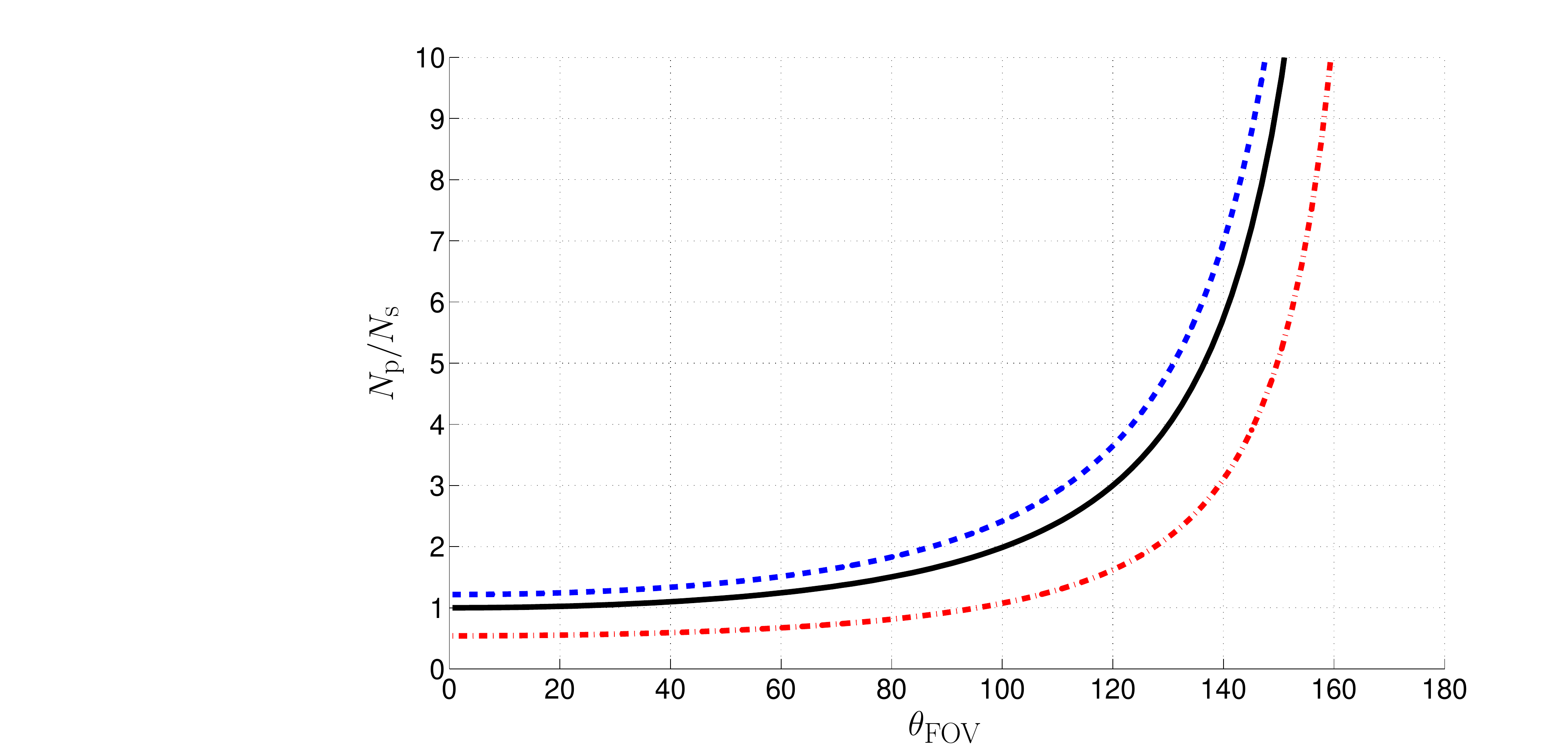}
\caption{Ratio of number of samples on the plane to the sphere
  ($\nplane/\nsphere$) to sufficiently sample a band-limited signal
  defined on the sphere when projected onto the plane (\tfov\ is specified in degrees).  Curves are
  plotted for a \healpix\ pixelisation of the sphere, with $\elmax = k
  \nside$, for values $k=3$ (blue/dashed line); $k=\sqrt{3}\,\pi/2$
  (black/solid line); $k=2$ (red/dot-dashed line).  To ensure a ratio
  of unity is obtained as \tfov\ tends to zero, we select
  $k=\sqrt{3}\,\pi/2$ throughout this work.}
\label{fig:nratio}
\end{figure}

Once the band-limit of a signal is defined, on both the sphere and the plane, sampling considerations then dictate the resolution of the discrete grid required to accurately represent the signal.  Consequently, once the band-limit of the signal on the sphere and the \fov\ is set, the required band-limit on the plane is determined through \eqn{\ref{eqn:bandlimit}}, and the sampling resolutions of both the sphere and the plane follow.  The Nyquist-Shannon sampling theorem on the plane states that $\nplane^{1/2} = 2 \: \bumax \: \pfov$, where $\nplane$ is the number of samples on the planar grid.  The relationship between the number of samples within the \fov\ on the sphere $\nsphere$ and the harmonic band-limit \elmax, depends on the pixelisation of the sphere adopted.  We choose the \healpix\footnote{\url{http://healpix.jpl.nasa.gov/}} pixelisation of the sphere \citep{gorski:2005} due to the equal area of each pixel element and its ubiquitous use
in the astrophysical community.  The resolution of the \healpix\ pixelisation is controlled by the parameter $\nside$, where the number of pixels at a given resolution is specified by $\npix=12\nside^2$.  Functions that are band-limited on the sphere at $\elmax$ can be resolved on a \healpix\ pixelisation at a resolution corresponding to $\elmax=3\nside$, albeit with integration errors\footnote{Integration errors may be reduced by running the analysis iteratively.} \citep{hivon:2010} (since no exact quadrature rule exists for \healpix).  For the correspondence $\elmax=2\nside$ integration errors are minimal \citep{hivon:2010}.  Consequently, harmonic transforms are typically performed for $\elmax=k \nside$, with $k \sim 2$--$3$.  In order to ensure that we do not favour either the plane or sphere in our analysis, we choose $k$ such that in the limit of a small \fov, the number of samples on the plane and within the \fov\ on the sphere are the same.  It can be shown trivially that 
\begin{equation*}
\frac{\nplane}{\nsphere} = \frac{2 k^2 \tan^2 (\tfov/2)}{3 \pi^2 \bigl( 1 - \cos(\tfov/2) \bigr)}
\spcend ,
\end{equation*}
where $\elmax$ drops out of the expression.  Consequently, we choose the value $k=\sqrt{3}\pi/2$ to ensure
\begin{equation*}
\lim_{\tfov\rightarrow 0} \nplane / \nsphere= 1
\spcend .
\end{equation*}
The ratio $\nplane/\nsphere$ is plotted in \fig{\ref{fig:nratio}} for various values of $k$.  Notice how the ratio increases rapidly with the \fov.  Naively, one might expect the ratio of sparsities between the plane and sphere $\sparsity_\pind / \sparsity_\sind$ to evolve in a similar manner, which would highlight the superiority of spherical reconstructions compared to planar ones (this is necessarily a very approximate prediction and the exact ratio of sparsities will depend highly on the signal examined).

%=============================================================================
\subsection{Projection operators}
\label{sec:wfov:projection}

Now that discrete grids are defined on the sphere and the plane, it is necessary to determine how to project a signal between these grids.  In the continuous setting this is trivial and simply corresponds to a change of variable, however the matter is complicated in practice due to the discrete nature of the sampled signals.

\begin{figure}
\centering
\includegraphics[clip=,viewport=100 0 370 235,width=75mm]{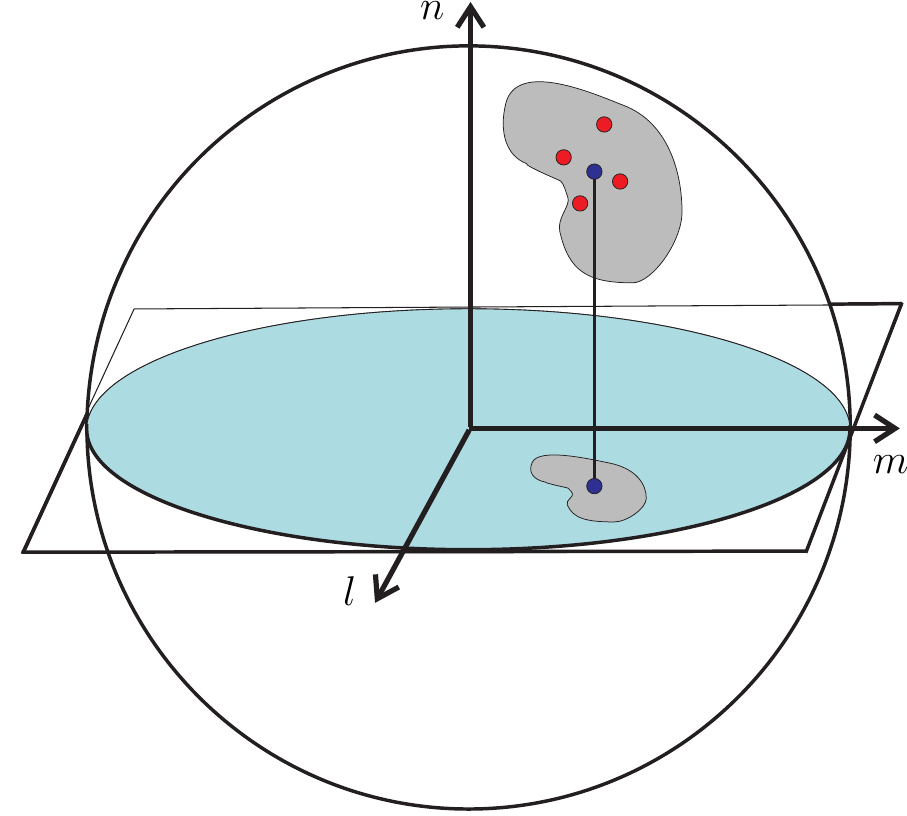}
\caption{Projection of a sampled signal from the sphere to the plane.
  In order to project onto a regular grid on the plane (to reduce
  significantly the computational load of subsequent analyses
  through the use of FFTs), a re-gridding operation is required.  The
  value of the point on the plane (blue/dark dot) is determined by
  convolving the points on the sphere (red/grey dots) with a suitable
  kernel.
  }
\label{fig:projection}
\end{figure}

In order to project onto a regular grid on the plane, it is necessary to re-grid the pixelisation on the sphere to recover sample values at spherical positions that project directly onto the planar grid, as illustrated in \fig{\ref{fig:projection}}.  We perform a convolution on the sphere to achieve this re-gridding.  This convolutional re-gridding on the sphere is similar to the re-gridding often performed when mapping the visibilities observed at continuous coordinates to a regular grid, also to afford the use of FFTs \citep{thompson:2001}.  The interferometric measurement operator on the sphere is therefore augmented by prepending a projection operator $\opproj$ that includes a spherical convolution followed by a project from the sphere to the tangent plane $\lxvect=(\lx,\mx)$.  

We consider three convolution kernels on the sphere: a box kernel, corresponding to a nearest-neighbour interpolation; a sinc-like kernel; and a Gaussian kernel.  The box kernel is well localised on the sphere but has infinite support in harmonic space.  The sinc-like kernel corresponds to evaluating the spherical harmonic transform of the sampled signal on the sphere, followed by evaluating the signal value at the new sample position from its spherical harmonic coefficients (we refer to this as a sinc-like kernel due to analogy with the plane).  This procedure results in a kernel that is well localised in harmonic space but has support in real space that entends over the entire sphere.  These two kernels represent the extremes between spatial and harmonic space localisation and each produces ringing in the projected signal in the domain in which the kernel is not well localised.  We therefore seek a kernel that provides a compromise between this trade-off and is reasonably well localised in both space and frequency.  In this article we settle on a simple Gaussian kernel, however alternative kernels such as the spherical equivalent of the Gaussian-sinc or spheroidal functions could also be considered \citep{thompson:2001}.

%=============================================================================
\subsection{Gradient operators on the sphere}

A discrete gradient operator must be defined on the sphere in order to compute the magnitude of the gradient of a function to solve the TV minimisation problem of \eqn{\ref{eqn:min_tv}} on the sphere.  The discrete gradient operator on the plane is defined simply through finite differences \citep{rudin:1992,chambolle:2004}.  However, it is not possible to define a discrete spherical gradient operator on the \healpix\ pixelisation in this manner since sample positions do not lie on both rings of constant latitude {\rm and} rings of constant longitude (only equiangular pixelisations of the sphere satisfy this property).  One alternative is to consider the continuous gradient operator on the sphere, for which the magnitude of the gradient of a function on the sphere $f$ reads 
\begin{equation*}
| \: \nabla_\sind f  \:| =
\sqrt{
\Biggl ( \frac{\partial f}{\partial \saa} \Biggr)^2
+
\frac{1}{\sin^2\saa}\Biggl (  \frac{\partial f}{\partial \sab} \Biggr )^2
}
\spcend ,
\end{equation*}
which may be computed in harmonic space \citep{wandelt:1998} to eliminate pixelisation concerns.  However, such an approach requires a spherical harmonic transform, which necessarily requires global support on the sphere and hence is non-local in nature, and also is of high computational cost.  

We define a discrete gradient operator on the sphere by analogue with the continuous gradient but computed through finite differences. Convolutional re-gridding is performed to obtain samples on the sphere at the positions required to compute finite difference values.  The same Gaussian convolution kernel that is used to project the sphere to the plane (as discussed in \sectn{\ref{sec:wfov:projection}}) is used.  Such an approach actually corresponds to computing the gradient of a smoothed version of the original signal.  However, the smoothing is minimal in practice and numerical experiments have shown that the discrete gradient defined in this manner is a good approximation to the continuous gradient computed in harmonic space but does not suffer from the global nature and high computational cost of the spherical harmonic transform.

%=============================================================================
\subsection{Interferometric imaging}
\label{sec:wfov:imaging}

We are now in a position to define explicitly the \wfov\ interferometric imaging framework developed in this work.  We attempt to recover the sky intensity directly on the sphere $\imsvect \in \complex^{\nsphere}$ by solving the inverse problem
\begin{equation}
\label{eqn:vis_linear_sphere}
\visvect = \sensmats \imsvect + \nimvect
\spcend ,
\end{equation}
with 
\begin{equation*}
\sensmats  = \opwhite \, \opmask \, \opfourier \, \opchirp \, \opbeam
\, \opproj
\spcend .
\end{equation*}
The measurement operator on the sphere $\sensmats \in \complex ^{\nmeas \times {\nsphere}}$ simply consists of augmenting the operator on the plane $\sensmatp$ by prepending a projection from the sphere to the plane $\opproj$, which incorporates a convolutional re-gridding as discussed in \sectn{\ref{sec:wfov:projection}}.  Careful consideration is given to samplings on the sphere and plane to ensure that the planar grid is sampled sufficiently to accurately represent the projection of a  band-limited signal defined on the sphere (as discussed in \sectn{\ref{sec:wfov:bandlimited}}).  For incomplete visibility coverage the inverse problem defined by \eqn{\ref{eqn:vis_linear_sphere}} becomes ill-posed.  We solve this ill-posed inverse problem in a compressed sensing framework by solving the BP and TV minimisation problems defined by \eqn{\ref{eqn:min_bp}} and \eqn{\ref{eqn:min_tv}} respectively.  In this work we consider the Dirac sparsity basis on the sphere for BP reconstructions.  

We expect the performance of compressed sensing reconstructions to be enhanced by recovering the sky intensity in the space where it naturally lives (\ie\ on the sphere), since we expect sparsity to be reduced in this space.  Furthermore, the effectiveness of the spread spectrum phenomenon is enhanced by going to a \wfov.  As discussed in \sectn{\ref{sec:background:imaging}}, the greater the frequency content of the \wterm, the more effective the spread spectrum phenomenon.  By examining the \wterm\ plotted in \fig{\ref{fig:chirps}}, we see that its frequency content increases with distance from the origin.  Consequently, the larger the \fov, the higher the frequency content achieved by the \wterm\ and the more effective the spread spectrum phenomenon.  Moreover, a secondary enhancement arises by eliminating the first order assumption $\| \lxvect \|_2 ^4 \: \bw \ll 1$ made previously by \citet{wiaux:2009:ss}, since this assumption reduces the maximum frequency content of the \wterm\ achieved on a given \fov\ (the band-limits of the various \wterm s are defined explicitly in \sectn{\ref{sec:recon:setup}}).
Finally, we comment on the impact of mutual coherence on reconstruction performance.  Since we consider the Dirac sparsity basis, coherence is optimal on the plane (recall that the Dirac and Fourier bases are maximally incoherent).  However, the projection of Dirac basis functions defined on the sphere to the plane does not result in Dirac functions on the plane (due to the convolutional re-gridding).  Consequently, coherence in the spherical setting is suboptimal.  The spread spectrum phenomenon acts to increase incoherence only in the case where it is not already maximally incoherence, thus we expect the spread spectrum phenomenon to be ineffective in the planar setting but to improve reconstruction performance in the spherical setting.  
For TV reconstructions, although a sparsity basis does not exist, one may gain some intuition regarding the impact of coherence by the following argument.  For a piecewise constant signal that is sparse in the magnitude of its gradient (\ie\ has a gradient defined by Dirac functions), the spectrum of the magnitude of its gradient must be flat.  The gradient operator in space essentially corresponds to a multiplication by frequency in Fourier space, hence the original spectrum of the piecewise constant signal must evolve as the inverse of frequency.  Since this differs to the optimally incoherent spectrum which is flat, the coherence must be suboptimal for TV reconstructions.  We therefore expect the spread spectrum phenomenon to provide improvements both on the sphere and plane for TV reconstructions, with a greater enhancement expected on the sphere due to the spatial spreading of the projection operator.
In any case, sparsity typically has a greater impact on the performance of compressed sensing reconstructions than coherence.  
These considerations pertaining to sparsity, mutual coherence and the spread spectrum phenomenon lead us to expect an improvement in the performance of compressed sensing reconstructions when recovering interferometric images in the \wfov\ framework developed here.

\newlength{\chirpplotwidth}
\setlength{\chirpplotwidth}{40mm}

\begin{figure}
\centering
%\mbox{
%\subfigure[Assuming $\| \lxvect \|_2 ^4 \: \bw \ll 1$]{\includegraphics[clip=,viewport=260 30 550 330,width=\chirpplotwidth]{figures/chirp1_real} \quad
%\includegraphics[clip=,viewport=260 30 550 330,width=\chirpplotwidth]{figures/chirp1_imag}}
%}\\
%\mbox{
%\subfigure[No small \fov\ assumption]{\includegraphics[clip=,viewport=260 30 550 330,width=\chirpplotwidth]{figures/chirp2_real} \quad
%\includegraphics[clip=,viewport=260 30 550 330,width=\chirpplotwidth]{figures/chirp2_imag}}
%}
\subfigure[Real part]{\includegraphics[clip=,viewport=260 30 550 330,width=\chirpplotwidth]{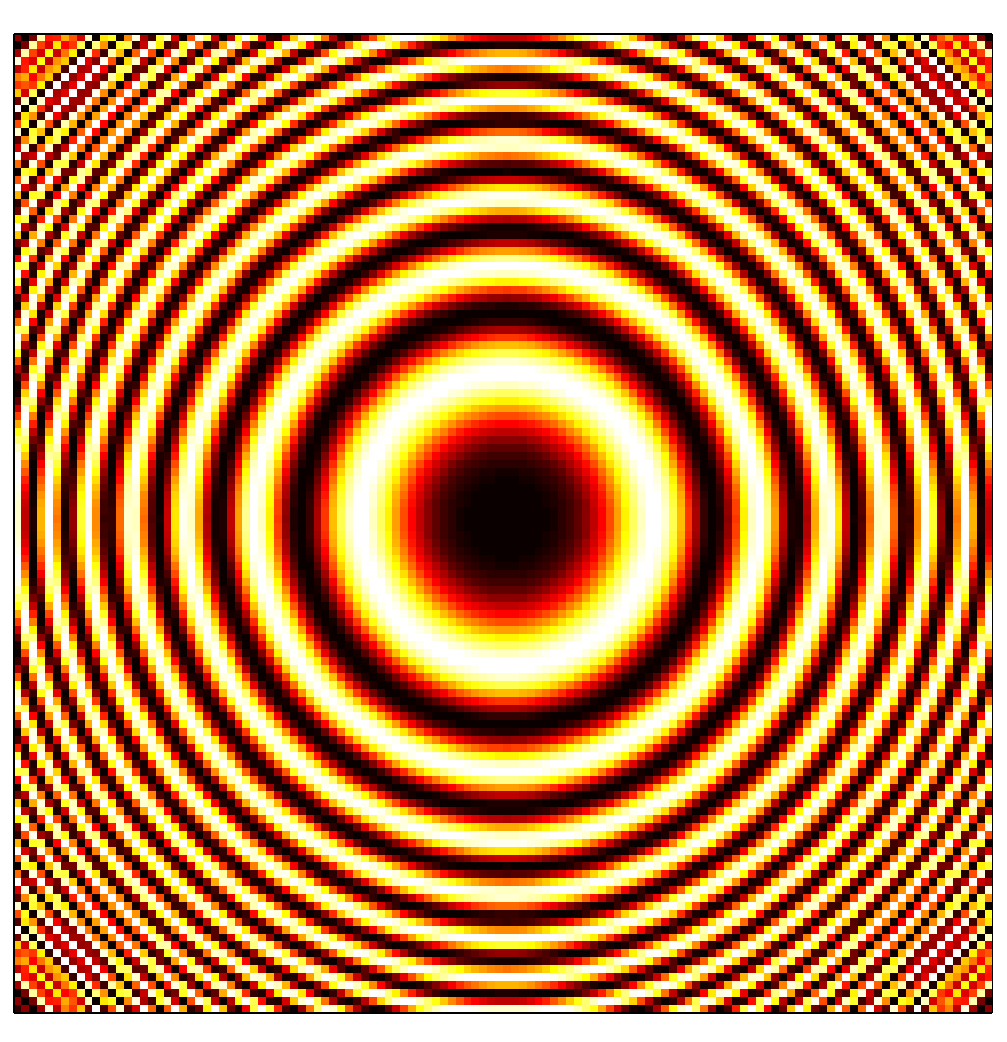}} 
\quad
\subfigure[Imaginary part]{\includegraphics[clip=,viewport=260 30 550 330,width=\chirpplotwidth]{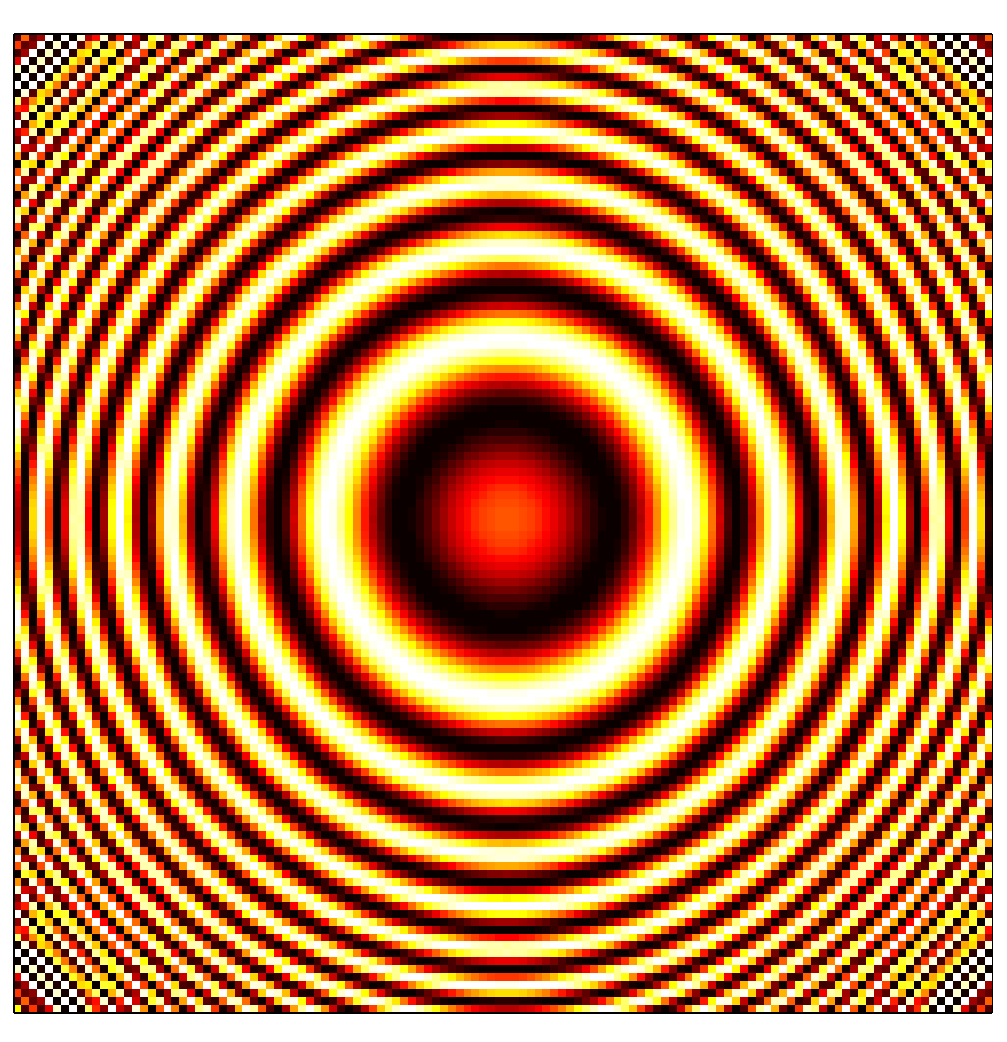}}
\caption{Real and imaginary parts of the \wterm\ modulation $\chirpfull$ (for $\bwd=1 / \sqrt{2}$; see text \sectn{\ref{sec:recon:setup}}).  Notice that the frequency content of the \wterm\ modulation increases with distance from the origin (image centre).  Consequently, for a \wfov\ the \wterm\ modulation spreads the spectrum more effectively, enhancing the performance of the spread spectrum phenomenon.  Dark and light regions correspond to positive and negative values respectively.  }
\label{fig:chirps}
\end{figure}

%=============================================================================
\section{Simulated reconstructions}
\label{sec:recon}
%=============================================================================

We evaluate the performance of the \wfov\ interferometric imaging framework defined on the sphere, as outlined in \sectn{\ref{sec:wfov}}, making a direct comparison with planar reconstructions.  After describing the observational set-up, performance is quantified thoroughly in a low-resolution setting on sets of simulations of sources with a Gaussian profile.  A more realistic setting at a higher resolution is then considered, where reconstruction performance is evaluated on a single simulated observation of Galactic dust emission.

%=============================================================================
\subsection{Observational set-up}
\label{sec:recon:setup}

Simulated observations of real signals are made on the \fov\ defined by the angular opening $\tfov=90^\circ$, corresponding to a planar \fov\ of \mbox{$\pfov=\sqrt{2}$}.  A real Gaussian primary beam is assumed, with \fwhmtext\ (\fwhm) of one half of the field of view.
%, \ie\ \mbox{${\rm \fwhm}=45^\circ$}.  
Random visibility coverage is considered, with visibility measurements falling on the discrete planar grid of spatial frequencies $\buvect_\ispar$.  We consider incomplete visibility coverage, with only \mbox{$2$--$25$}\% of the discrete visibilities measured (since we consider real signals,
measuring 50\% of the complex visibilities corresponds to a number of measurements identical to the number of unknowns in the real signal that we attempt to recover).

As discussed previously, although we consider a \wfov\ we restrict ourselves to the case of constant \bw.  We parameterise the continuous $\bw$ in terms of the discrete component $\bwd$, following the parameterisation used by \citet{wiaux:2009:cs} of $\bw=\bwd \: \nplane^{1/2} / \pfov^2$.  The band-limit of the modulating \wterm\ $\chirpfull$ is given approximately by
\begin{equation*}
\chirpfullbl \simeq \frac{\bwd \: \nplane^{1/2}}{2 \pfov \sqrt{ 1- (L/2)^2}}
\spcend ,
\end{equation*}
corresponding to its maximum instantaneous frequency.  Under the first order small \fov\ assumption $\| \lxvect \|_2 ^4 \: \bw \ll 1$, the \wterm\ reduces to the linear chirp $\chirpfullb$ with approximate band-limit
\begin{equation*}
\chirpfullbbl \simeq \frac{\bwd \: \nplane^{1/2}}{2 \pfov}
\spcend .
\end{equation*}
Notice that the band-limit for given non-zero values of $\bwd$, $\nplane$ and $\pfov$ is greater in the absence of the small \fov\ assumption, justifying rigorously the secondary enhancement discussed in \sectn{\ref{sec:wfov:imaging}} of the spread spectrum phenomenon due to the \wfov.
The band-limit of the spread signal is given by the sum of the original band-limit $\bumax = \nplane^{1/2} / 2 \pfov$ and the band-limit of the \wterm.  Previously \citet{wiaux:2009:ss} considered the linear chirp $\chirpfullb$ and the value $\bwd=1$, corresponding to spreading by a factor of two.  We also consider spreading by a factor of two, but since we consider the exact \wterm\ $\chirpfull$, this corresponds to a value of $\bwd=1/\sqrt{2}$.  The corresponding continuous $\bw$ is of the same order as the maximum visibility measurements in $\bu$ and $\bv$, \ie\ $\bw = \bumax$, hence it is an appropriate value to consider when studying the spread spectrum phenomenon.  Note that in the absence of a \wfov, the value of $\bw$ considered by \citet{wiaux:2009:ss} to achieve the same spreading was a factor of $2/\pfov$ greater than $\bumax$.  Since the band-limit of the original signal is doubled due to modulation by the \wterm, to avoid aliasing we apply an upsampling operator to increase the resolution of the planar grid by a factor of two prior to application of the modulation (upsampling is performing by zero-padding in the Fourier domain).  In the subsequent analysis we consider the exact \wterm\ \chirpfull, with values of $\bwd= \{ 0, 1/\sqrt{2}\}$ to highlight the effectiveness of the spread spectrum phenomenon.  
%Note that we present results for the \wterm\ \chirpfull\ only, and not the approximate term \chirpfullb, since we are concerned with the \wfov\ setting and similar reconstruction performance can be achieve by applying \chirpfullb\ with a greater value of \bwd\ (which we verifying on numerical simulations).

Instrumental noise is also added to the simulated visibilities.  Independent identically distributed Gaussian noise is assumed with variance $\sigma_{\nimvect}^2 = 10^{-3} \sigma_{\visvect}^2$, where $\sigma_{\visvect}^2$ is the variance of the visibilities in the absence of noise and \wterm\ modulation.  The added instrumental noise results in observed visibilities at a signal-to-noise ratio of $\snr_{\rm n} = 10 \log_{10} (\sigma_{\visvect}^2/\sigma_{\nimvect}^2)= 30$dB.

%=============================================================================
\subsection{Gaussian sources}
\label{sec:recon:gaussians}

The \wfov\ interferometric imaging framework is evaluated thoroughly in this section on simulated observations of Gaussian sources.  These simulations are first described, before we analyse their sparsity properties.  Reconstruction performance is then evaluated both in the spherical and planar settings.  Finally, reconstruction performance is compared to calculations of cumulative coherence. 

\subsubsection{Simulations}

In order to perform a thorough evaluation, we analyse sets of simulations of Gaussian sources of various size, with each set containing 30 simulations, for all variations of reconstruction procedures (BP and TV reconstructions on the plane and the sphere, both with and without application of the spread spectrum phenomenon) and for various visibility coverages.  Due to the large number of simulated reconstructions, we restrict these simulations to a low resolution.  We consider a \healpix\ resolution parameter of $\nside = 32$, corresponding to a harmonic band-limit of $\elmax=88$.  The remaining parameters defining the resolution of the simulations then follow from the considerations discussed in \sectn{\ref{sec:wfov:bandlimited}}; we find $\nsphere=1740$, $\bumax=19.6$ and $\nplane = 58 \times 58=3364$.  
The size of the Gaussian kernel in the convolutional re-gridding of the projection operator, defined by its standard deviation  $\sigma_{\rm P}$, is chosen to ensure that the kernel is well sampled.  Since a kernel with small support is required, a local tangent plane approximation is made to relate its Fourier size to its spatial size through $\sigma_{\rm F} = (2\pi\sigma_{\rm P})^{-1}$.  We ensure $2\sigma_{\rm F}$ is within the band-limit supported by the planar grid, resulting in the kernel size $\sigma_{\rm P} = 0.02$ radians.

For each simulation, 10 Gaussian sources of the same size are laid down at random positions within the \fov\ and with random amplitudes $A_{\rm S} \in [0.5, 1.0]$.  For each set of 30 simulations, source objects of a different size are considered, with sizes defined by the standard deviation of the Gaussian source \mbox{$\objsize \in \{0.01, 0.02, 0.04, 0.10 \}$ radians}.  For some cases the standard deviation of the source is smaller than the pixel size, hence the Gaussian sources are not necessarily well sampled on the spherical grid.  However, this is of no concern since the simulated signals are defined by their discrete version and no contact is subsequently made with the continuous representation from which they originate.  To simulate compact Gaussian objects on the sphere we use the \comb\ \citep{mcewen:2006:filters} and \stwo\ \citep{mcewen:2006:fcswt} packages\footnote{\url{http://www.jasonmcewen.org/}}.  Random visibility coverages are considered with only $\kappa \in \{ 2, 4, 6, 10, 15, 20, 25 \}$\% of the discrete visibilities measured.

\subsubsection{Sparsity}

We perform tests to determine whether our hypothesis holds that sparsity is reduced by going to the space in which the signal inherently lives.  Although sparsity levels will depend highly on the signal considered, since each set of Gaussian simulations has similar properties we expect sparsity levels to be reasonably consistent over the set of simulations.  We therefore average the sparsities measured over the set of 30 simulations for each source size and also provide an indication of their spread.  

Since the signals analysed are compressible rather than exactly sparse, we require a measure of compressibility.  We construct such a measure as follows.  For the BP problem, we first order the coefficients of the signal in the sparsity basis, which in the case of the Dirac sparsity basis that we adopt corresponds to simply ordering the sampled signal values.  We then set the smallest coefficient to zero and measure both the sparsity of the resultant signal and also the \snr\ of the original signal relative to the error between the original and resultant signal.  We set the next smallest coefficient to zero and repeat these two measurements, repeating the procedure until all signal values are set to zero.  Following this approach we build a curve of sparsity against \snr.  For the TV problem, we repeat the same procedure but in the space of the magnitude of the gradient of the signal, rather than a sparsity basis.  

\begin{figure}
\centering
\mbox{
\subfigure[Dirac sparsity for $\objsize=0.01$]{\includegraphics[clip=,viewport=110 0 610 480,width=\chirpplotwidth]{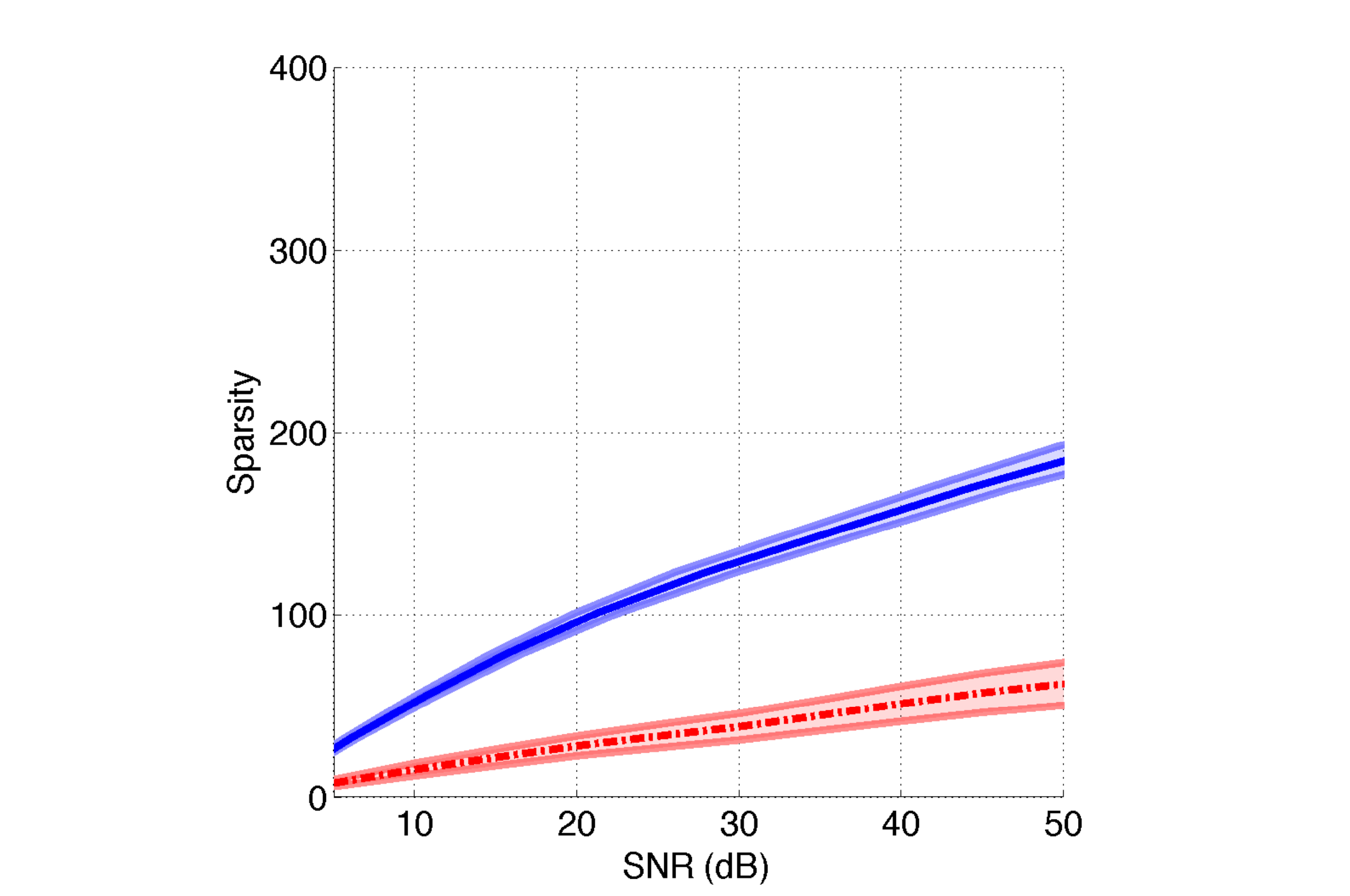}} \quad
\subfigure[TV sparsity for $\objsize=0.01$]{\includegraphics[clip=,viewport=110 0 610 480,width=\chirpplotwidth]{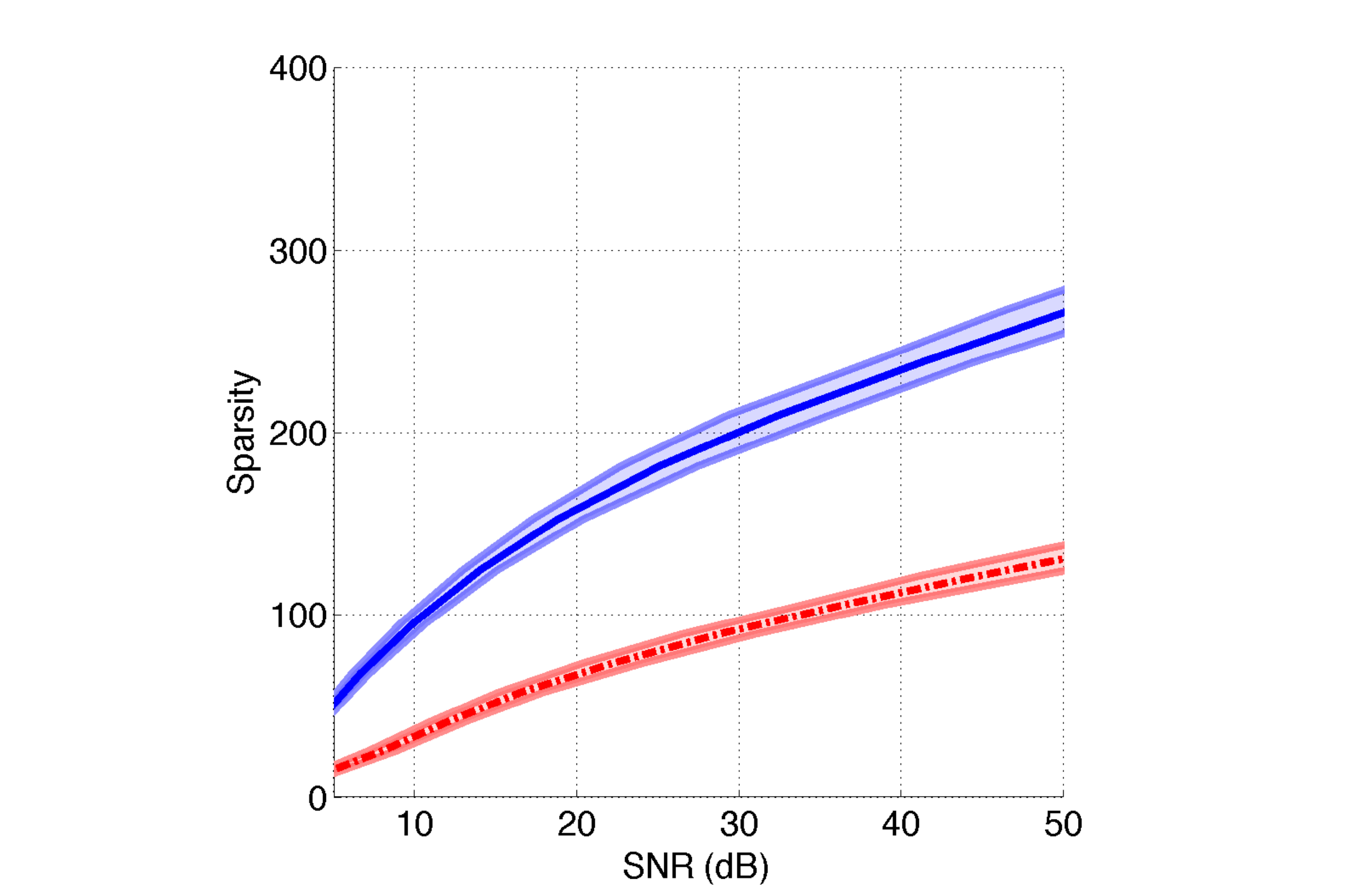}}
}\\
\mbox{
\subfigure[Dirac sparsity for $\objsize=0.10$]{\includegraphics[clip=,viewport=110 0 610 480,width=\chirpplotwidth]{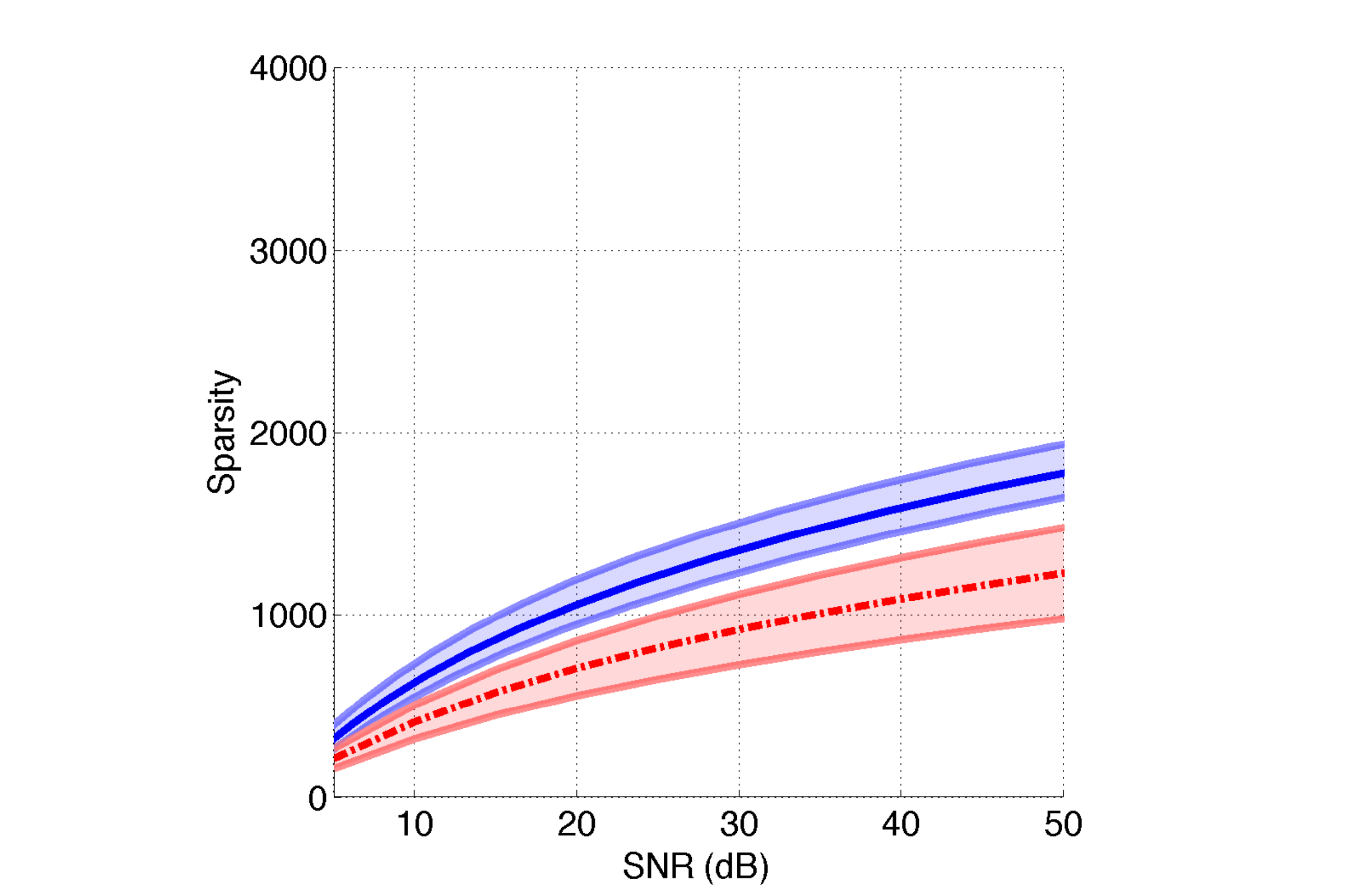}} \quad
\subfigure[TV sparsity for $\objsize=0.10$]{\includegraphics[clip=,viewport=110 0 610 480,width=\chirpplotwidth]{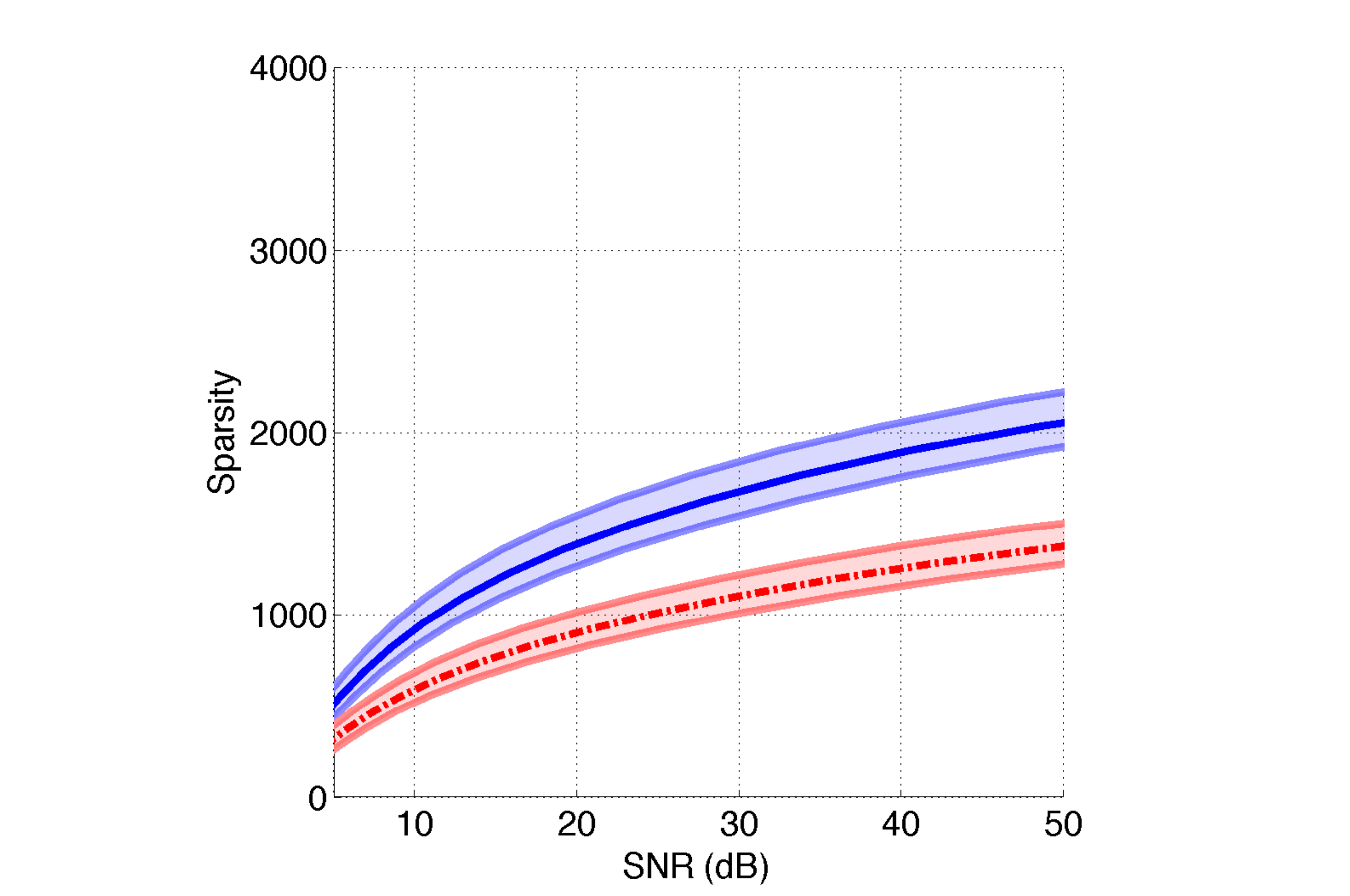}}
}
\caption{Imposed sparsity of simulations of Gaussian sources as a function of the \snr\ of the original compressible signal compared to the imposed sparse signal (as explained in the text).  Curves correspond to the mean of 30 simulations, while shaded regions show one standard deviation confidence intervals.  Curves are plotted for the signal defined on the sphere (red/dot-dashed line) and for its projected version on the plane (blue/solid line).  Sparsity is clearly enhanced on the sphere.}
\label{fig:sparsity}
\end{figure}

Sparsity measurements are computed for the signal on the sphere and its projected version on the plane in order to make comparisons.  Curves are plotted in \fig{\ref{fig:sparsity}} for the two extreme source sizes (curves for intermediate source sizes are essentially interpolations between these extremes).  
Sparsity is clearly enhanced on the sphere.  Moreover, the ratio of sparsities between the plane and sphere ($\sparsity_\pind / \sparsity_\sind$) does indeed approximately follow the ratio of the number of samples between the plane and sphere ($\nplane / \nsphere \sim 1.9$ for $\tfov=90^\circ$), as predicted naively in \sectn{\ref{sec:wfov:bandlimited}}.  Although sparsity will always be highly dependent on the signal under investigation, we have at least demonstrated for Gaussian sources that the sparsity of the signal is indeed enhanced on the sphere.

\subsubsection{Reconstruction performance}

We evaluate reconstruction performance in the \wfov\ setting  by recovering interferometric images directly on both the sphere and the plane, in order to made a direct comparison.  Furthermore, we consider BP and TV reconstructions, both with and without application of the spread spectrum phenomenon.  It is of particular interest to see whether our predictions are demonstrated through reconstruction quality. 

The measurement contraint $\epsilon$ is defined by the level $\plevel=0.99$ (as described in \sectn{\ref{sec:background:cs}}) and we solve the BP and TV minimisation problems using the algorithms derived by \citet{combettes:2007} and \citet{durand:2010} respectively.  For the low-resolution simulations performed here, the computation time required to solve the optimisation problems are typically of the order of one minute on a standard laptop with a 2.66GHz Intel Core 2 Duo processor with 4GB of memory.  

The quality of reconstruction is measured by the \snr\ of the original signal relative to the difference between the original and reconstructed signal, defined explicitly by
\mbox{$\snrs=10 \log_{10} (\sigma_{\imsvect}^2 / \sigma_{\imsvect - \imsvectrecon}^2)$},
where $\sigma_{\imsvect}^2$ is the variance of the original signal $\imsvect$ and $\sigma_{\imsvect - \imsvectrecon}^2$ is the variance of the discrepancy signal $\imsvect - \imsvectrecon$.  Since the sky intensity to be recovered lives inherently on the sphere, we consider the \snr\ defined on the sphere.  It is therefore necessary to lift the image reconstructed on the plane to the sphere in order to make a comparison.  We do this through a projection operator that is the direct analogue of the operator $\opproj$ that projects the sphere to the plane, using an identical kernel in the convolutional re-gridding.  We then compare \snrs, measured on the sphere, for both the spherical and planar based reconstructions.
\fig{\ref{fig:recon_sphere}} shows reconstruction quality measured by \snrs\ for various visibility coverages, averaged over the 30 simulations for each source size.  Reconstruction quality in the spherical setting is clearly superior to the quality of planar reconstructions.  However, for sources of small size, lifting the planar reconstruction to the sphere introduces error and limits the effectiveness of the reconstruction.  Before discussing reconstruction performance in more detail, we consider the \snr\ defined on the plane.

We also examine the \snr\ defined on the plane by
\mbox{$\snrp=10 \log_{10} (\sigma_{\impvect}^2 / \sigma_{\impvect - \impvectrecon}^2)$},
where $\sigma_{\impvect}^2$ and $\sigma_{\impvect - \impvectrecon}^2$ are the variances of the original and discrepancy signals on the plane respectively.  To compute \snrp\ for interferometric images recovered on the sphere, the spherical reconstructions are projected onto the plane by the projection operator $\opproj$.
\fig{\ref{fig:recon_plane}} shows reconstruction quality measured by \snrp\ for various visibility coverages, averaged over the 30 simulations for each source size.  The superiority of reconstructions on the sphere is again clear.  Even if one were interested in planar reconstructions, superior reconstruction quality is achieved by first recovering interferometric images on the sphere, before projecting the recovered spherical image to the plane.  In any case, we advocate the direct use of spherical reconstructions since signal content is not distorted by any projection.

The reconstruction performance observed in \fig{\ref{fig:recon_sphere}} and \fig{\ref{fig:recon_plane}} is now discussed in more detail and related to the predictions that we made through intuitive reasoning.  For BP reconstructions, the improvement in spherical reconstruction quality due to the spread spectrum phenomenon is apparent, whereas the spread spectrum phenomenon is clearly ineffective on the plane.  For TV reconstructions, the spread spectrum phenomenon is effective both on the sphere and the plane, although the enhancement on the sphere is slightly larger than that on the plane.  Also note that the variance of reconstruction quality (as indicated by the size of the error bars) is reduced by the spread spectrum phenomenon in the cases where it is effective.  This effect was also observed by \citet{wiaux:2009:ss}.  Notice that the performance of BP reconstructions drops as the size of the Gaussian sources increase due to the corresponding increase in sparsity value.  However, this is not the case for TV reconstructions.  Although TV reconstructions do not perform as well for signals that are extremely sparse in the Dirac basis, they are more stable as sparsity varies and are certainly superior for reconstructing diffuse signal content.  In summary, all of our intuitive predictions are indeed manifest in the reconstruction quality observed and the superiority of reconstructing the sky intensity directly on the sphere in the \wfov\ interferometric imaging framework that we develop is clear.

\newlength{\reconplotwidth}
\setlength{\reconplotwidth}{43mm}

\begin{figure*}
\centering
\mbox{
\subfigure[Typical simulation ($\objsize=0.01$)]{\includegraphics[clip=,viewport=180 30 600 480,width=\reconplotwidth]{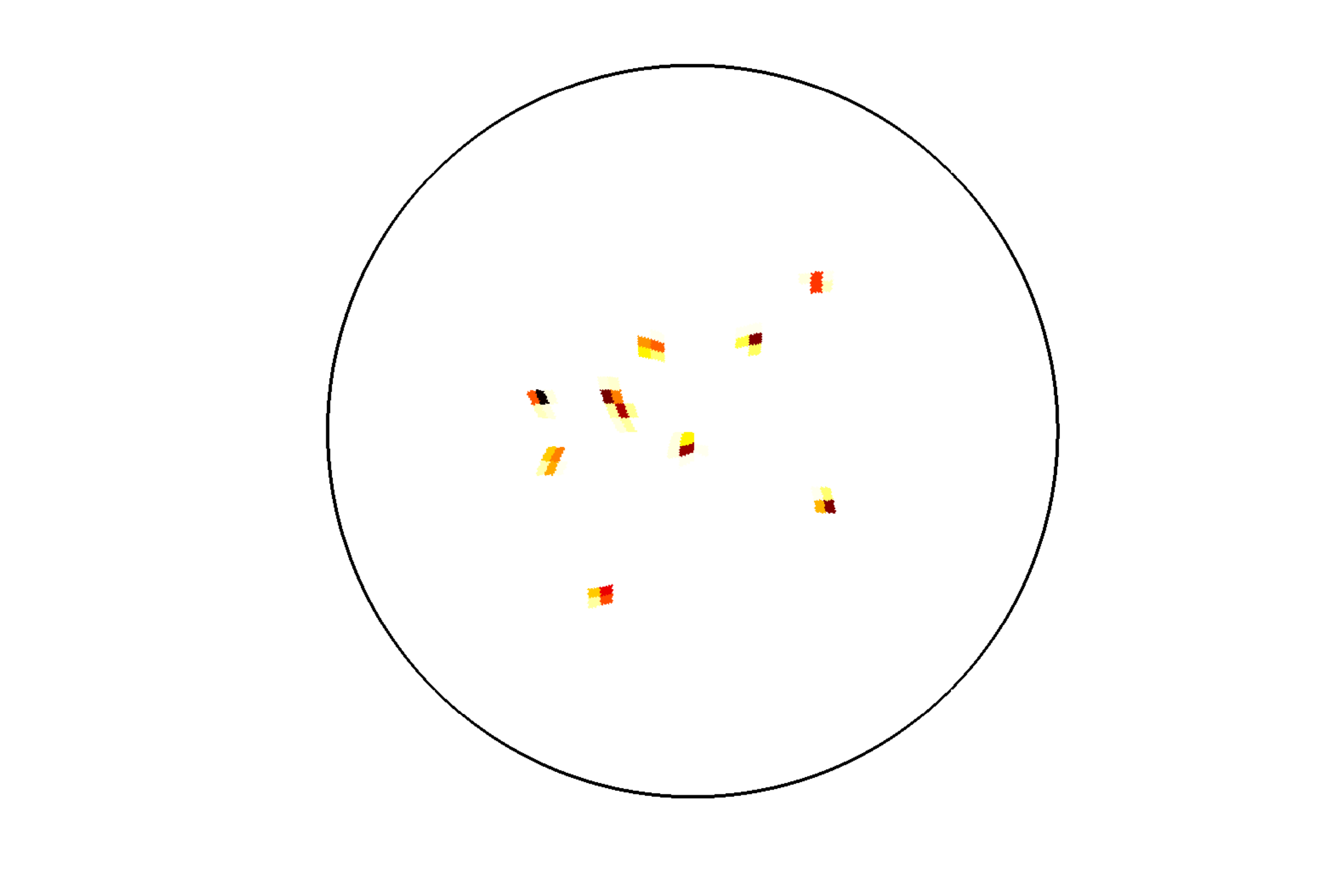}}
\quad\quad
\subfigure[BP reconstruction ($\objsize=0.01$)]{\includegraphics[clip=,viewport=130 0 610 480,width=\reconplotwidth]{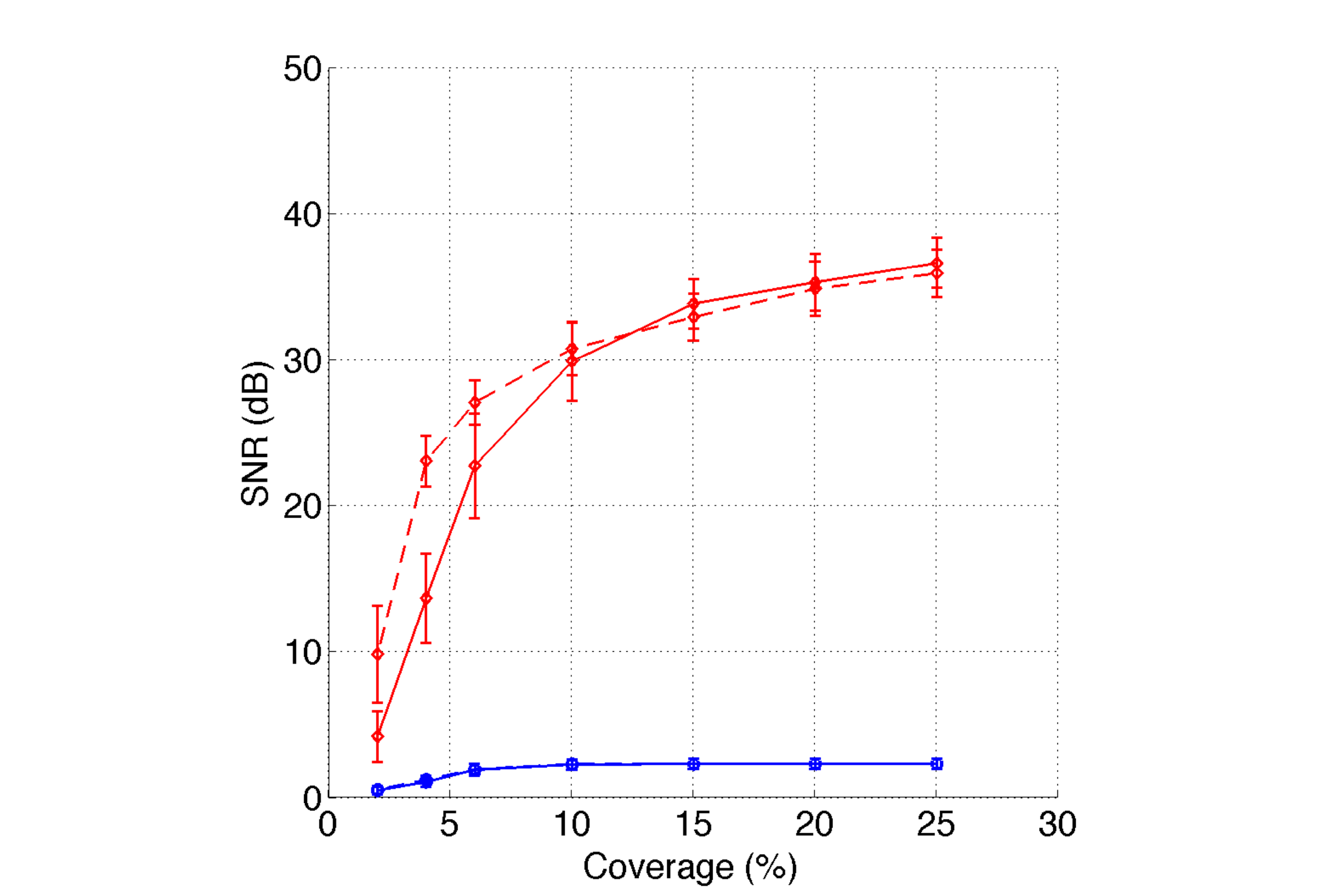}}
\quad\quad
\subfigure[TV reconstruction ($\objsize=0.01$)]{\includegraphics[clip=,viewport=130 0 610 480,width=\reconplotwidth]{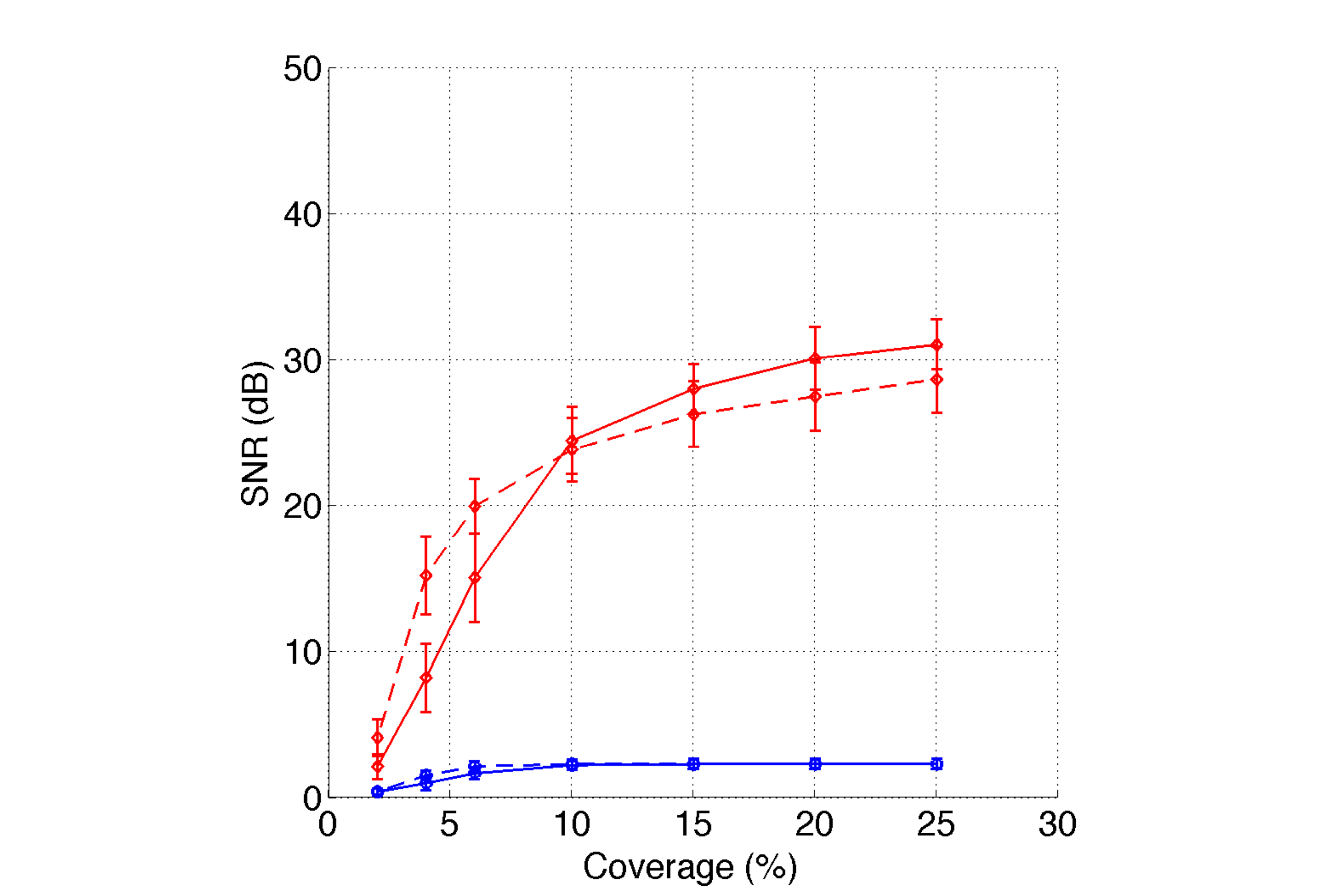}}
}\\
\mbox{
\subfigure[Typical simulation ($\objsize=0.02$)]{\includegraphics[clip=,viewport=180 30 600 480,width=\reconplotwidth]{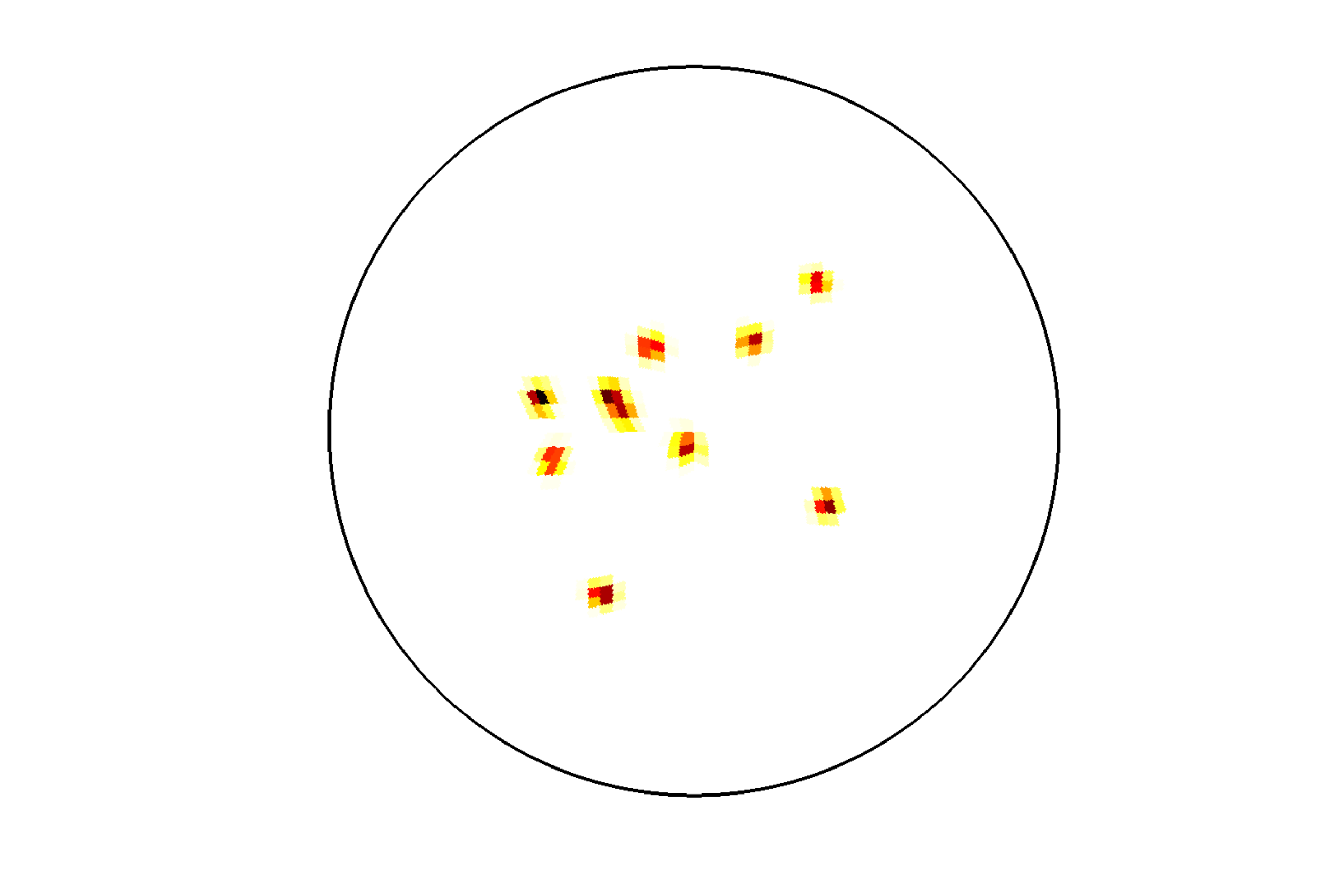}}
\quad\quad
\subfigure[BP reconstruction ($\objsize=0.02$)]{\includegraphics[clip=,viewport=130 0 610 480,width=\reconplotwidth]{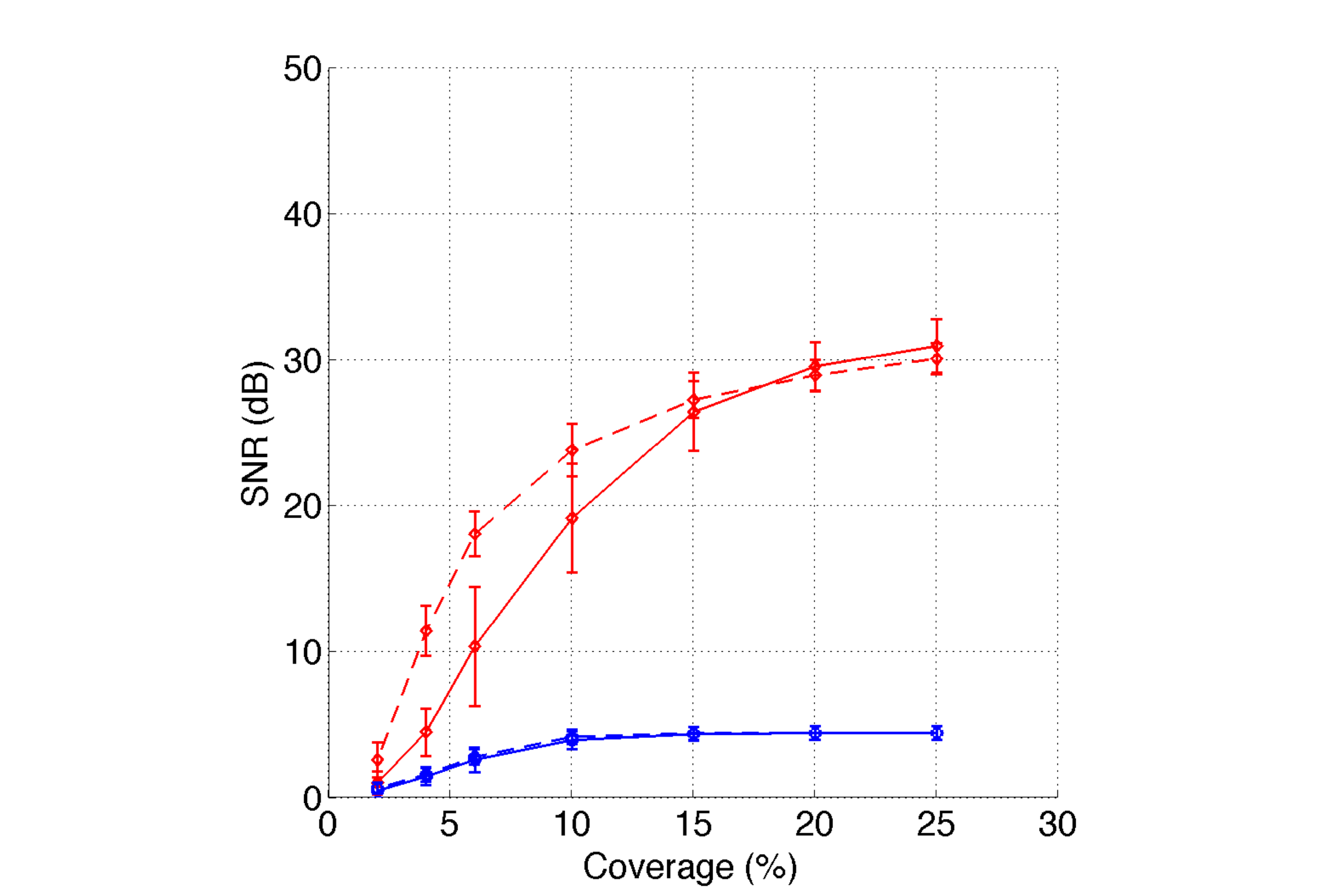}}
\quad\quad
\subfigure[TV reconstruction ($\objsize=0.02$)]{\includegraphics[clip=,viewport=130 0 610 480,width=\reconplotwidth]{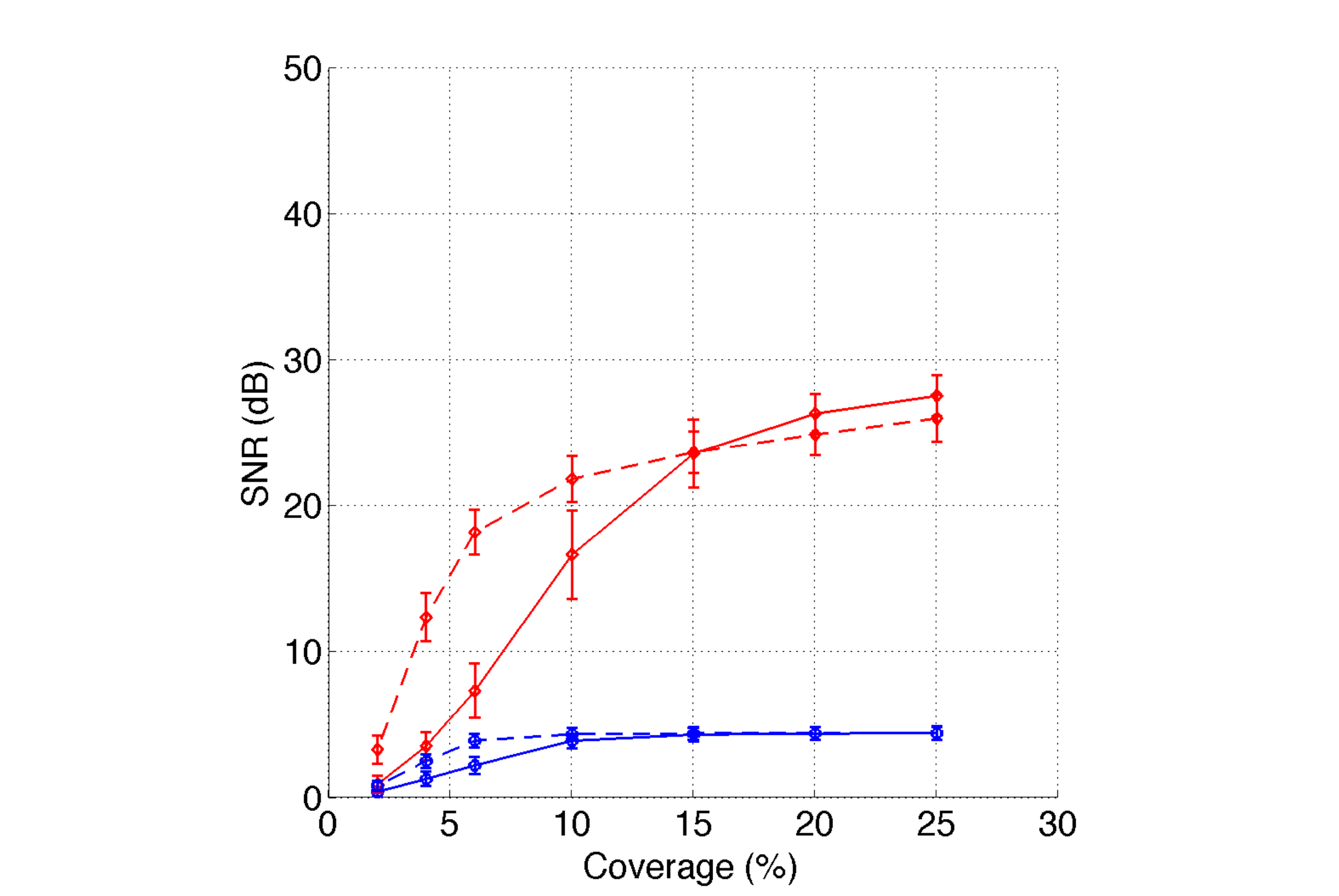}}
}\\
\mbox{
\subfigure[Typical simulation ($\objsize=0.04$)]{\includegraphics[clip=,viewport=180 30 600 480,width=\reconplotwidth]{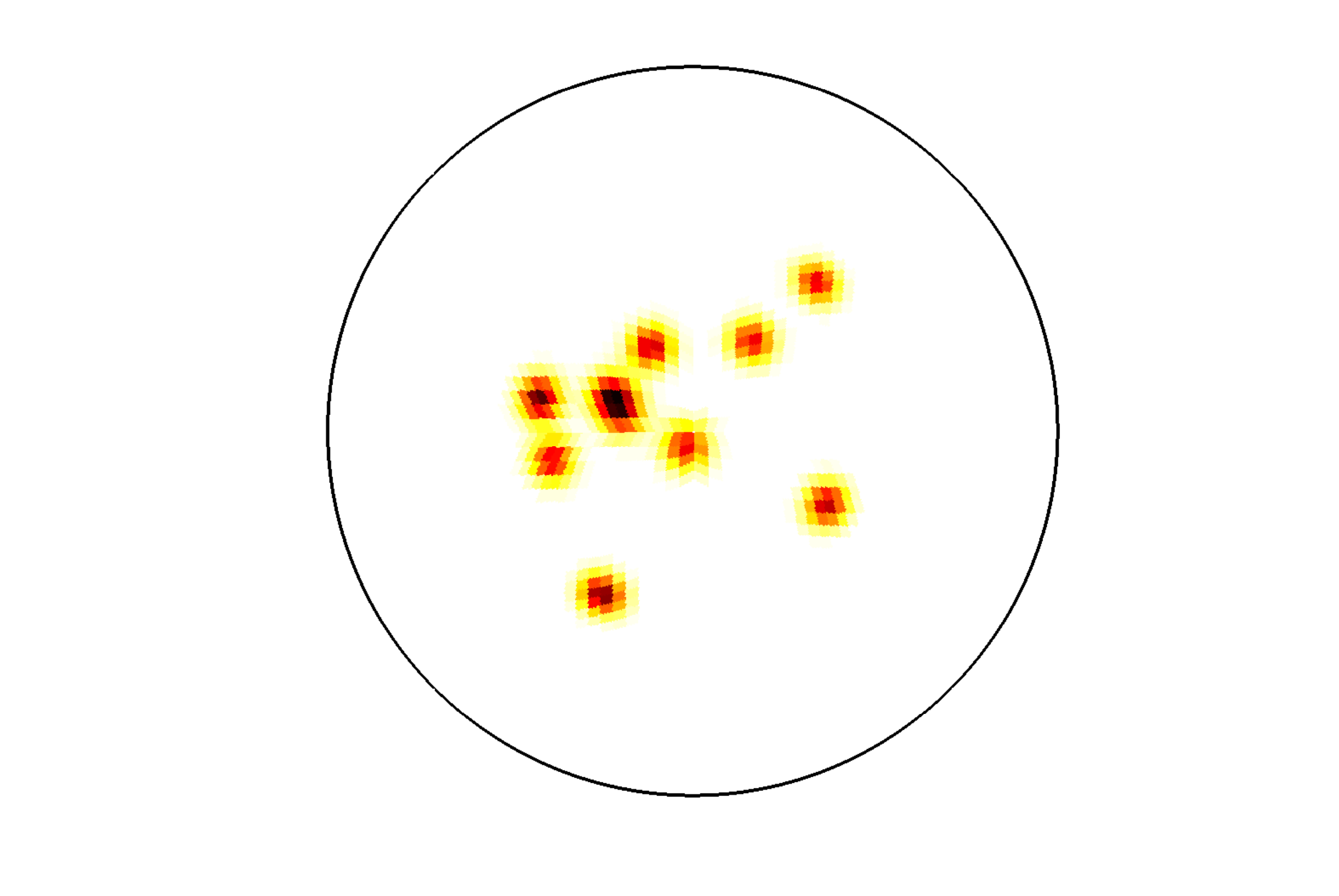}}
\quad\quad
\subfigure[BP reconstruction ($\objsize=0.04$)]{\includegraphics[clip=,viewport=130 0 610 480,width=\reconplotwidth]{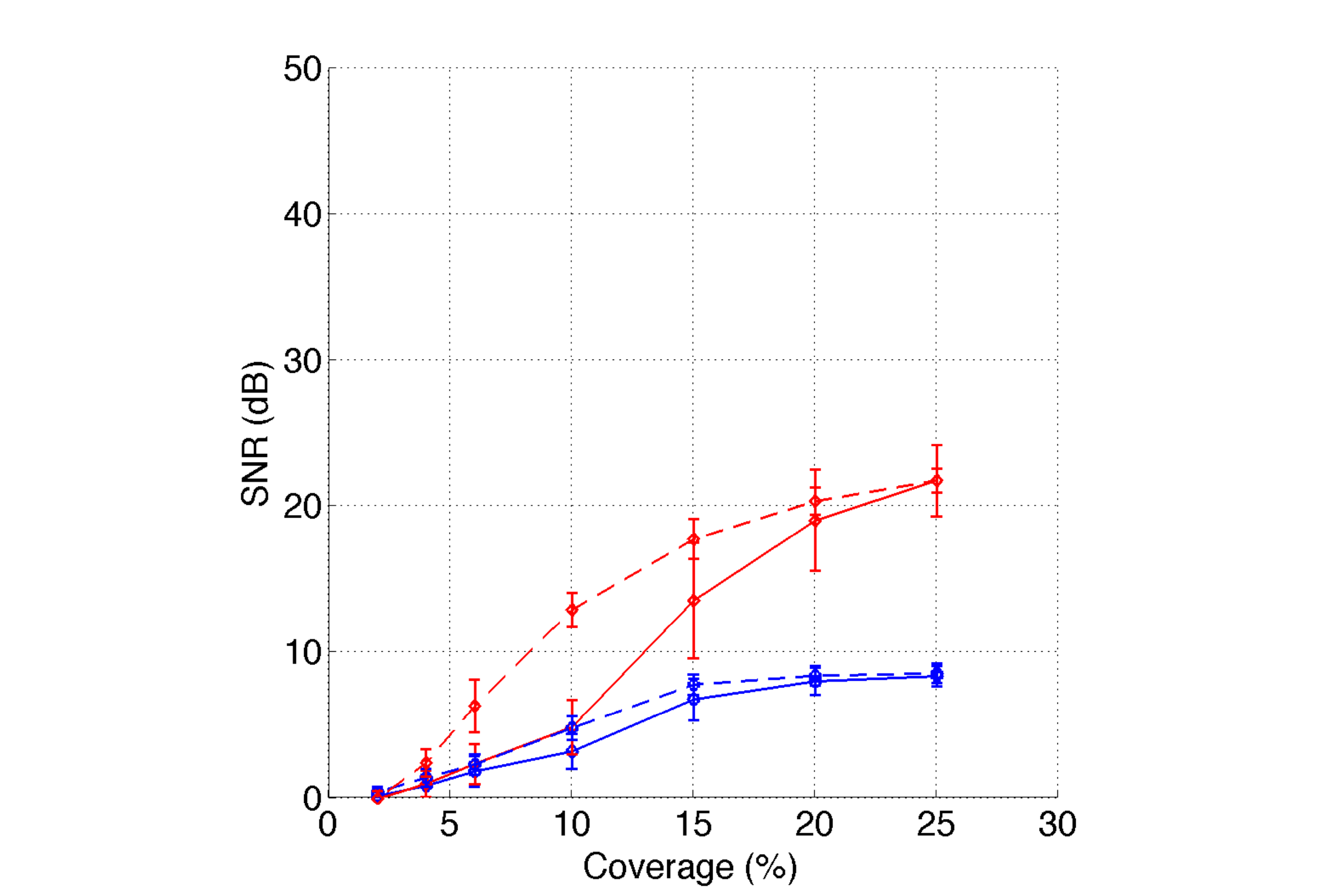}}
\quad\quad
\subfigure[TV reconstruction ($\objsize=0.04$)]{\includegraphics[clip=,viewport=130 0 610 480,width=\reconplotwidth]{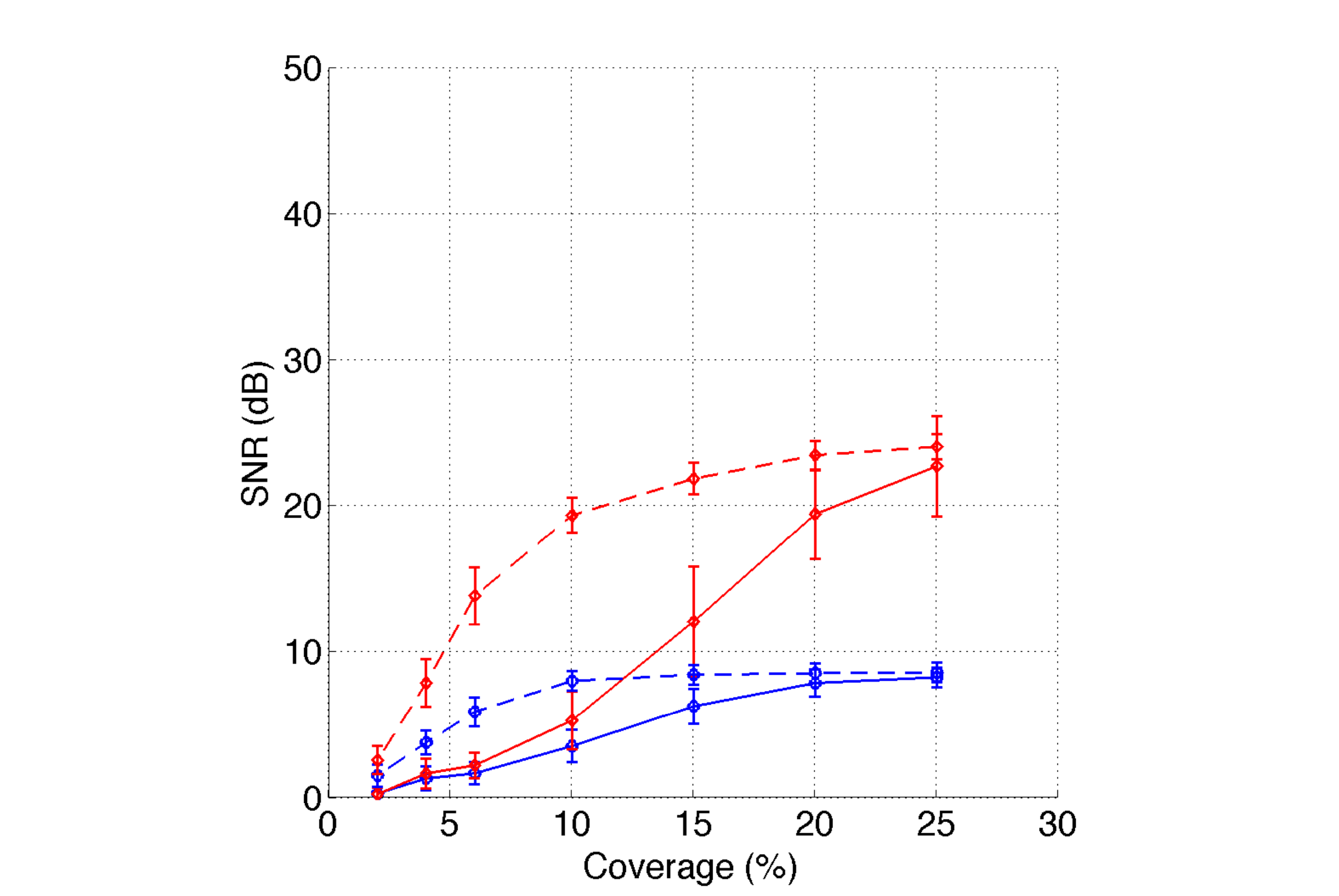}}
}\\
\mbox{
\subfigure[Typical simulation ($\objsize=0.10$)]{\includegraphics[clip=,viewport=180 30 600 480,width=\reconplotwidth]{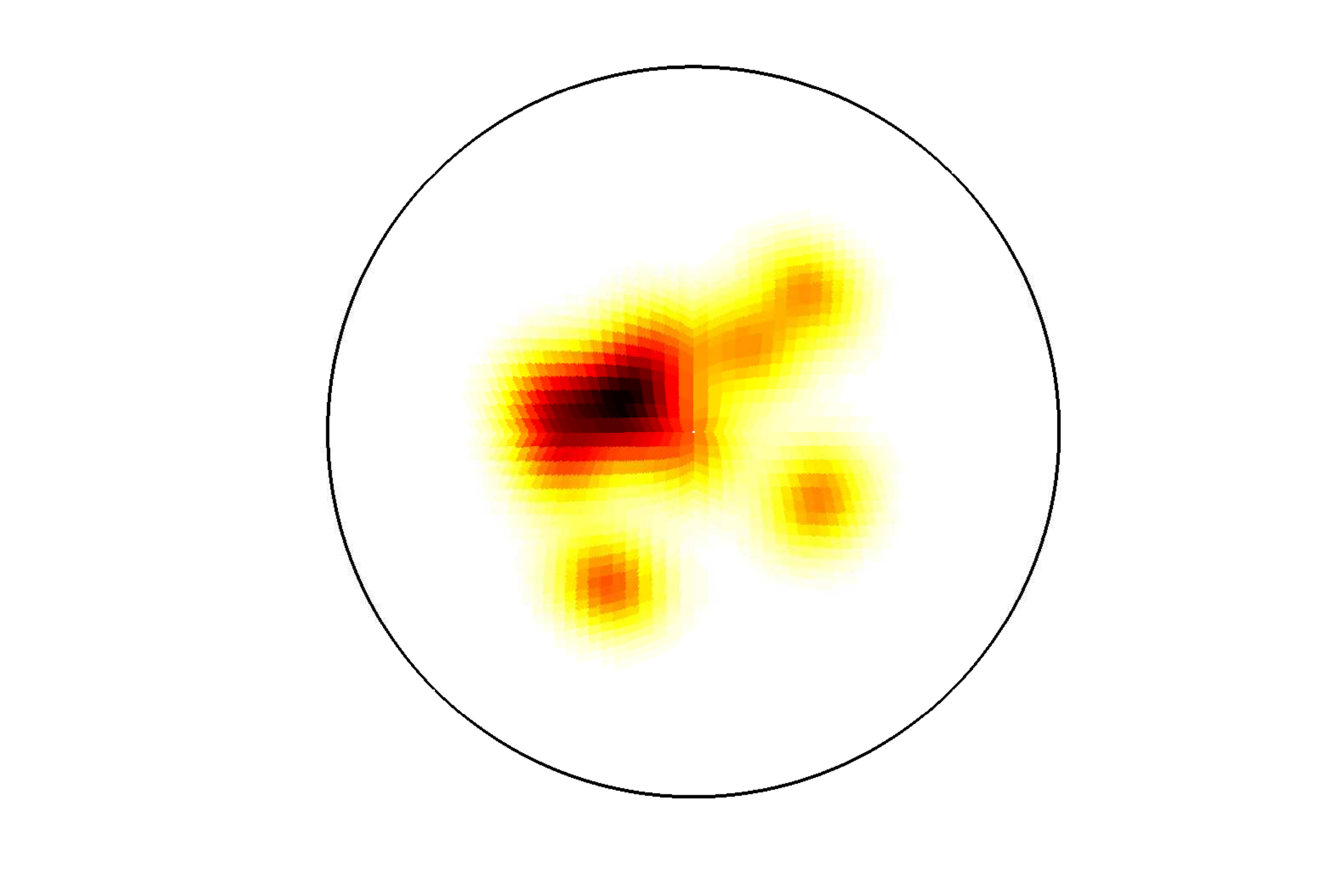}}
\quad\quad
\subfigure[BP reconstruction ($\objsize=0.10$)]{\includegraphics[clip=,viewport=130 0 610 480,width=\reconplotwidth]{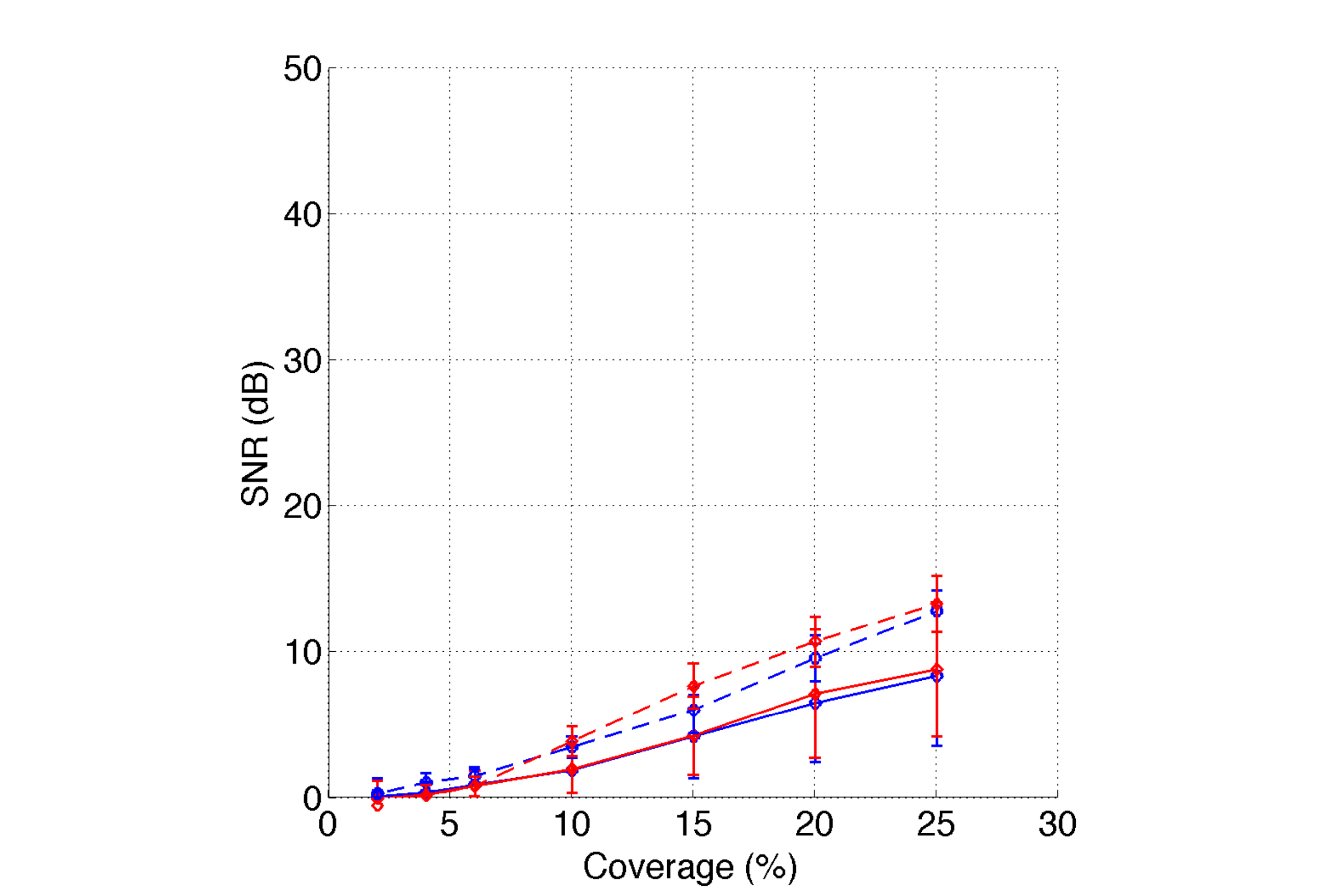}}
\quad\quad
\subfigure[TV reconstruction ($\objsize=0.10$)]{\includegraphics[clip=,viewport=130 0 610 480,width=\reconplotwidth]{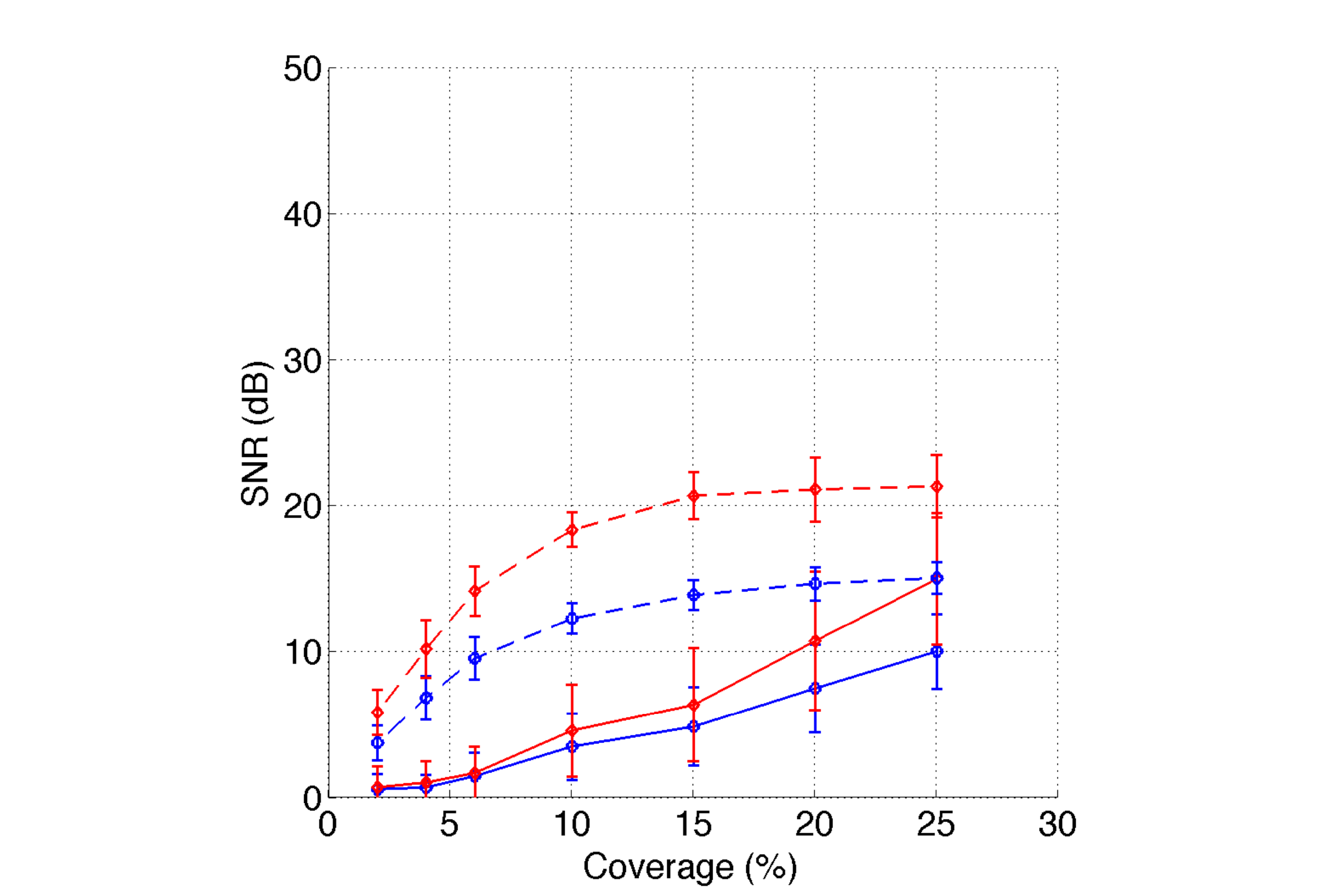}}
}
\caption{Reconstruction quality for simulated Gaussian sources measured by the \snr\ defined on the sphere (\snrs).  Each row of panels corresponds to Gaussian sources of a given size.  The first column of panels shows a typical simulation of Gaussian sources on the sphere.  The remaining columns of panels show reconstruction quality, with the second column illustrating the performance of BP reconstructions, while the third column illustrates the performance of TV reconstructions.  Curves are plotted for reconstructions performed on the sphere (red/diamonds) and on the plane (blue/circles), both in the absence (solid lines) and presence (dashed lines) of the spread spectrum phenomenon.  Reconstruction quality is averaged over 30 simulations for each source size and error bars corresponding to one standard deviation are shown.
}
\label{fig:recon_sphere}
\end{figure*}

\begin{figure*}
\centering
\mbox{
\subfigure[Typical simulation ($\objsize=0.01$)]{\includegraphics[clip=,viewport=220 20 700 530,width=\reconplotwidth]{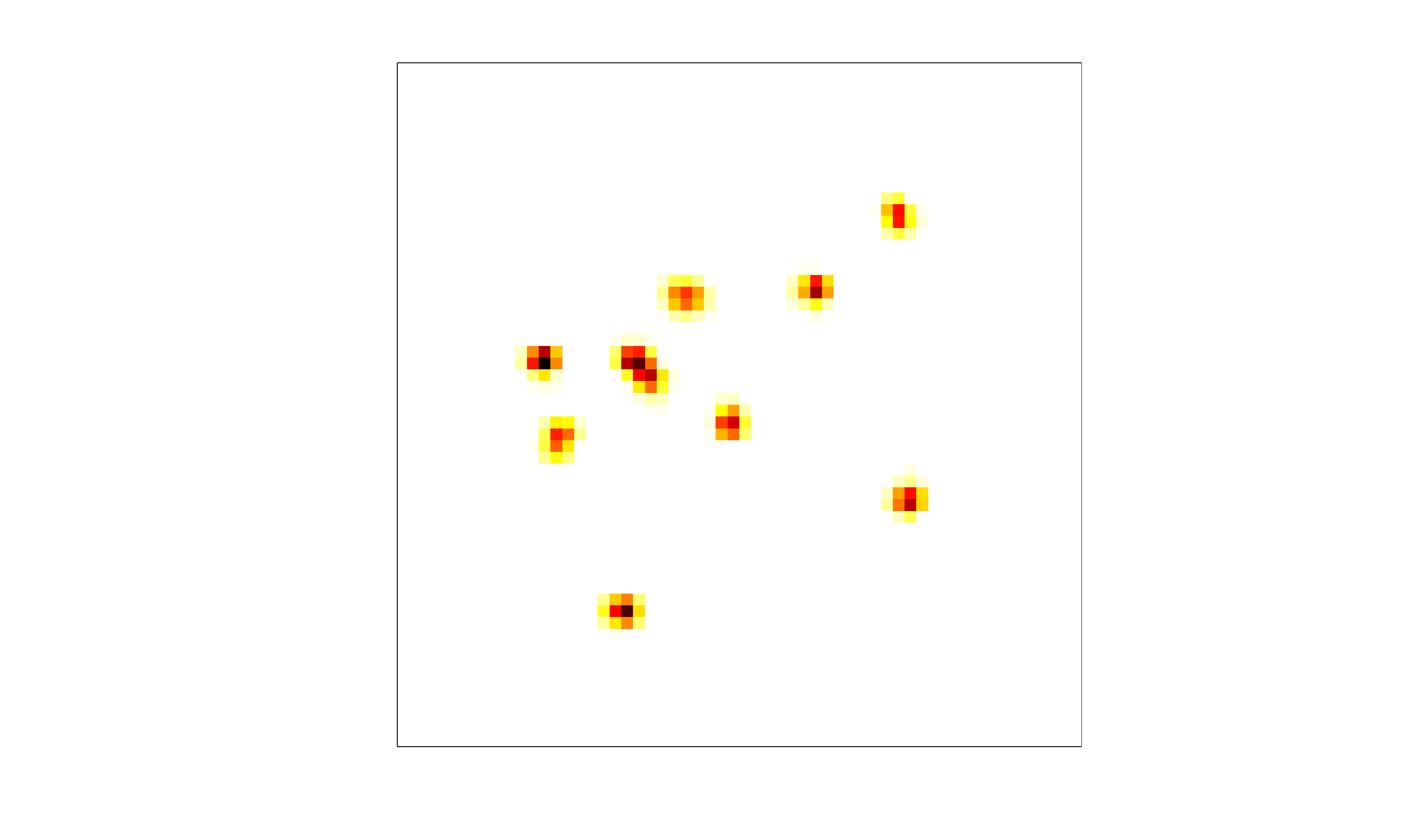}} 
\quad \quad
\subfigure[BP reconstruction ($\objsize=0.01$)]{\includegraphics[clip=,viewport=130 0 610 480,width=\reconplotwidth]{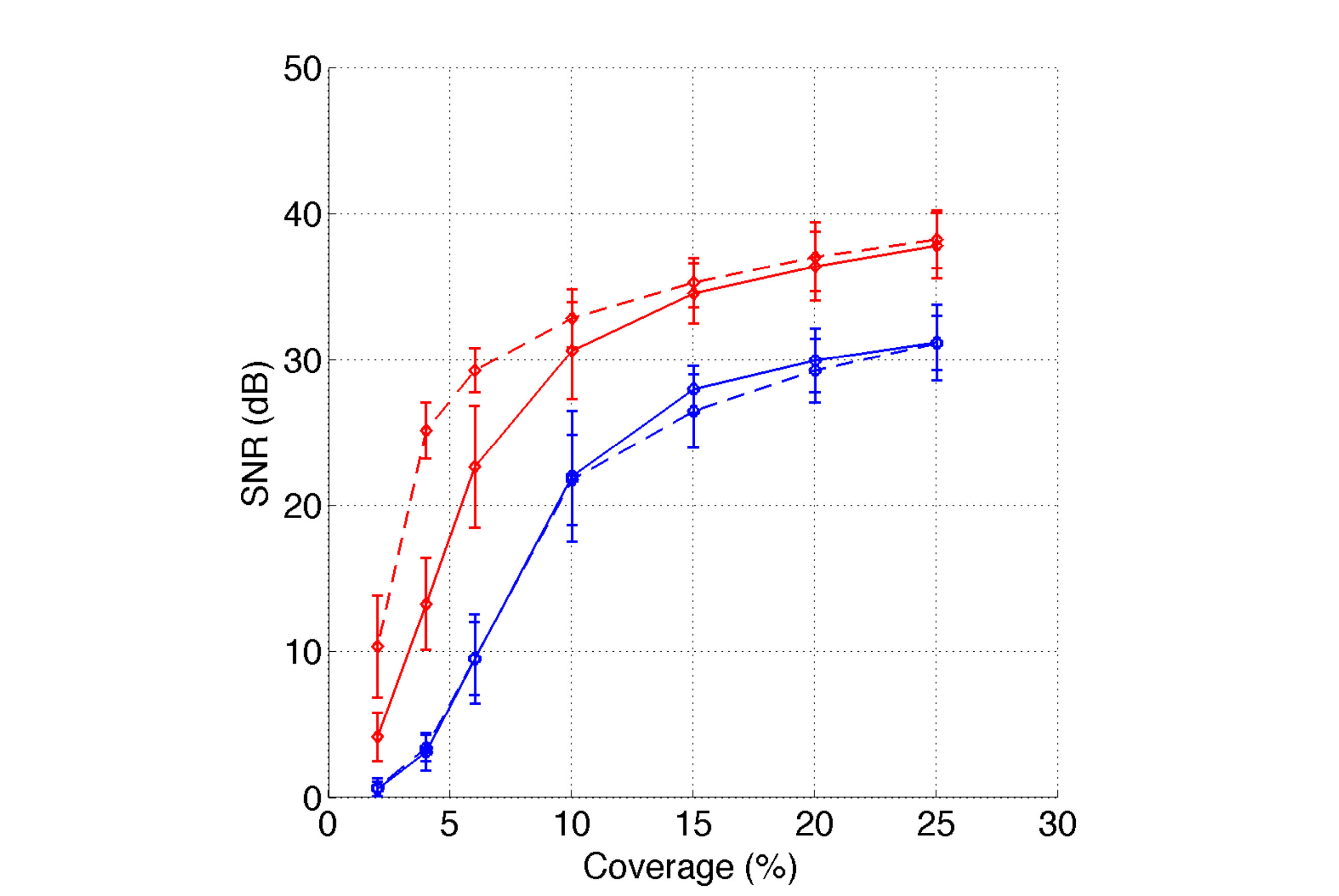}}
\quad \quad
\subfigure[TV reconstruction ($\objsize=0.01$)]{\includegraphics[clip=,viewport=130 0 610 480,width=\reconplotwidth]{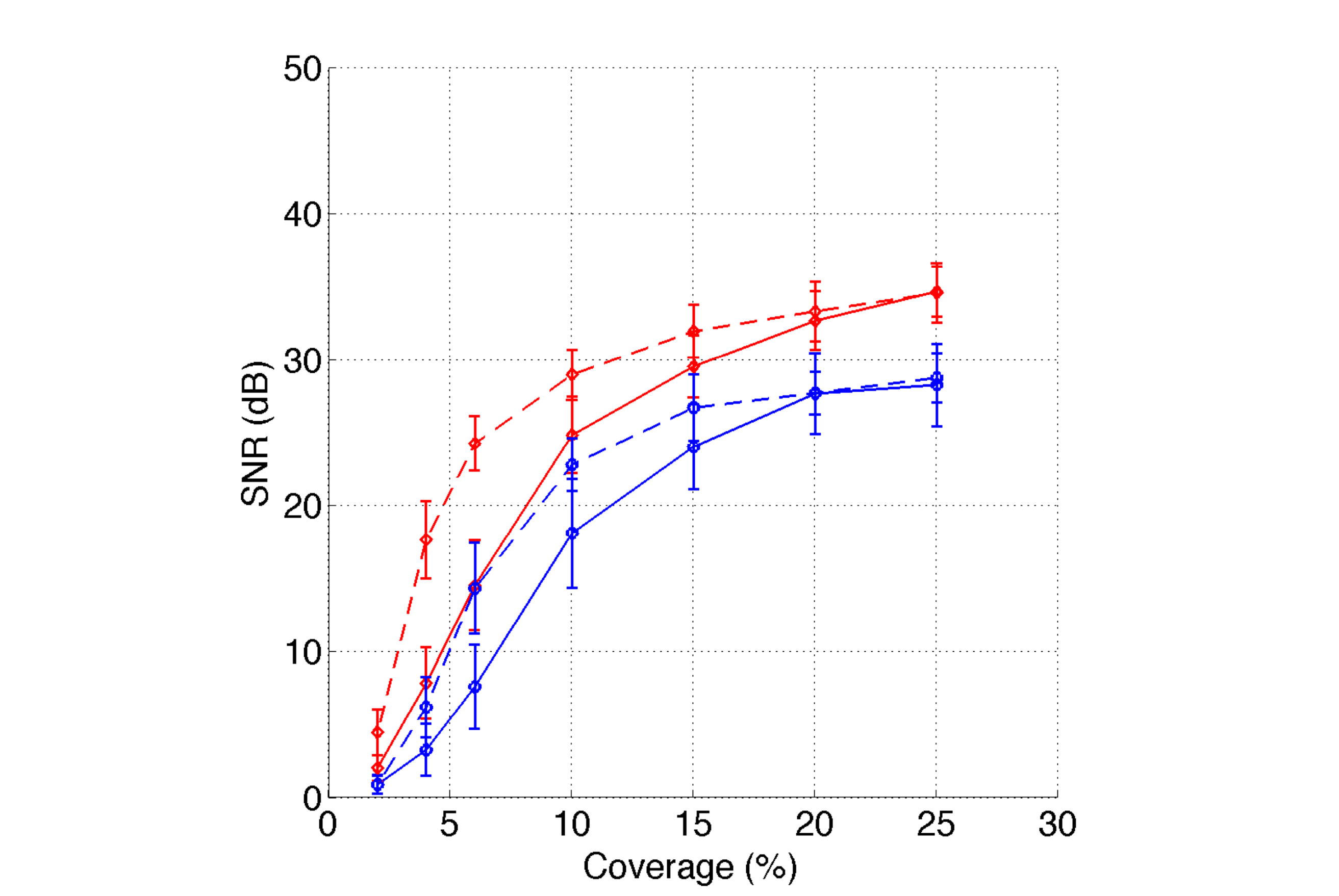}}
}\\
\mbox{
\subfigure[Typical simulation ($\objsize=0.02$)]{\includegraphics[clip=,viewport=220 20 700 530,width=\reconplotwidth]{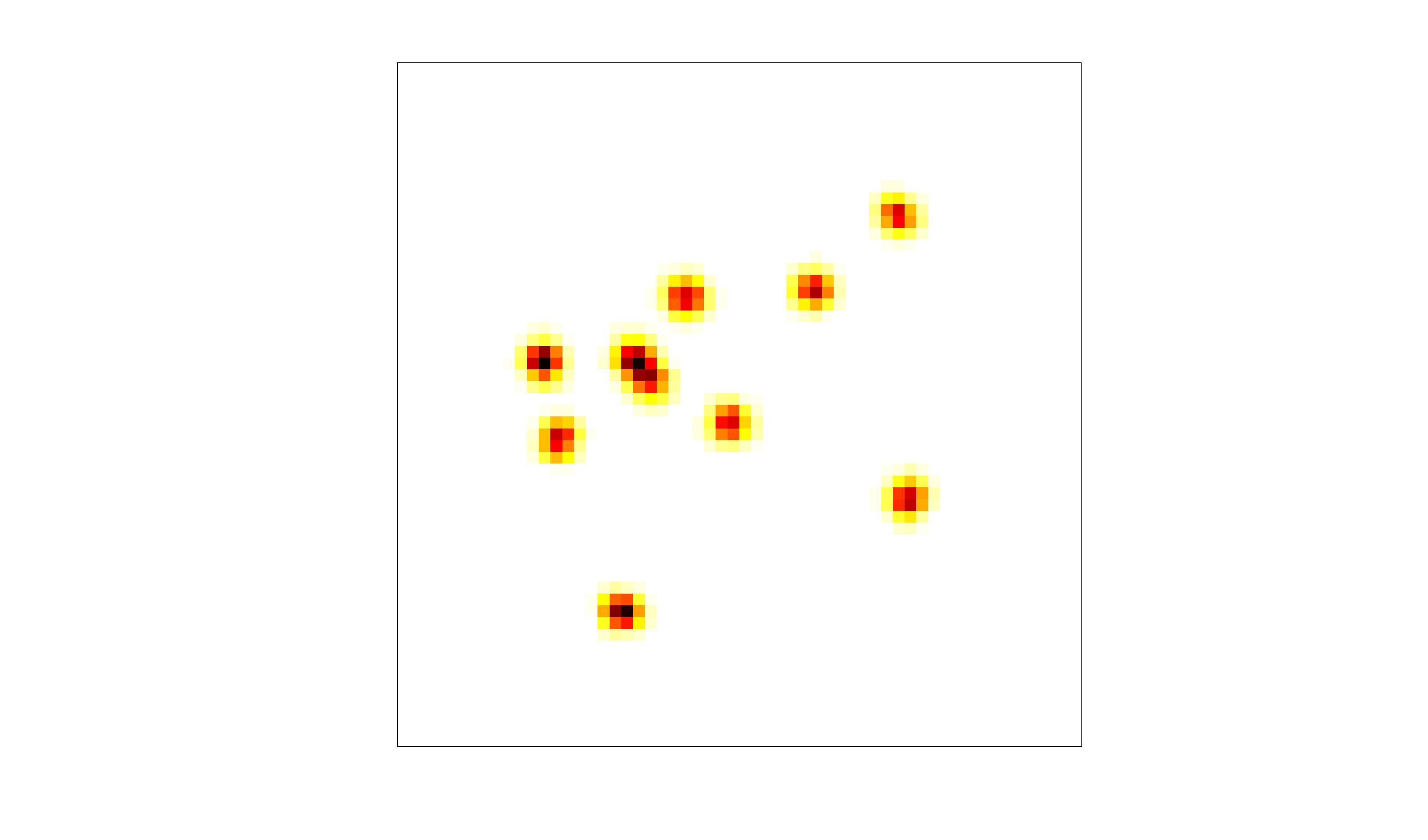}}
\quad \quad
\subfigure[BP reconstruction ($\objsize=0.02$)]{\includegraphics[clip=,viewport=130 0 610 480,width=\reconplotwidth]{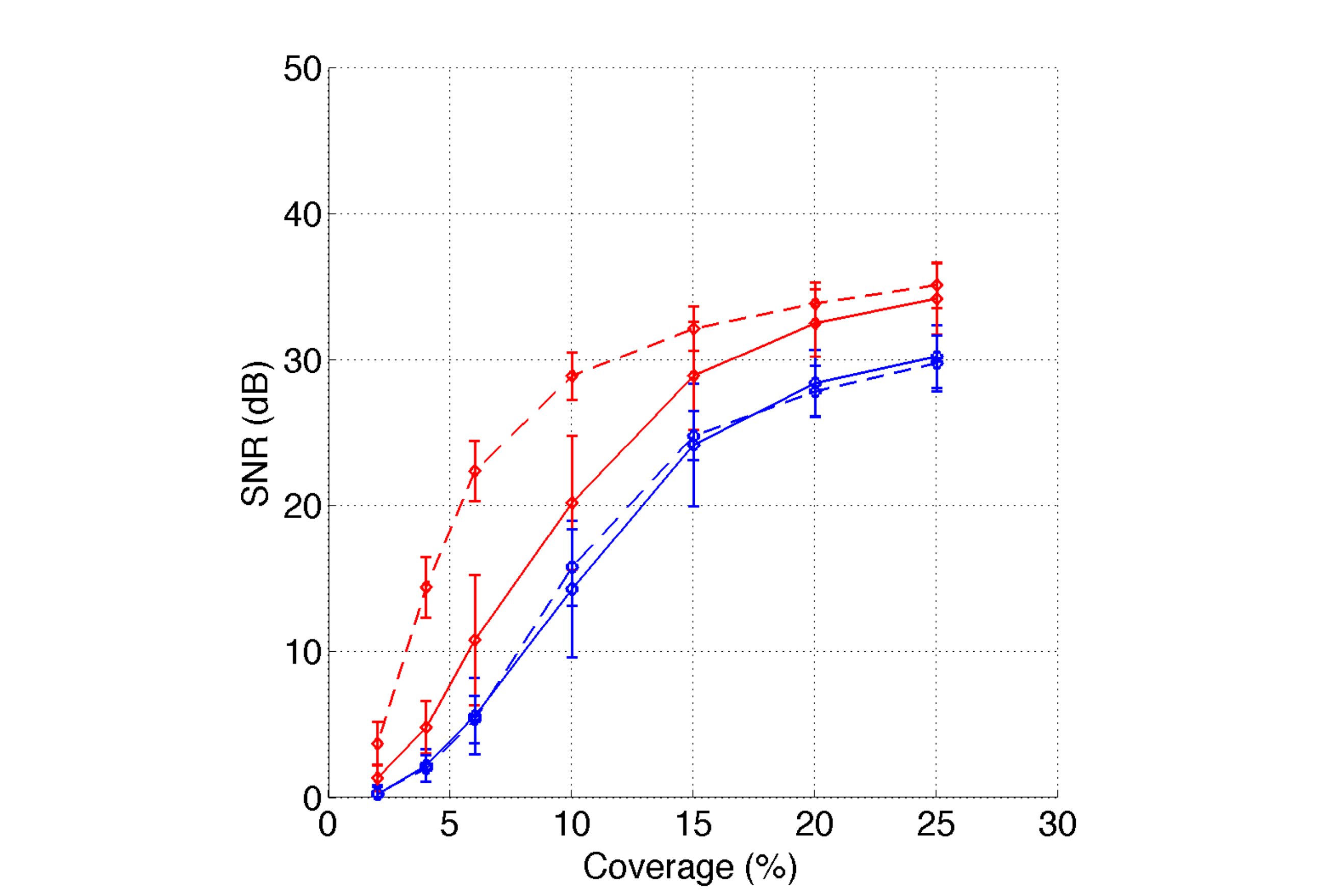}}
\quad \quad
\subfigure[TV reconstruction ($\objsize=0.02$)]{\includegraphics[clip=,viewport=130 0 610 480,width=\reconplotwidth]{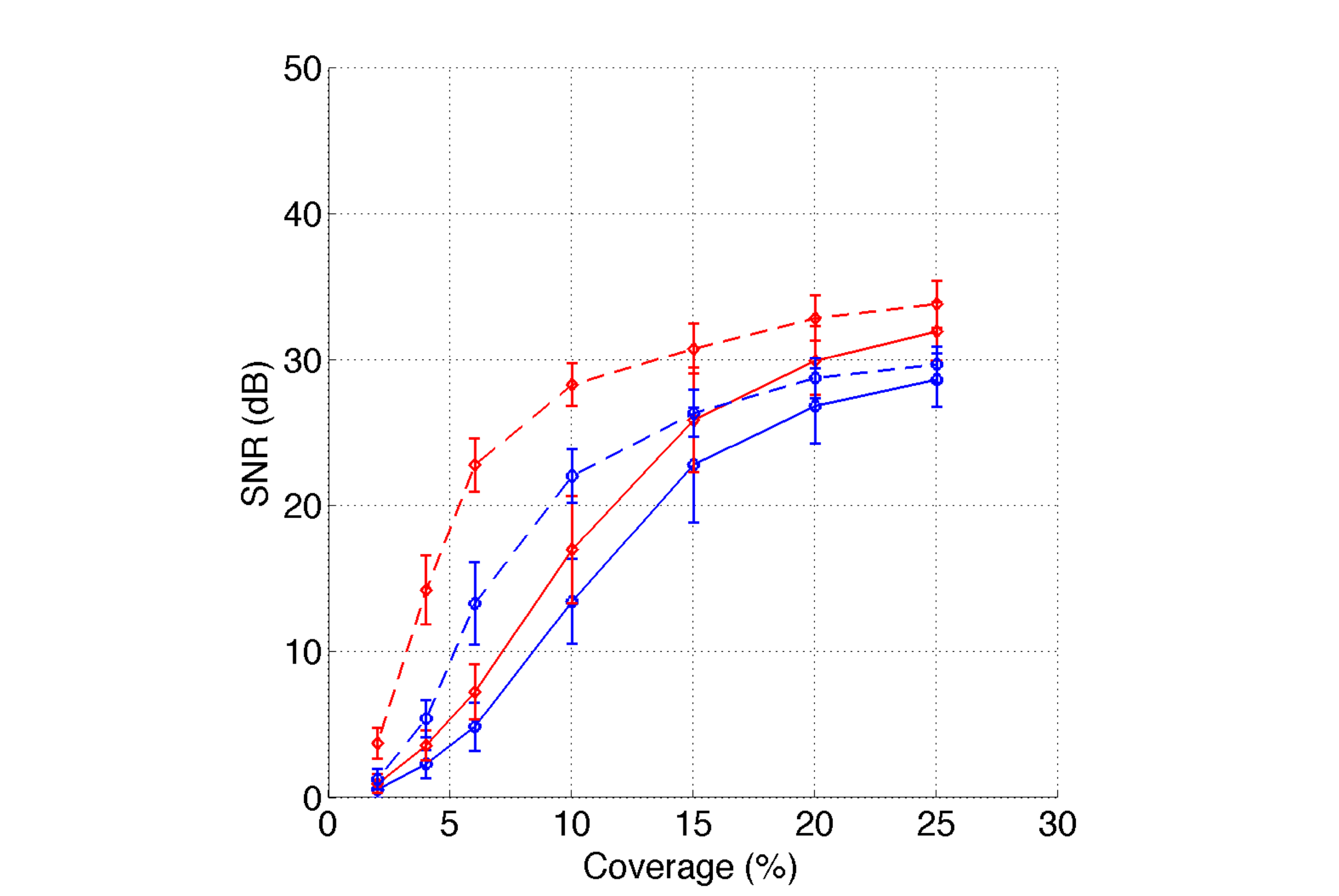}}
}\\
\mbox{
\subfigure[Typical simulation ($\objsize=0.04$)]{\includegraphics[clip=,viewport=220 20 700 530,width=\reconplotwidth]{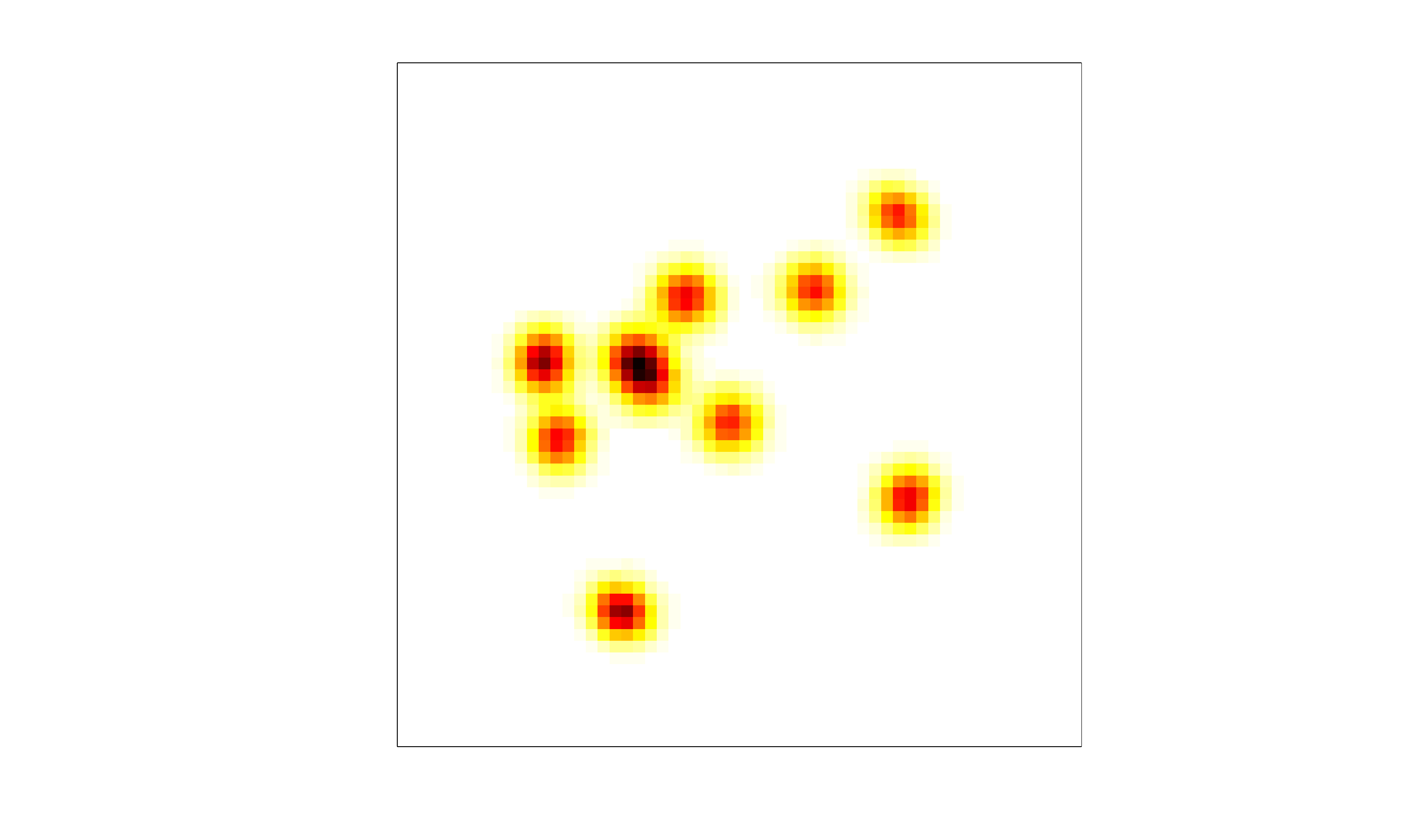}}
\quad \quad
\subfigure[BP reconstruction ($\objsize=0.04$)]{\includegraphics[clip=,viewport=130 0 610 480,width=\reconplotwidth]{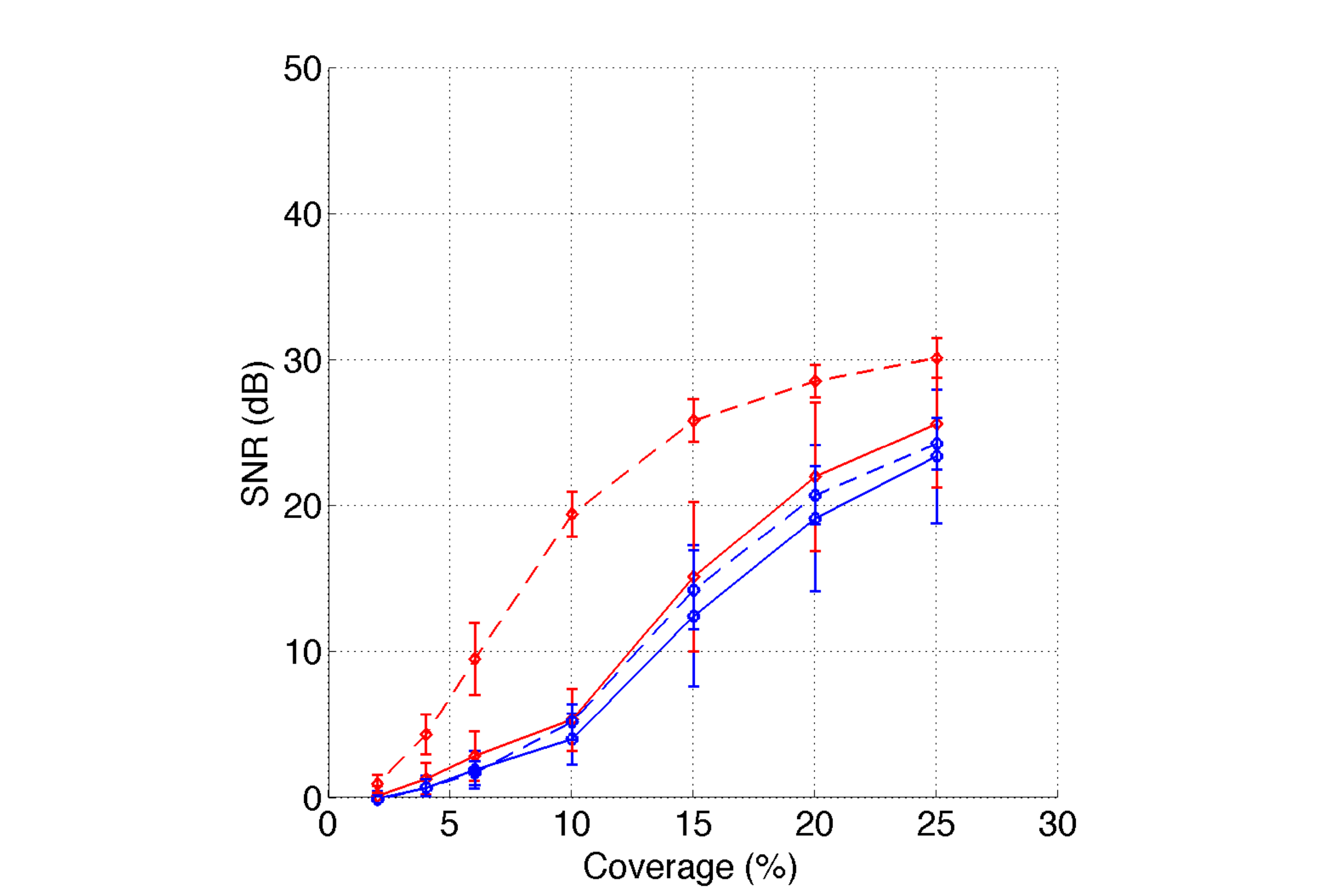}}
\quad \quad
\subfigure[TV reconstruction ($\objsize=0.04$)]{\includegraphics[clip=,viewport=130 0 610 480,width=\reconplotwidth]{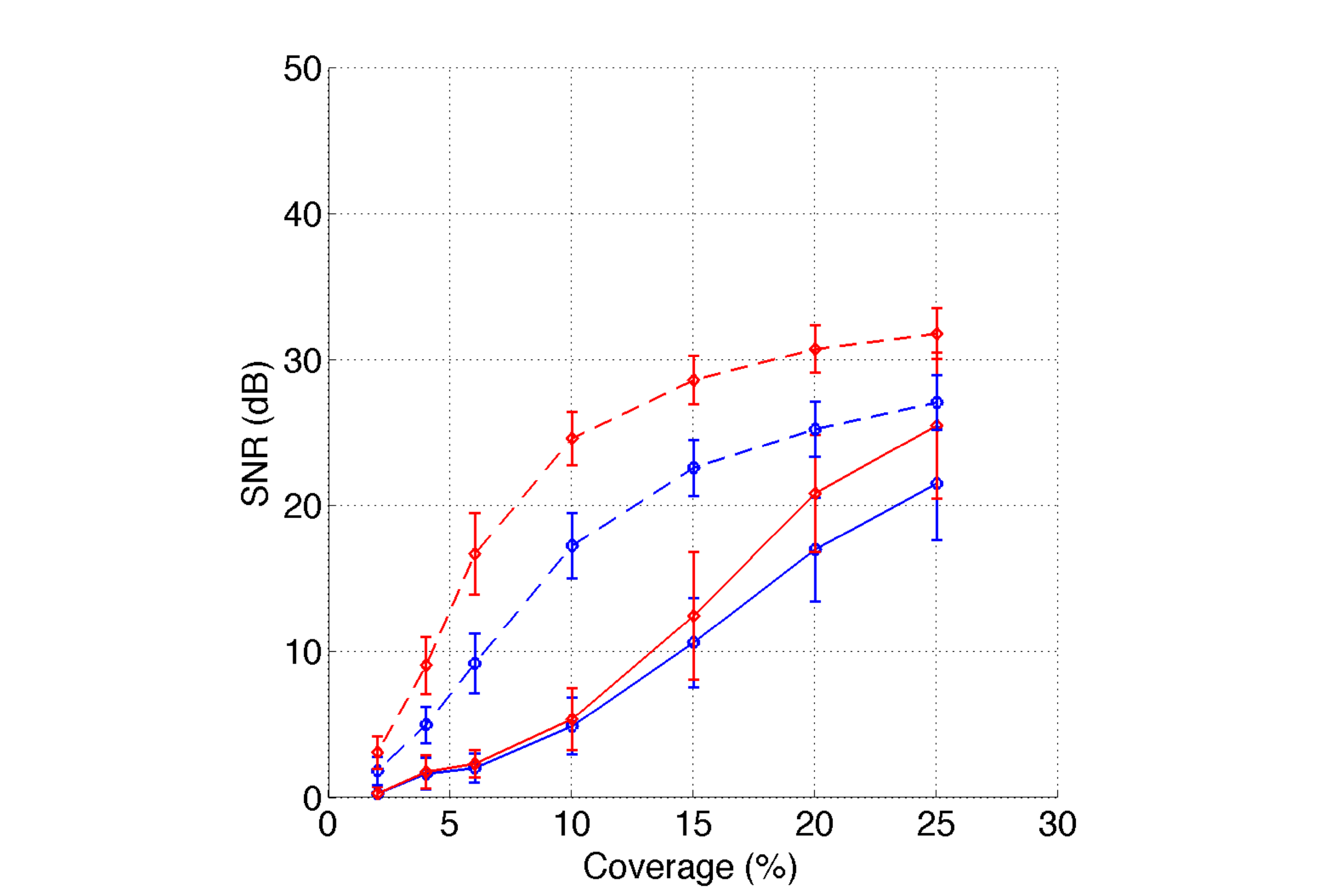}}
}\\
\mbox{
\subfigure[Typical simulation ($\objsize=0.10$)]{\includegraphics[clip=,viewport=220 20 700 530,width=\reconplotwidth]{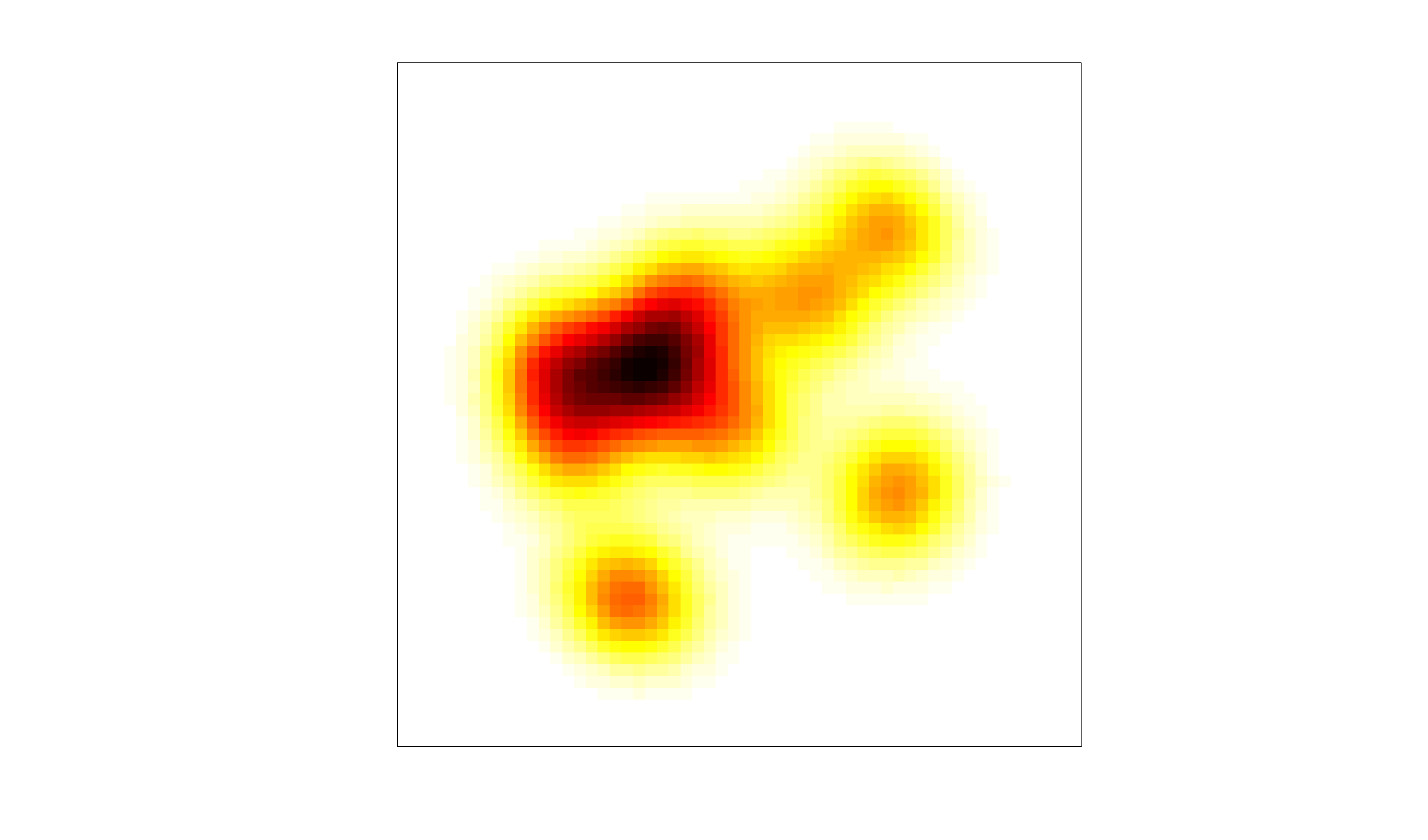}}
\quad \quad
\subfigure[BP reconstruction ($\objsize=0.10$)]{\includegraphics[clip=,viewport=130 0 610 480,width=\reconplotwidth]{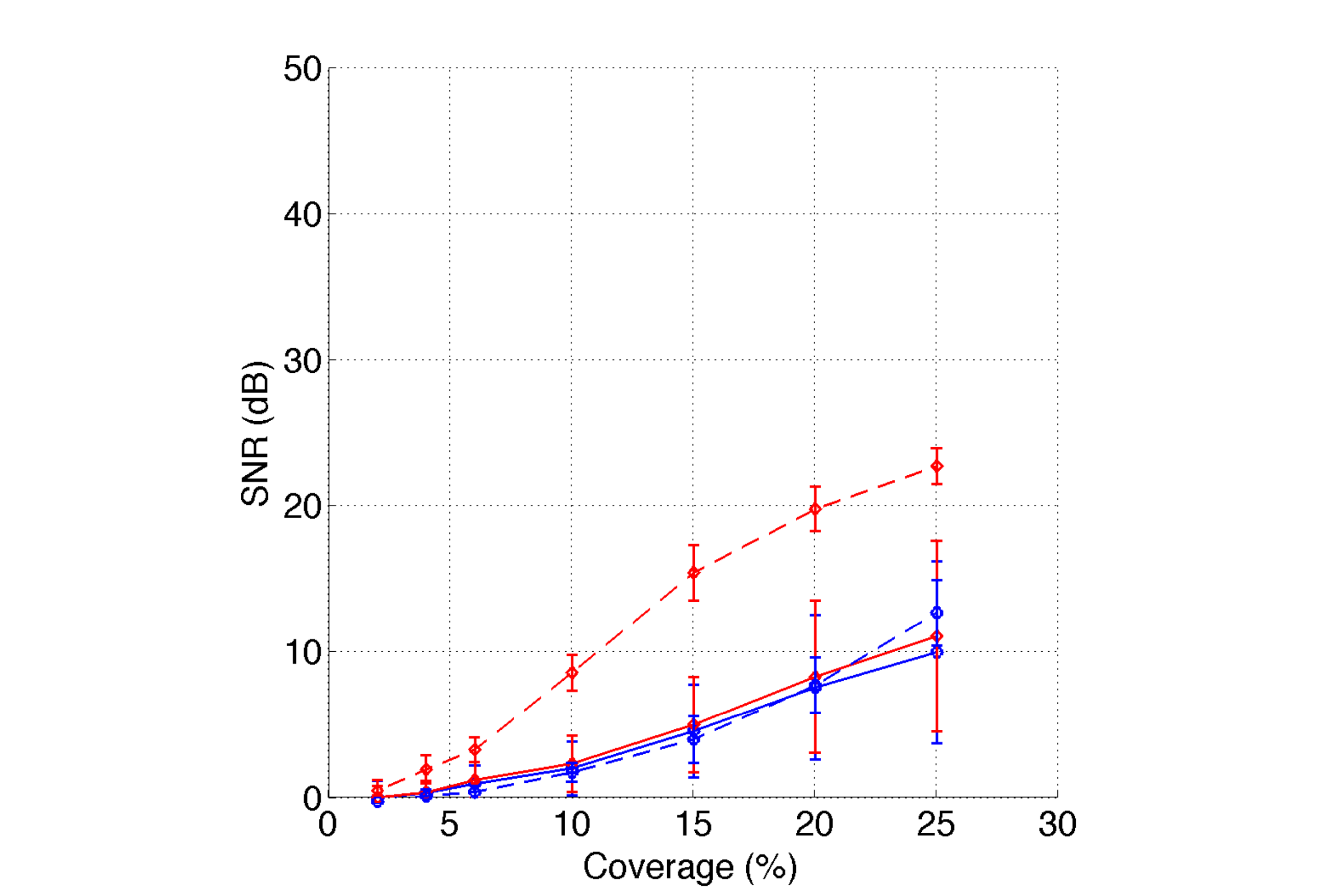}}
\quad \quad
\subfigure[TV reconstruction ($\objsize=0.10$)]{\includegraphics[clip=,viewport=130 0 610 480,width=\reconplotwidth]{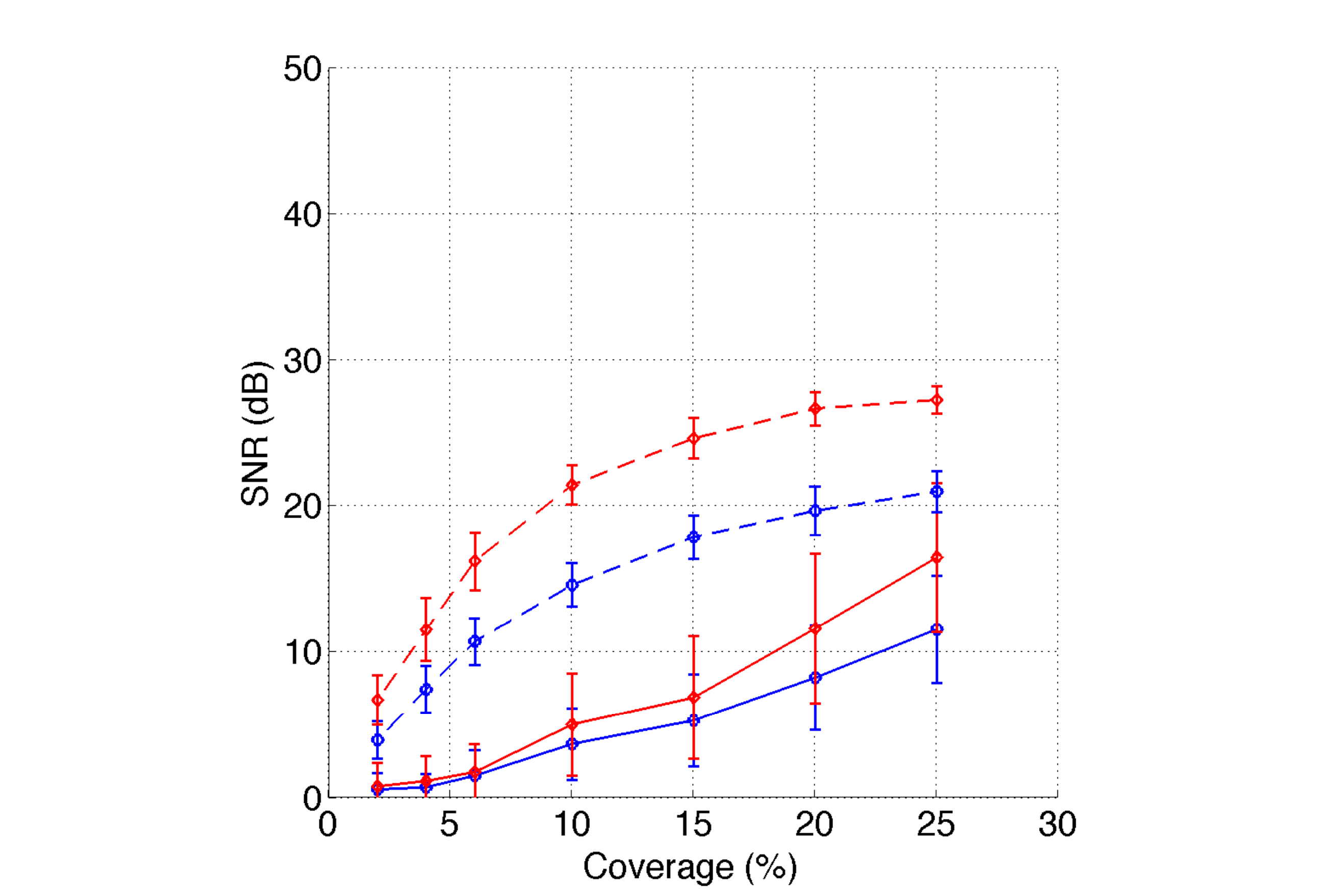}}
}
\caption{Reconstruction quality for simulated Gaussian sources measured by the \snr\ defined on the plane (\snrp).  Each row of panels corresponds to Gaussian sources of a given size.  The first column of panels shows a typical simulation of Gaussian sources projected onto the plane.  The remaining columns of panels show reconstruction quality, with the second column illustrating the performance of BP reconstructions, while the third column illustrates the performance of TV reconstructions.  Curves are plotted for reconstructions performed on the sphere (red/diamonds) and on the plane (blue/circles), both in the absence (solid lines) and presence (dashed lines) of the spread spectrum phenomenon.  Reconstruction quality is averaged over 30 simulations for each source size and error bars corresponding to one standard deviation are shown.
}
\label{fig:recon_plane}
\end{figure*}

\subsubsection{Comparison of performance with sparsity and coherence}

Although mutual coherence is a useful measure for theoretical and intuitive considerations, it is not well suited to numerical computation (as discussed in \sectn{\ref{sec:background:cs}}).  Cumulative coherence is a more robust measure of coherence and thus is more suitable for numerical evaluation. Furthermore, cumulative coherence is signal dependent and incorporates sparsity information.  We therefore use cumulative coherence to study the combined sparsity and coherence of the interferometric imaging framework in the context of Gaussian sources and relate this to the reconstruction performance presented in the previous section.  However, we may only study cumulative coherence in the presence of a sparsity basis, thus the analysis and discussion here is restricted to BP reconstructions only. 

In a strict compressed sensing framework with orthogonal measurement and sparsity bases and in the absence of noise, the number of measurements required to reconstruct a signal accurately from incomplete random measurements evolves as \eqn{\ref{eqn:ccoherence}}.  The square of the normalised cumulative coherence $\ccoherence \sqrt{\ndim}$ provides an approximate relative measure of the number of measurements required to recover a signal, or similarly, the quality of the reconstruction performance for a given number of measurements.  Although the interferometric imaging framework we consider differs from the theoretical compressed sensing framework, the normalised cumulative coherence nevertheless provides a measure of the impact of sparsity and coherence on reconstruction performance.

We compute cumulative coherences for all of the sets of Gaussian simulations, averaging over the 30 simulations for each source size.  Normalised cumulative coherences are computed on the plane and sphere, denoted $\ccoherencep(\bwd)\sqrt{\nplane}$ and $\ccoherences(\bwd)\sqrt{\nsphere}$ respectively, both in the presence ($\bwd=1/\sqrt{2}$) and absence ($\bwd=0$) of the spread spectrum phenomenon.  Results are displayed in \tbl{\ref{tbl:coherences}}.  Four insights can be made by studying the coherence values reported in \tbl{\ref{tbl:coherences}}.  Firstly, the coherences on the plane are not significantly altered in the presence of the spread spectrum phenomenon, indicating that the spread spectrum phenomenon is likely to be ineffective in improving reconstruction performance in this setting.  Secondly, the coherences on the sphere are reduced by application of the spread spectrum phenomenon, indicating that the spread spectrum phenomenon is likely to be effective in this setting. Thirdly, we note that coherences on the sphere are always lower than the corresponding values on the plane, highlighting the superiority of spherical reconstruction.  Fourthly, we note that cumulative coherences increase with source size, indicating that reconstruction performance should reduce.  All of these findings verify our intuitive expectations and are consistent with the reconstruction performance presented in the previous section.

% \begin{table}
% \begin{center}
% \begin{tabular}{ccccccc} \hline
% Source && \multicolumn{2}{c}{Plane} && \multicolumn{2}{c}{Sphere} \\ 
% $\objsize$ && \multicolumn{2}{c}{$\ccoherencep(\bwd)\sqrt{\nplane}$} && \multicolumn{2}{c}{$\ccoherences(\bwd)\sqrt{\nsphere}$} \\ 
% %$\objsize$ && \multicolumn{2}{c}{$\ccoherencep(\bwd)\sqrt{\nplane}$} && \multicolumn{2}{c}{$\ccoherences(\bwd)\sqrt{\nsphere}$} \\ 
% && $\bwd=0$ & $\bwd=\frac{1}{\sqrt{2}}$                          && $\bwd=0$ & $\bwd=\frac{1}{\sqrt{2}}$\\ \hline
% $0.01$      && $8.1\pm0.7$          & $8.3\pm0.8$              && $5.1\pm0.5$            & $4.7\pm0.6$  \\
% $0.02$      && $11.1\pm1.0$        & $11.2\pm1.0$            && $8.2\pm0.8$            & $7.6\pm0.9$  \\
% $0.04$      && $16.3\pm1.2$        & $16.4\pm1.2$            && $13.9\pm1.2$          & $12.8\pm1.3$  \\
% $0.10$      && $22.3\pm0.3$        & $22.3\pm0.3$            && $19.8\pm0.3$          & $18.2\pm0.3$  \\ \hline
% \end{tabular}
% \end{center}
% \caption{Normalised cumulative coherences on the plane and sphere of simulations of Gaussian sources for various size \objsize.  Coherences values computed both in the presence ($\bwd=1/\sqrt{2}$) and absence ($\bwd=0$) of the spread spectrum phenomenon are displayed.  Quoted values correspond to the mean of 30 simulations, with errors corresponding to one standard deviation.  These measures incorporate both the impact of sparsity and coherence on reconstruction performance (with lower values indicating superior performance).
% }
% \label{tbl:coherences}
% \end{table}

\begin{table*}
\begin{center}
\begin{tabular}{cccccccccccc} \hline
Source     && \multicolumn{3}{c}{Plane}                                        && \multicolumn{3}{c}{Sphere}                                   && \multicolumn{2}{c}{Difference} \\ 
$\objsize$ && \multicolumn{3}{c}{$\ccoherencep(\bwd)\sqrt{\nplane}$}           && \multicolumn{3}{c}{$\ccoherences(\bwd)\sqrt{\nsphere}$}      && \multicolumn{2}{c}{$\Delta(\bwd)$}\\ 
           && $\bwd=0$            & $\bwd=\frac{1}{\sqrt{2}}$ & $\Delta_\pind$   && $\bwd=0$      & $\bwd=\frac{1}{\sqrt{2}}$ & $\Delta_\sind$    && $\bwd=0$       & $\bwd=\frac{1}{\sqrt{2}}$ \\ \hline
$0.01$     && $8.14\pm0.72$       & $8.31\pm0.78$             & $0.17\pm0.10$  && $5.07\pm0.52$   & $4.69\pm0.55$           & $-0.38\pm0.06$   && $-3.08\pm0.23$ & $-3.62\pm0.25$ \\
$0.02$     && $11.05\pm0.95$      & $11.17\pm1.00$            & $0.12\pm0.08$  && $8.23\pm0.83$   & $7.64\pm0.88$           & $-0.59\pm0.08$   && $-2.82\pm0.15$ & $-3.53\pm0.16$ \\
$0.04$     && $16.32\pm1.19$      & $16.36\pm1.23$            & $0.05\pm0.05$  && $13.85\pm1.17$  & $12.84\pm1.25$          & $-1.01\pm0.11$   && $-2.46\pm0.07$ & $-3.52\pm0.13$ \\
$0.10$     && $22.32\pm0.25$      & $22.33\pm0.25$            & $0.01\pm0.01$  && $19.79\pm0.25$  & $18.23\pm0.25$          & $-1.73\pm0.03$   && $-2.53\pm0.02$ & $-4.12\pm0.05$ \\ \hline
\end{tabular}
\end{center}
\caption{Normalised cumulative coherences, on the plane and sphere, of simulations of Gaussian sources for various size \objsize.  Coherences values computed both in the presence ($\bwd=1/\sqrt{2}$) and absence ($\bwd=0$) of the spread spectrum phenomenon are displayed.  Normalised coherence differences between the presence and absence of the spread spectrum phenomenon are computed on the plane and sphere (with $\Delta_\pind$ and $\Delta_\sind$ denoting, respectively, the coherence for $\bwd=0$ subtracted from the coherence for $\bwd=1/\sqrt{2}$).  In addition, normalised coherence differences between the plane and sphere are also computed, with $\Delta(\bwd) = \ccoherencep(\bwd)\sqrt{\nplane} - \ccoherences(\bwd)\sqrt{\nsphere}$.  Quoted values correspond to the mean of 30 simulations, with errors corresponding to one standard deviation.  Notice that coherence differences are more stable than coherence values over the simulations.  These measures incorporate both the impact of sparsity and coherence on reconstruction performance (with lower values indicating superior performance).
}
\label{tbl:coherences}
\end{table*}

%=============================================================================
\subsection{Galactic dust}

Now that the \wfov\ interferometric imaging framework developed in this article has been evaluated thoroughly, in this section we simply illustrate the reconstruction, at a higher resolution, of a more realistic simulation of Galactic dust emission.  The simulation is first described, before reconstruction performance is evaluated.

\subsubsection{Simulation}

For this higher resolution simulation we consider the 94GHz map of predicted submillimeter and microwave emission of diffuse interstellar Galactic dust \citep{finkbeiner:1999}, hereafter referred to as the FDS map.  This predicted map is based on the merged Infrared Astronomy Satellite (IRAS) and Cosmic Background Explorer Diffuse Infrared Background Experiment (COBE-DIRBE) observations produced by \citet{schlegel:1998}.  An undersampled version of the FDS map is available from the \lambdaarchtext\ \footnote{\url{http://lambda.gsfc.nasa.gov/}} (\lambdaarch) in the \healpix\ pixelisation at resolution $\nside=512$.  

We consider the same observational set-up discussed in \sectn{\ref{sec:recon:setup}}, focusing on the \fov\ centred on Galactic coordinates $(l,b)=(210^\circ,-20^\circ)$.  The full-sky FDS map and the \fov\ considered are illustrated in \fig{\ref{fig:fds_original}}.
We downgrade the original FDS map to a resolution of $\nside=128$ for our simulated observations, corresponding to a harmonic band-limit of $\elmax=348$.  All other resolution parameters follow once the harmonic band-limit on the sphere and the size of the \fov\ are chosen; we find $\nsphere=28560$, $\bumax=78.4$ and $\nplane = 214 \times 214=45796$.  The size of the Gaussian kernel in the convolutional re-gridding of the projection operation is chosen by the same condition as before, resulting in the value $\sigma_{\rm P} = 0.004$ radians.  Random visibility coverage is again considered, with only 25\% of the discrete visibilities measured.  Note that the incomplete visibility coverage creates a very challenging setting for the recovery of a realistic, diffuse image.

\newlength{\fdsplotwidth}
\setlength{\fdsplotwidth}{74mm}

\begin{figure}
\centering
\subfigure[Mollweide projection of full-sky]{\includegraphics[clip=,viewport=0 0 800 400,width=70mm]{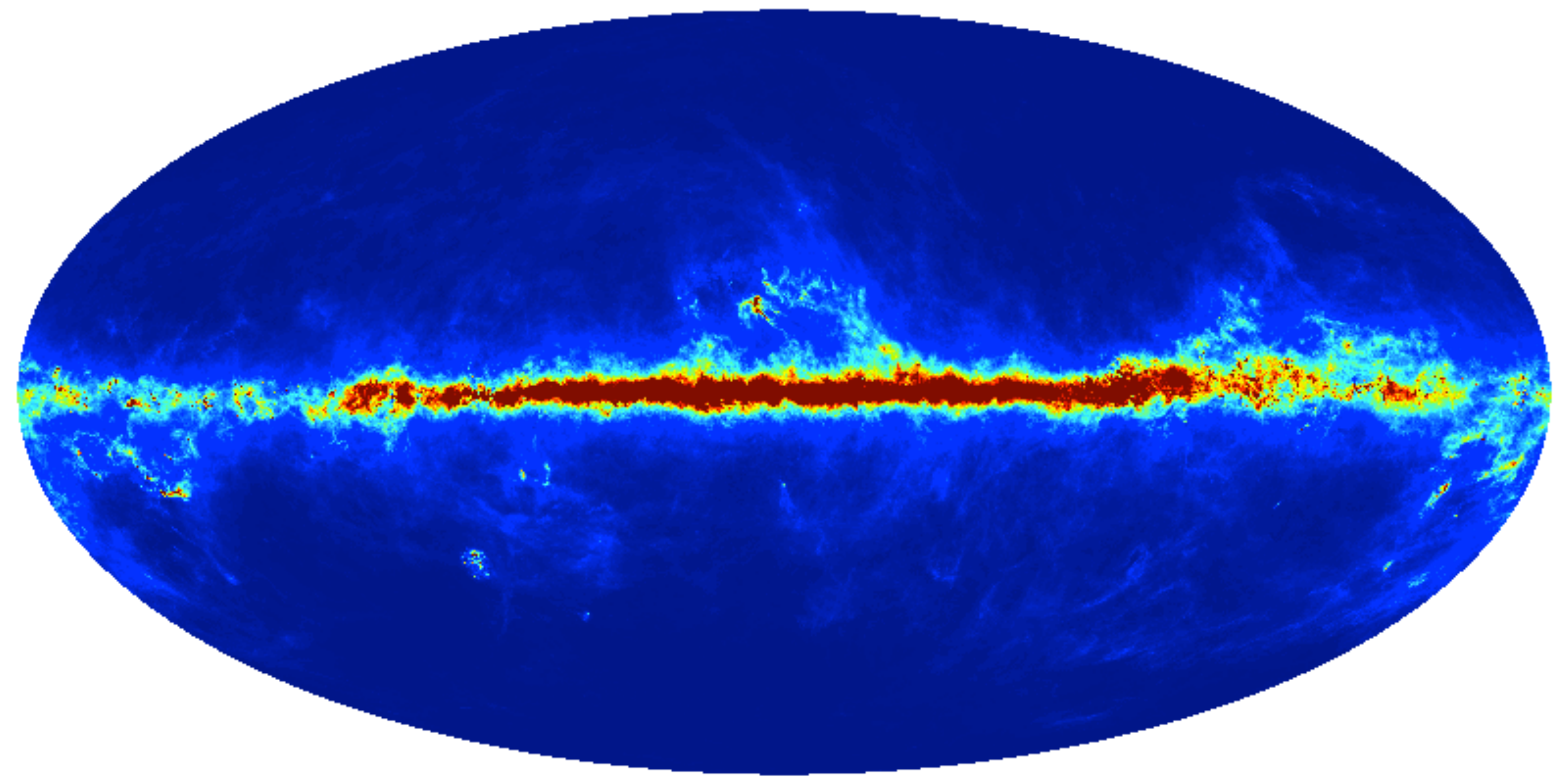}}\\
\subfigure[Orthographic projection of FOV]{\includegraphics[clip=,viewport=-20 30 420 320,width=\fdsplotwidth]{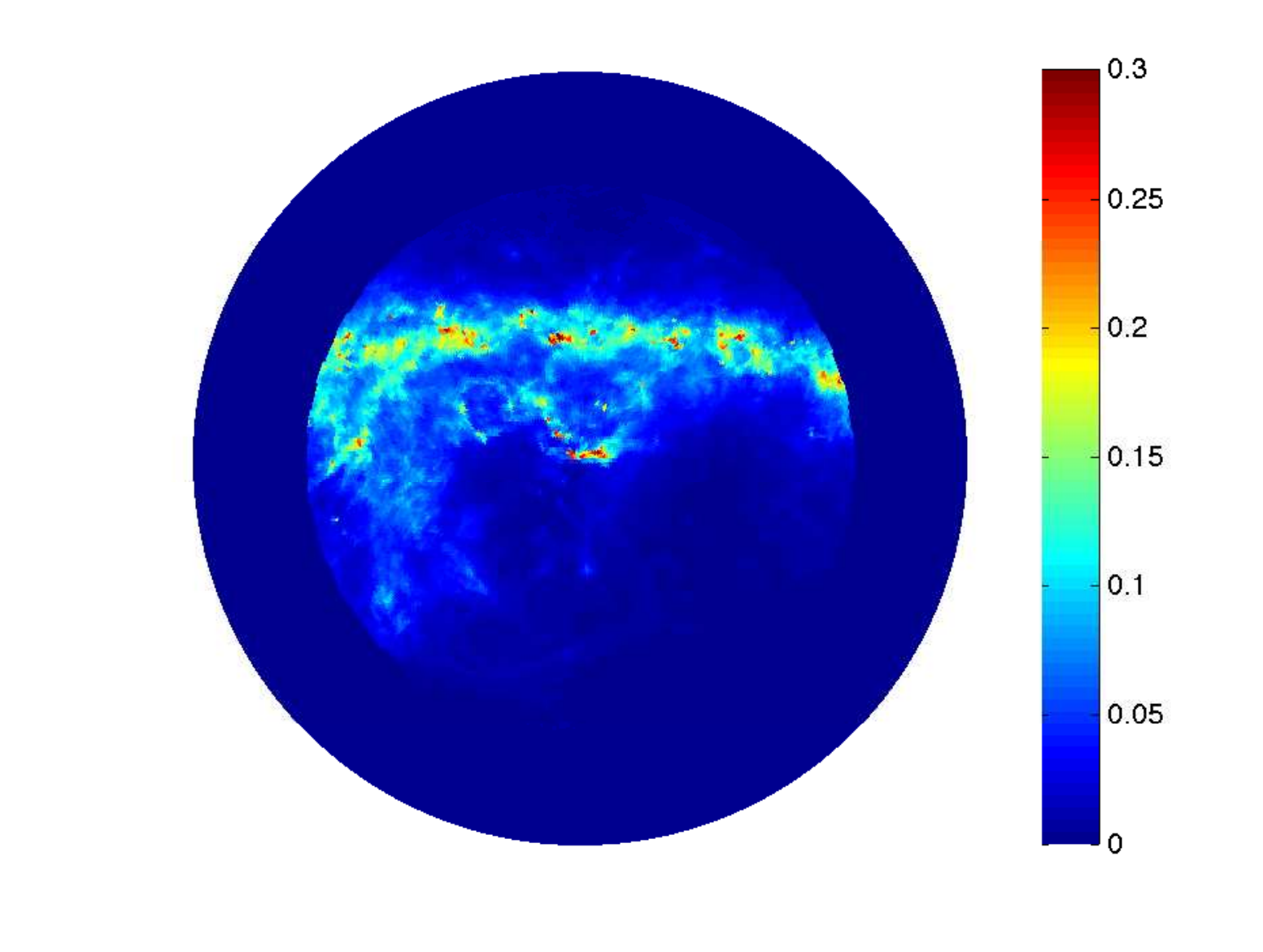}
}
\caption{FDS map of predicted submillimeter and microwave emission of diffuse interstellar Galactic dust.  The \fov\ of angular opening \mbox{$\tfov=90^\circ$} and centred on Galactic coordinates $(l,b)=(210^\circ,-20^\circ)$ is considered for simulating interferometric observations.  In panel~(a) the \fov\ is located near the Galactic plane towards the left edge of the image.
% (viewed from outside celestial sphere)
}
\label{fig:fds_original}
\end{figure}

\subsubsection{Reconstruction performance}

All of the reconstruction techniques discussed previously are applied to reconstruct the FDS map on the sphere (\ie\ BP and TV reconstructions in the spherical and planar settings, both with and without application of the spread spectrum phenomenon).  The same optimisation algorithms and measurement constraint level $\plevel$ considered previously are applied.
For these high-resolution simulations, the computation time required to solve the optimisation problems are typically of the order of 15 minutes on the same machine described previously.  The results presented in \sectn{\ref{sec:recon:gaussians}} indicate that for diffuse signals the \snr\ measured on the sphere (\snrs) is not adversely affected significantly by lifting planar reconstructions to the sphere.  Due to the diffuse nature of the underlying signal (and also since it is inherently defined on the sphere), we therefore measure reconstruction quality by \snrs\ only.\footnote{Although we only quote \snr\ measured on the sphere, the \snr\ measured on the plane (\snrp) were also examined.  Corresponding \snr\ values do change marginally, nevertheless the conclusions drawn remain the same regardless of whether \snrs\ or \snrp\ is examined.}  Note that for such signals the \snr\ metric is not a perfect measure of reconstruction quality and a visual inspection remains important.  The diffuse nature of the FDS map and the results presented previously suggest that the TV reconstruction on the sphere in the presence of spread spectrum phenomenon will be most effective.  

\begin{figure*}
\centering
\mbox{
\subfigure[Planar BP reconstruction with $\bwd=0$ ($\snrs=2.5$dB)%; $\snrp=2.9$dB)
]{\includegraphics[clip=,viewport=-20 30 420 320,width=\fdsplotwidth]{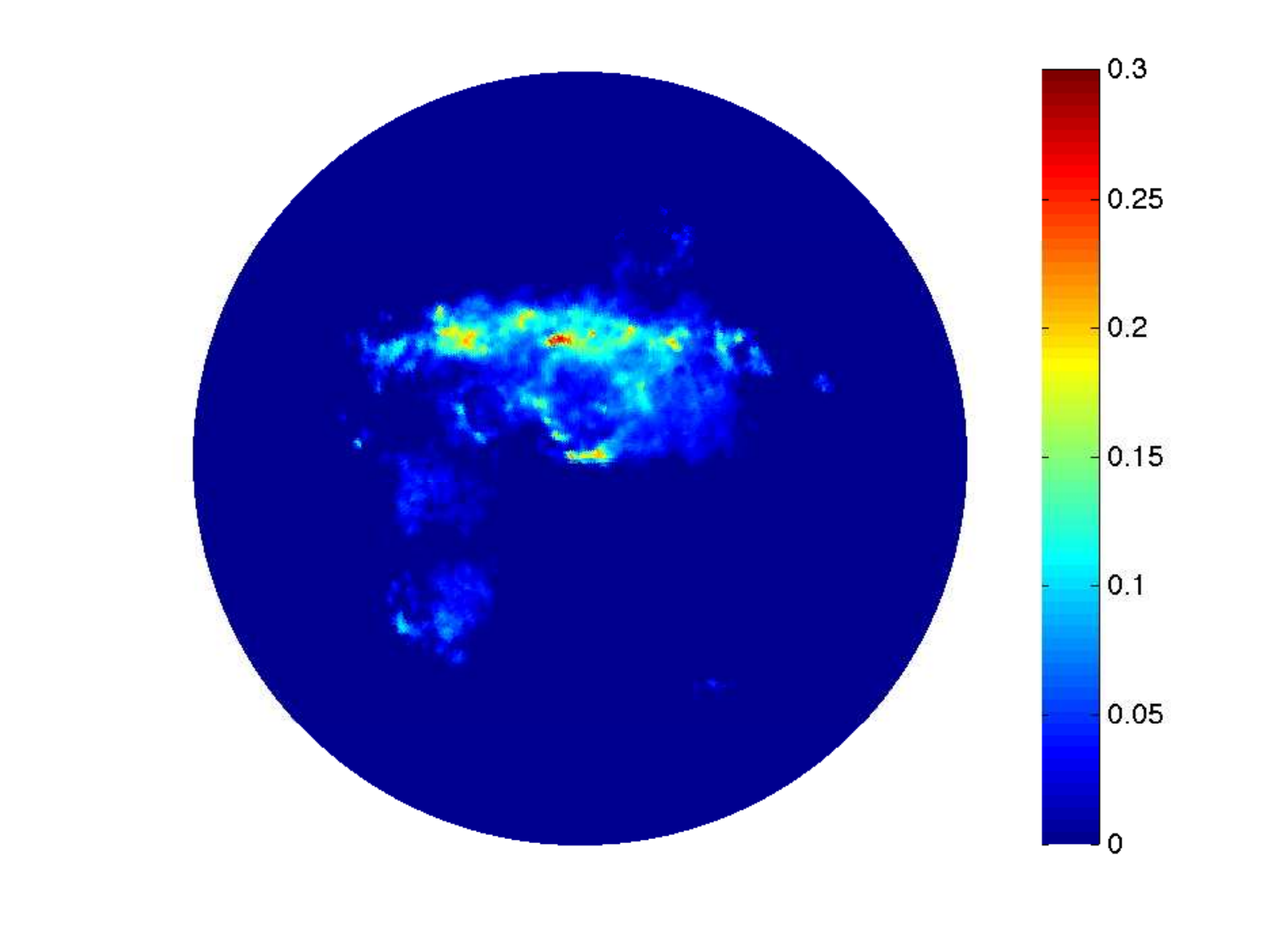}}
\quad
\subfigure[Spherical BP reconstruction with $\bwd=0$ ($\snrs=2.7$dB)%; $\snrp=3.1$dB)
]{\includegraphics[clip=,viewport=-20 30 420 320,width=\fdsplotwidth]{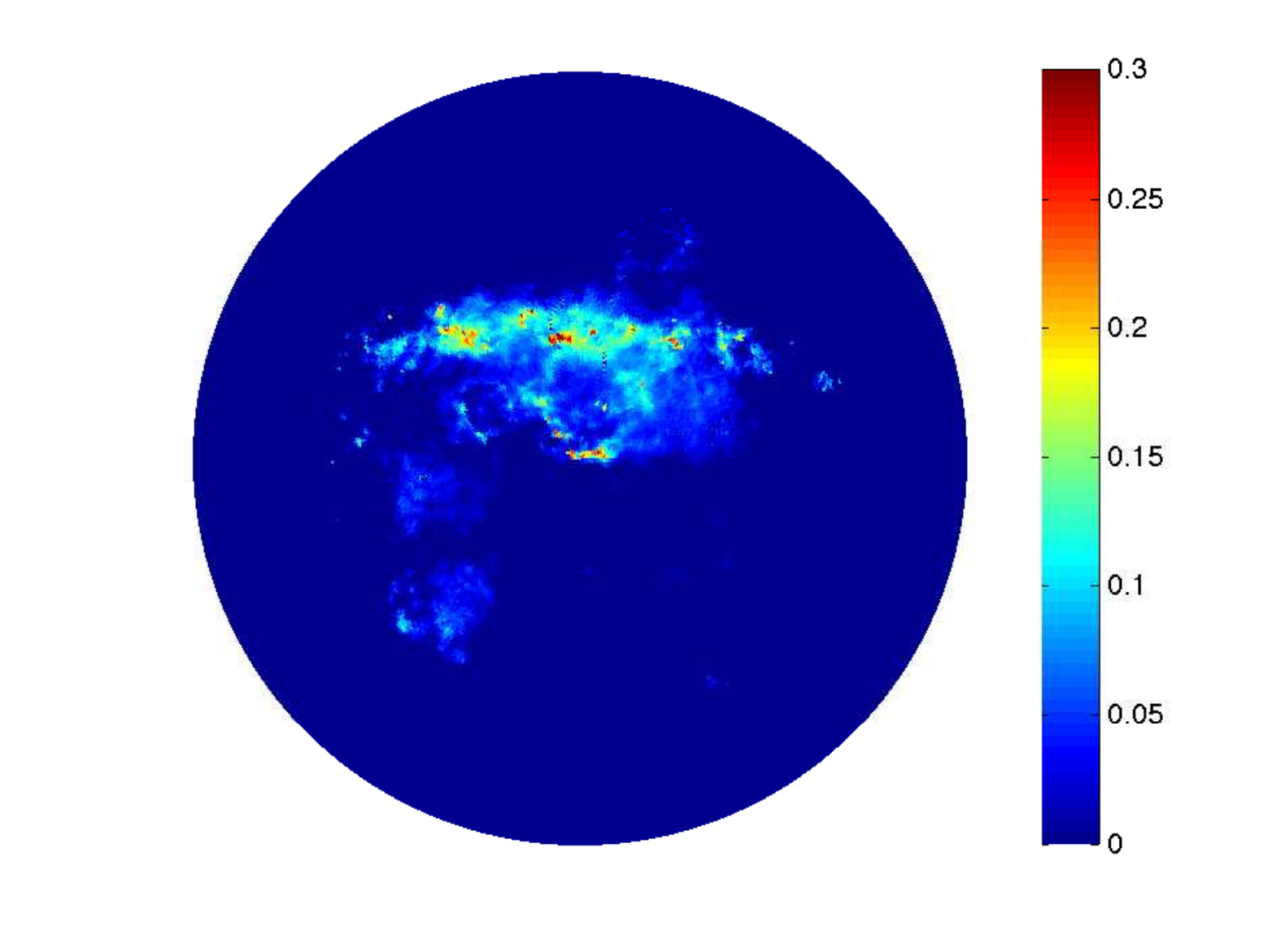}}
}\\
\mbox{
\subfigure[Planar BP reconstruction with $\bwd=\frac{1}{\sqrt{2}}$ ($\snrs=5.3$dB)%; $\snrp=3.6$dB)
]{\includegraphics[clip=,viewport=-20 30 420 320,width=\fdsplotwidth]{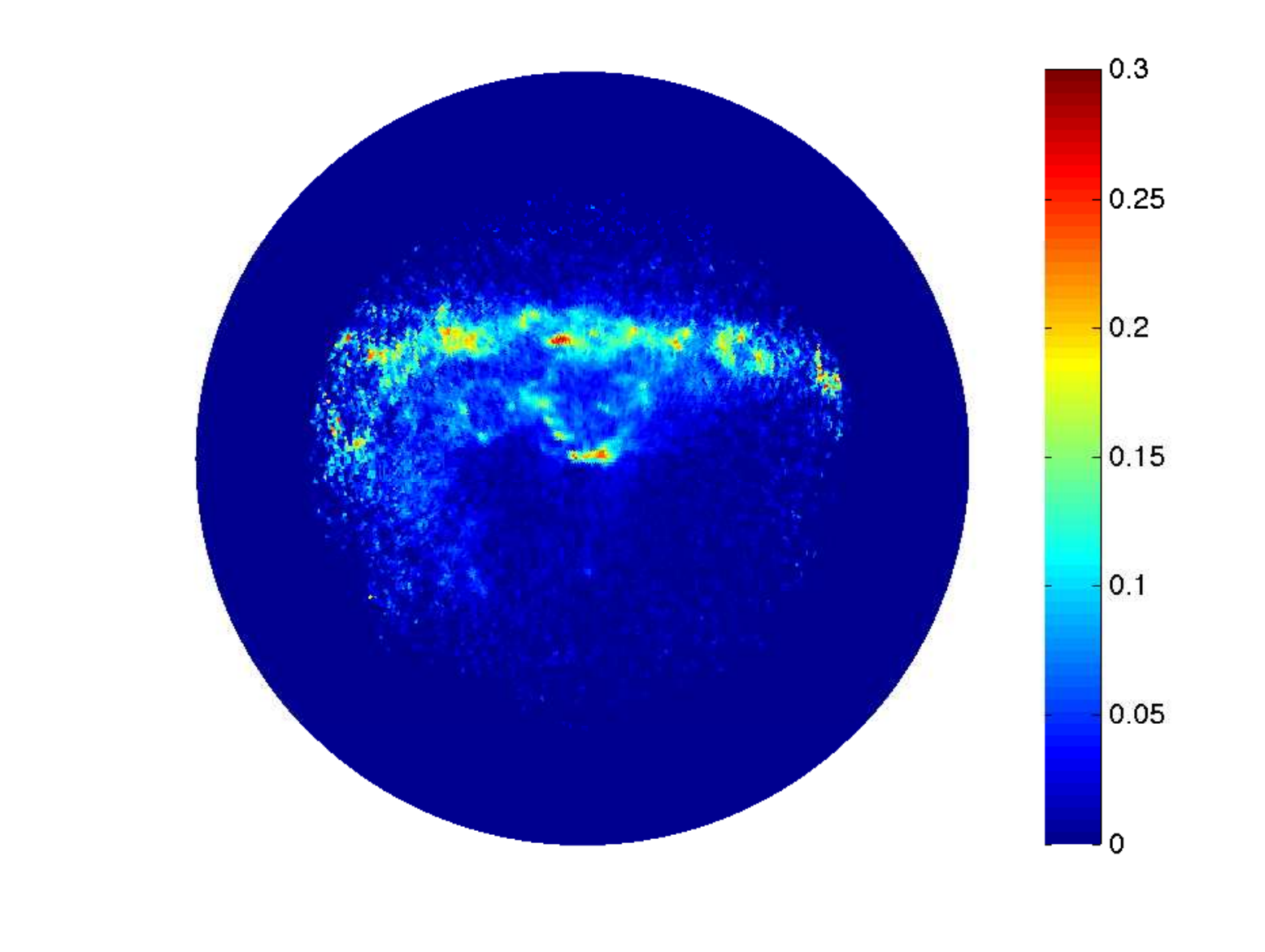}}
\quad
\subfigure[Spherical BP reconstruction with $\bwd=\frac{1}{\sqrt{2}}$ ($\snrs=5.0$dB)%; $\snrp=11.1$dB)
]{\includegraphics[clip=,viewport=-20 30 420 320,width=\fdsplotwidth]{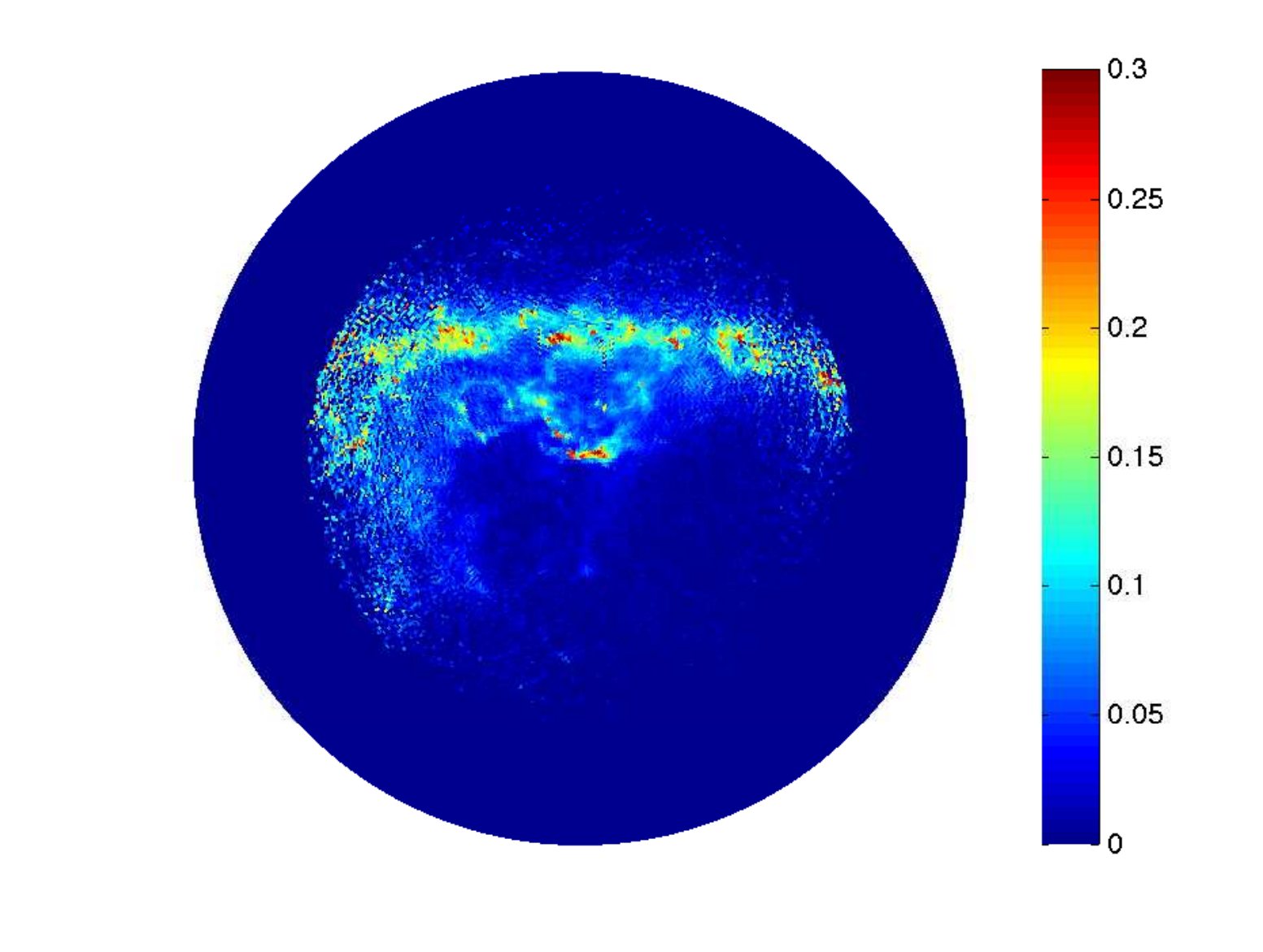}}
}\\
\mbox{
\subfigure[Planar TV reconstruction with $\bwd=0$ ($\snrs=7.4$dB)%; $\snrp=8.1$dB)
]{\includegraphics[clip=,viewport=-20 30 420 320,width=\fdsplotwidth]{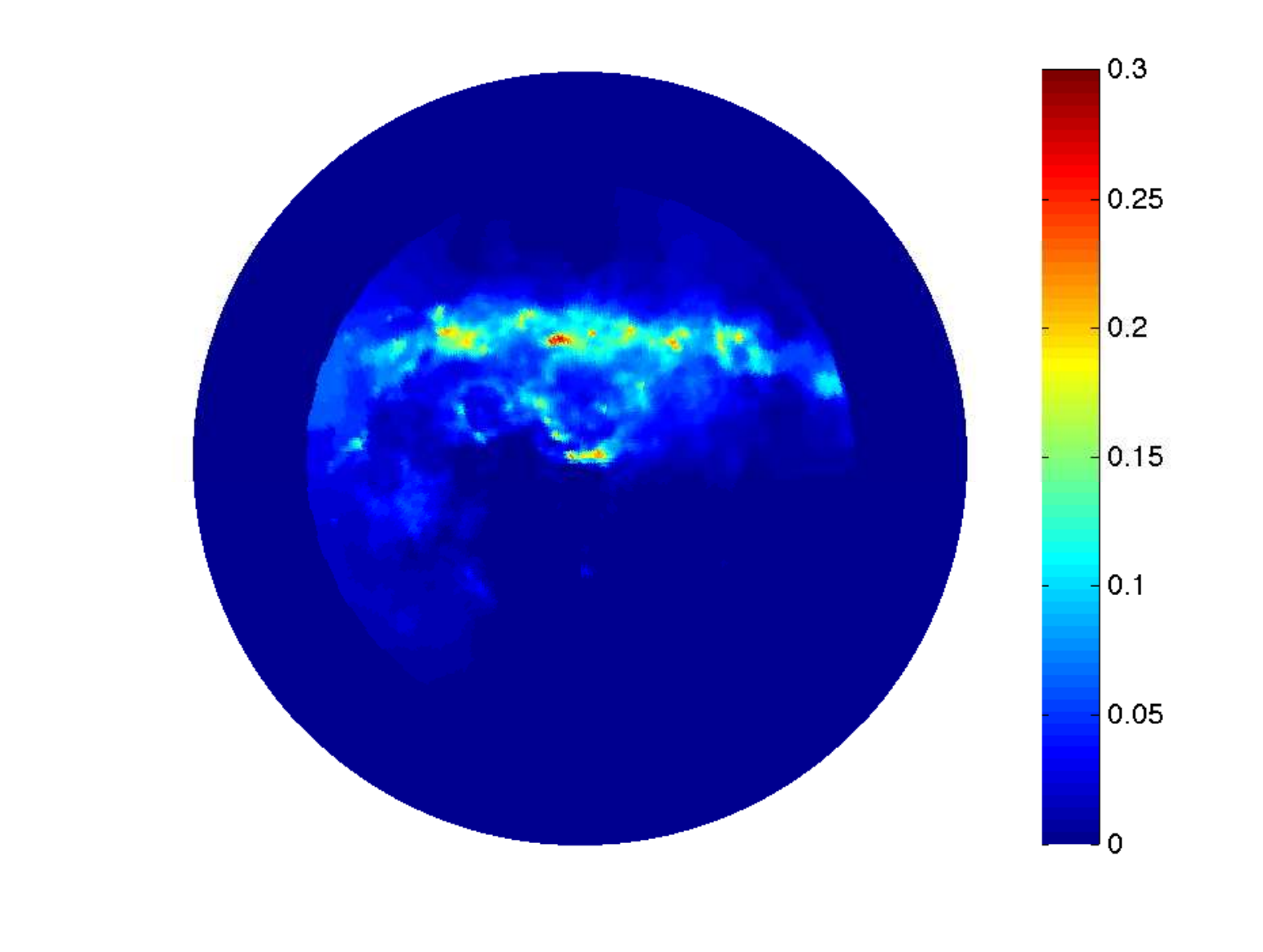}}
\quad
\subfigure[Spherical TV reconstruction with $\bwd=0$ ($\snrs=6.7$dB)%; $\snrp=6.9$dB))
]{\includegraphics[clip=,viewport=-20 30 420 320,width=\fdsplotwidth]{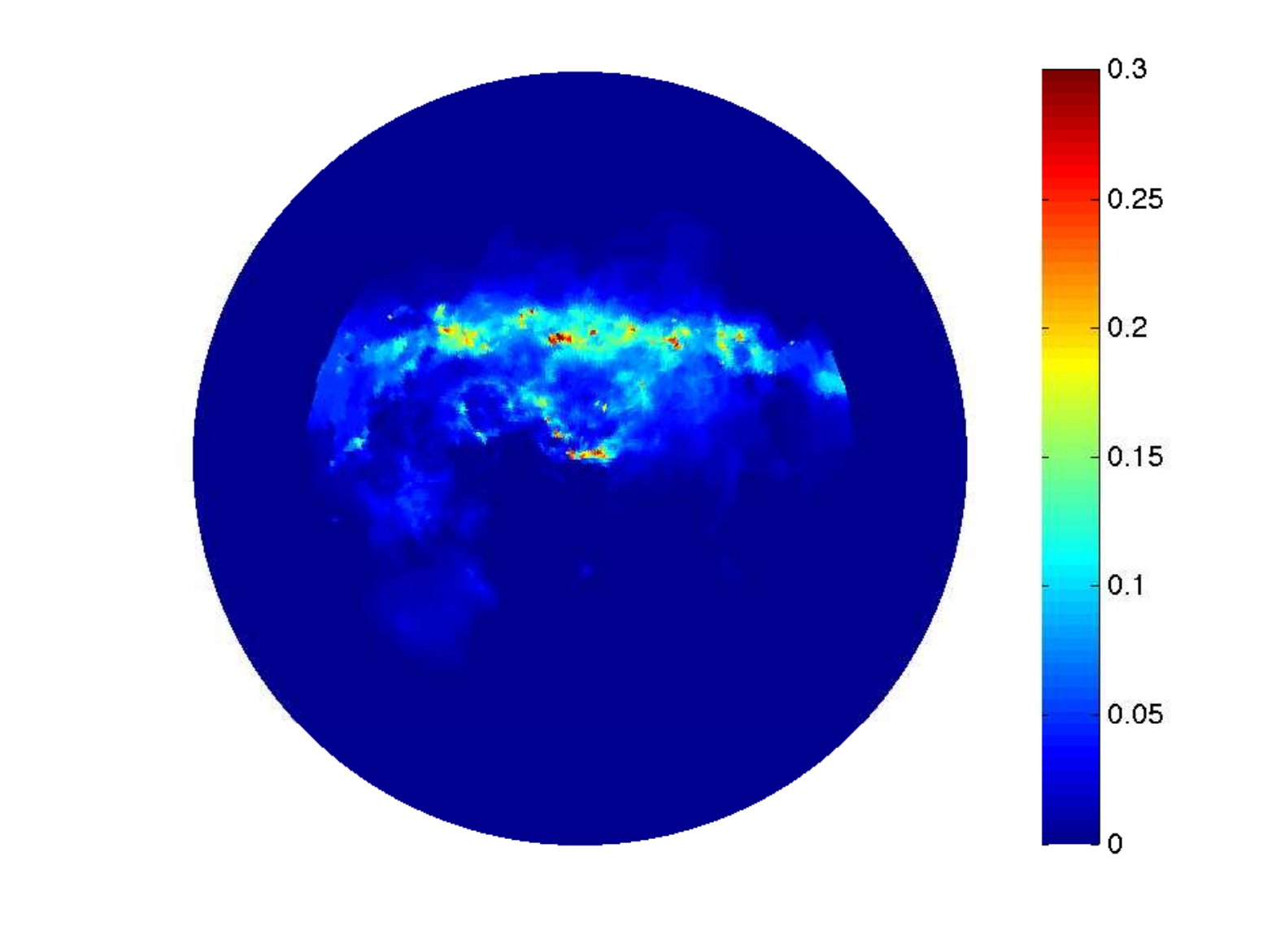}}
}\\
\mbox{
\subfigure[Planar TV reconstruction with $\bwd=\frac{1}{\sqrt{2}}$ ($\snrs=13.7$dB)%; $\snrp=19.5$dB)
]{\includegraphics[clip=,viewport=-20 30 420 320,width=\fdsplotwidth]{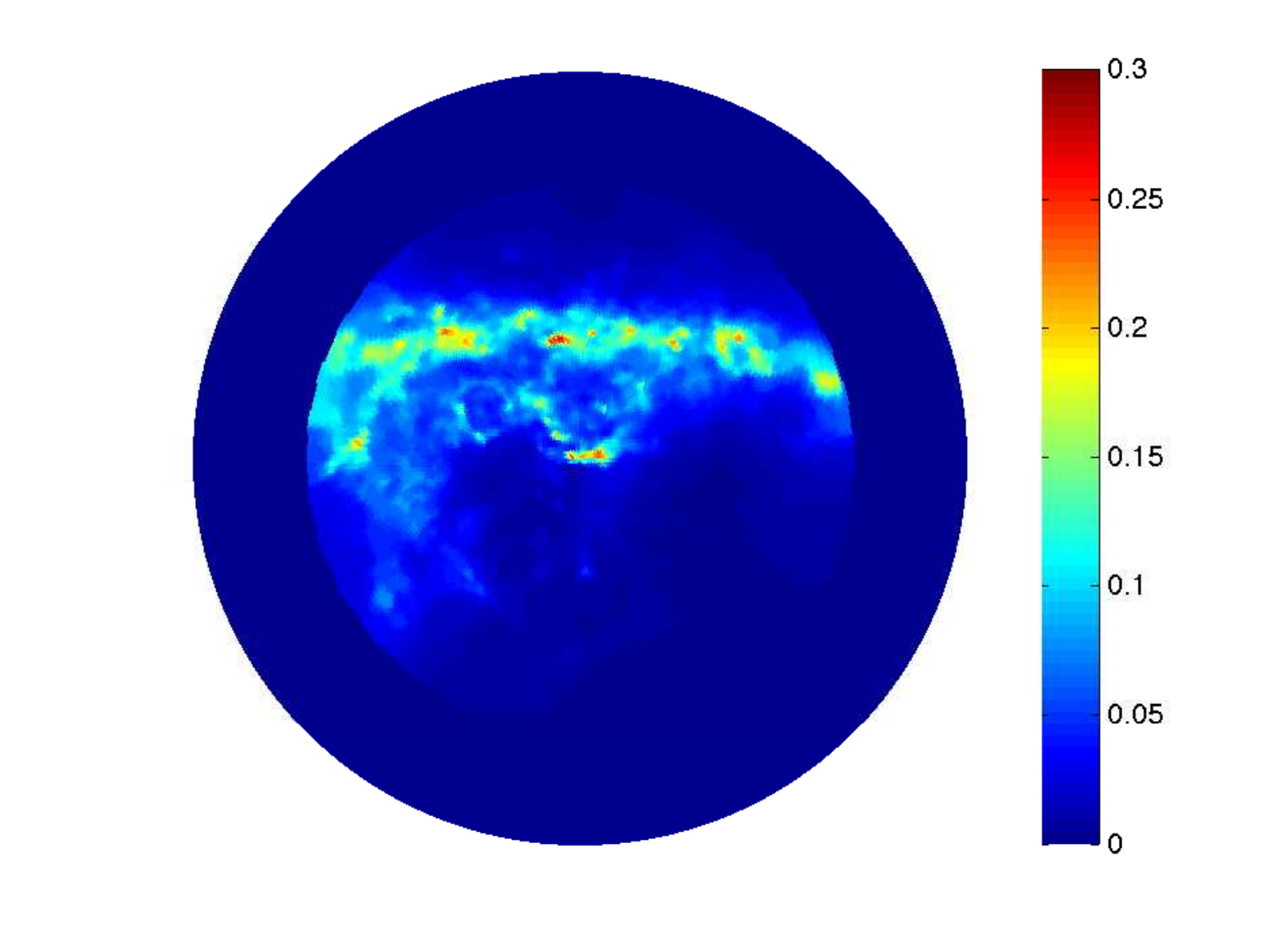}}
\quad
\subfigure[Spherical TV reconstruction with $\bwd=\frac{1}{\sqrt{2}}$ ($\snrs=19.3$dB)%; $\snrp=23.0$dB)
]{\includegraphics[clip=,viewport=-20 30 420 320,width=\fdsplotwidth]{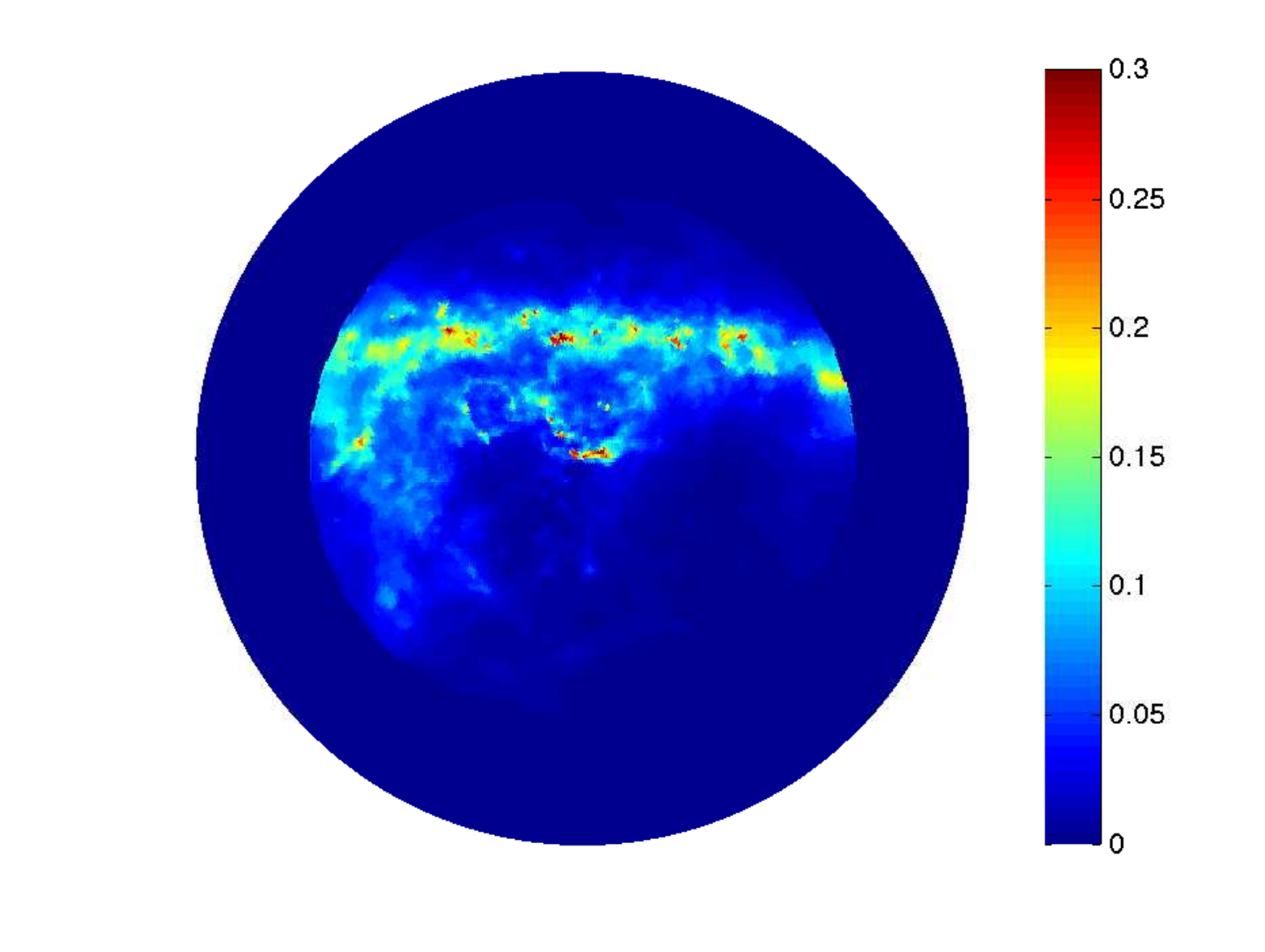}}
}
\caption{Reconstructed spherical interferometric images of the FDS map.  The first column of panels shows images of planar recoveries lifted to the sphere, while the second column shows images recovered on the sphere directly.  Reconstructions both in the presence ($\bwd=1/\sqrt{2}$) and absence ($\bwd=0$) of the spread spectrum phenomenon are displayed.}
\label{fig:fds_recon}
\end{figure*}

Reconstructed spherical interferometric images of the FDS map are displayed in \fig{\ref{fig:fds_recon}}, with \snrs\ of each recovery also specified.  
The quality of BP reconstructions is relatively poor (as expected since the FDS map is not particularly sparse in the Dirac basis), and a significant difference between planar and spherical reconstructions is not readily apparent.  Spherical reconstructions show greater detail than the planar reconstructions, particularly within the centre of the \fov, but suffer near the extremes of the \fov\ since the signal is poorly constrained here due to the low magnitude of the primary beam.  Planar reconstructions also suffer from this effect, however it is not as apparent in reconstructed spherical images since a small degree of smoothing is performed in the projection operator when lifting the reconstructed planar signal to the sphere.
%In fact, in the presence of the spread spectrum phenomenon \snrs\ for the spherical reconstruction is lower than the corresponding value for the planar reconstruction.  This is most likely due to the introduction of a small degree of smoothing of the speckled regions recovered near the extents of the \fov\ when lifting the planar reconstruction to the sphere.  
%
The quality of reconstruction is improved considerably for the TV case, again as expected since the diffuse FDS map is much sparser in the magnitude of its gradient than it is in the Dirac basis.  Spherical reconstructions again show greater detail than the planar reconstructions, however in the absence of the spread spectrum phenomenon \snrs\ is in fact lower for the spherical reconstruction.  On visual inspection the spherical reconstruction is clearly superior as structure of much finer detail is reconstructed, highlighting the weakness of the \snr\ metric.  
%\snrs\ most likely favours the planar reconstruction since large regions of small value are recovered away from the dominant signal content, while these regions are zero in the spherical reconstruction.  
In any case, the superiority of the spherical reconstruction is clearly apparent for the TV reconstruction when the spread spectrum phenomenon is present, both through visual inspection and comparison of \snrs.  This is the most effective reconstruction technique for the FDS map for both planar and spherical reconstructions, as expected.  Comparing the most effective reconstruction on the plane and sphere, recovering the sky intensity directly on the sphere improves reconstruction quality from $13.7$dB to $19.3$dB.

%=============================================================================
\section{Conclusions}
\label{sec:conclusions}
%=============================================================================

Incorporating \wfov\ contributions when reconstructing interferometric images is becoming increasingly important, particularly for next-generation interferometers, such as the SKA and upcoming pathfinder telescopes.  If these contributions are ignored, the fidelity of reconstructed images may suffer.  In this article we have extended the compressed sensing interferometric imaging techniques developed by \citet{wiaux:2009:cs} and \citet{wiaux:2009:ss} to a \wfov.  In doing so, we recover interferometric images defined directly on the celestial sphere.   We augment the usual measurement operator with a projection from the sphere to the plane, which essentially corresponds to a change from Cartesian to spherical coordinates.  Practically, however, the projection is complicated by the discrete setting and careful consideration is given to the sampling of signals on the sphere and plane to ensure that the planar grid is sufficiently sampled to support the maximum projected frequency content of a band-limited signal on the sphere.  Although a projection is incorporated in this framework, it is included in the measurement operator and thus is regularised when solving the interferometric inverse problem.  Moreover, it is the space where recovery is performed that drives the performance of compressed sensing reconstructions, through sparsity and coherence considerations.  The framework remains general and does not rely on any small \fov\ assumptions.  

%By avoiding distorting projections, one would expect the sparsity of a signal to be enhanced by going to the space in which it naturally lives.  This indeed proves to be the case, as we have demonstrated through numerical experiments.  Sparsity is improved by recovering interferometric images directly on the sphere, enhancing the quality of compressed sensing reconstructions.  Coherence, however, is increased by going to the sphere, due to the convolutional re-gridding performed in the projection operator, which acts to reduce the quality of reconstruction.   Since coherence is optimal on the plane but suboptimal on the sphere, the spread spectrum phenomenon is ineffective in improving reconstruction quality in the former setting but effective in the latter.  The higher coherence on the sphere is therefore mitigated by the spread spectrum phenomenon.  Furthermore, the spread spectrum phenomenon is more effective in the \wfov\ setting due to the higher frequency content of the \wterm\ when no small \fov\ assumptions are made.

The effectiveness of the spread spectrum phenomenon is enhanced when going to a \wfov, while sparsity is promoted by recovering images directly on the sphere.  These predictions have been verified by numerical tests and are also manifest in reconstruction performance.  Low-resolution simulations of Gaussian sources were considered to quantify reconstruction performance thoroughly.  Interferometric images were recovered directly on the sphere and the plane in order to make comparisons. For simulated images extremely sparse in the Dirac basis, BP reconstructions were shown to perform very well.  However, as Dirac sparsity was reduced the quality of BP reconstructions fall, while the quality of TV reconstructions remained relatively stable.  For diffuse images, TV reconstructions were shown to be superior since such signals are much more sparse in the magnitude of their gradient than in the Dirac basis.  In all cases, the superior quality of recovering interferometric images directly on the sphere was clear.  A simulation of diffuse interstellar Galactic dust was then performed to demonstrate \wfov\ reconstruction techniques in a more realistic, higher resolution setting.  For the diffuse FDS map considered, TV reconstruction on the sphere in the presence of the spread spectrum phenomenon was most effective, as expected.  For this case, reconstruction quality was improved from 13.7dB for the planar reconstruction to 19.3dB when recovering the interferometric image directly on the sphere.

The compressed sensing techniques developed for interferometric imaging by \citet{wiaux:2009:cs} and \citet{wiaux:2009:ss}, and extended here to a \wfov, remain somewhat idealised.  Random visibility coverage is assumed, with the spatial frequencies probed by the interferometer also assumed to fall on discrete grid points.  Furthermore, to study the spread spectrum phenomenon a constant $\w$ is assumed.  These restrictions have been necessary to remain as close to the theory of compressed sensing as possible during the development and evaluation of interferometric imaging techniques.  In reality, however, \bw\ will not be constant and the performance of the spread spectrum phenomenon will lie between the extreme cases that we have considered of $\bw=0$ and $\bw=\bumax$.  Extensions to realistic and continuous visibility coverage and their impact on compressed sensing based interferometric imaging are now of considerable importance.  In general, compressed sensing addresses imaging by optimising both reconstruction and acquisition, while we have essentially focused on reconstruction only.  The possibility of optimising the configuration of interferometers to enhance the spread spectrum phenomenon for compressed sensing reconstruction is an exciting avenue of research at the level of acquisition.  In addition, direction dependent beam effects may also provide an alternative source of the spread spectrum phenomenon.  All of these issues should be studied in future works. 
Furthermore, the performance of other sparsity bases on the sphere, such as scale discretised wavelets \citep{wiaux:2007:sdw}, should also be studied. 
Next-generation radio interferometers will inherently observe very large fields of view, thus \wfov\ interferometric imaging techniques, such as the compressed sensing techniques developed in this article and the future research outlined here, are of increasing importance to ensure that the fidelity of reconstructed images keeps pace with the capabilities of new instruments.

%=============================================================================
\section*{Acknowledgements}
%=============================================================================

We thank Gilles Puy for providing some code and for useful discussions.  We also thank Pierre Vandergheynst, Jean-Philippe Thiran and Dimitri Van De Ville for providing the infrastructure to support our research.  JDM is supported by the Swiss National Science Foundation (SNSF) under grant 200021-130359.  YV is supported in part by the Center for Biomedical Imaging (CIBM) of the Geneva and Lausanne Universities, EPFL, and the Leenaards and Louis-Jeantet foundations, and in part by the SNSF under grant PP00P2-123438.  We acknowledge the use of the \comb\ \citep{mcewen:2006:filters}, \stwo\ \citep{mcewen:2006:fcswt} and \healpix\ \citep{gorski:2005} packages.  We also acknowledge the use of the \lambdaarchtext\ (\lambdaarch).  Support for \lambdaarch\ is provided by the NASA Office of Space Science.

%=============================================================================
% Bibliography
\bibliographystyle{mymnras_eprint}
\bibliography{bib}

%=============================================================================
\label{lastpage}
\end{document}